\documentclass[reqno,11pt]{amsart}
\usepackage[utf8]{inputenc}
\usepackage{graphicx}
\usepackage{slashed}
\usepackage{amscd}
\usepackage{amssymb}
\usepackage{esint}
\usepackage{enumitem}
\usepackage{comment}
\usepackage{amsmath}
\usepackage{dsfont}
\usepackage[mathscr]{eucal}
\usepackage[linktocpage]{hyperref}
\usepackage{pstricks}
\usepackage{pst-all}

\numberwithin{equation}{section}
\allowdisplaybreaks[1]
\definecolor{labelkey}{gray}{.65}

\title[A Positive Quasilocal Mass for Causal Variational Principles]{A Positive Quasilocal Mass for \\
Causal Variational Principles}

\author[F.\ Finster]{Felix Finster}
\address{Fakult\"at f\"ur Mathematik \\ Universit\"at Regensburg \\ D-93040 Regensburg \\ Germany}
\email{finster@ur.de}

\author[N.\ Kamran]{Niky Kamran \\ \\ October 2023}
\address{Department of Mathematics and Statistics \\ McGill University \\ Montr{\'e}al \\ Canada}
\email{niky.kamran@mcgill.ca}

\newtheorem{Def}{Definition}[section]
\newtheorem{Thm}[Def]{Theorem}
\newtheorem{Prp}[Def]{Proposition}
\newtheorem{Lemma}[Def]{Lemma}
\newtheorem{Remark}[Def]{Remark}
\newtheorem{Corollary}[Def]{Corollary}

\newcommand{\Thanks}{\vspace*{.5em} \noindent \thanks}
\newcommand{\beq}{\begin{equation}}
\newcommand{\eeq}{\end{equation}}
\newcommand{\Proof}{\begin{proof}}
	\newcommand{\QED}{\end{proof} \noindent}
\newcommand{\QEDrem}{\ \hfill $\Diamond$}

\newcommand{\la}{\langle}
\newcommand{\ra}{\rangle}

\newcommand{\C}{\mathbb{C}}
\newcommand{\R}{\mathbb{R}}

\newcommand{\N}{\mathbb{N}}

\DeclareMathOperator{\tr}{tr}
\renewcommand{\O}{{\mathscr{O}}}
\renewcommand{\L}{{\mathcal{L}}}
\newcommand{\Sact}{{\mathcal{S}}}

\newcommand{\Dir}{{\mathcal{D}}}

\newcommand{\D}{\mathscr{D}}

\DeclareMathOperator{\supp}{supp}

\newcommand{\scrM}{\mycal M}

\newcommand{\scrN}{\mycal N}

\newcommand{\J}{\mathfrak{J}}
\newcommand{\s}{\mathfrak{s}}
\newcommand{\Jdiff}{\mathfrak{J}^\text{\rm{\tiny{diff}}}}
\newcommand{\Jtest}{\mathfrak{J}^\text{\rm{\tiny{test}}}}
\newcommand{\Jvary}{\mathfrak{J}^\text{\rm{\tiny{vary}}}}

\newcommand{\Jlin}{\mathfrak{J}^\text{\rm{\tiny{lin}}}}
\newcommand{\Gdiff}{\Gamma^\text{\rm{\tiny{diff}}}}
\newcommand{\Gtest}{\Gamma^\text{\rm{\tiny{test}}}}
\newcommand{\Ctest}{C^\text{\rm{\tiny{test}}}}

\newcommand{\macro}{\text{\tiny{\rm{macro}}}}
\renewcommand{\u}{\mathfrak{u}}
\renewcommand{\v}{\mathfrak{v}}

\newcommand{\bu}{{\mathbf{u}}}
\newcommand{\bv}{\mathbf{v}}

\newcommand{\bitem}{\begin{itemize}[leftmargin=2em]}
\newcommand{\eitem}{\end{itemize}}
\renewcommand{\div}{{\rm{div}}\,}

\newcommand{\G}{{\mathscr{G}}}
\newcommand{\x}{\mathbf{x}}
\newcommand{\y}{\mathbf{y}}
\newcommand{\z}{\mathbf{z}}
\newcommand {\bxi}{\boldsymbol\xi}
\newcommand {\bzeta}{\boldsymbol\zeta}

\newcommand{\Mass}{{\mathfrak{M}}}

\DeclareFontFamily{OT1}{rsfso}{}
\DeclareFontShape{OT1}{rsfso}{m}{n}{ <-7> rsfso5 <7-10> rsfso7 <10-> rsfso10}{}
\DeclareMathAlphabet{\mycal}{OT1}{rsfso}{m}{n}

\newcommand\Niky[1]{}
\newcommand\Felix[1]{}

\begin{document}
\maketitle
\begin{abstract}
A new inequality for a nonlinear surface layer integral is proved for minimizers of causal
variational principles. This inequality is applied to obtain a new proof of the positive mass theorem with volume
constraint. Next, a positive mass theorem without volume
constraint is stated and proved by introducing and using the concept of asymptotic alignment.
Moreover, a positive quasilocal mass and a synthetic definition of scalar curvature are introduced
in the setting of causal variational principles. Our notions and results are illustrated by the explicit examples of
causal fermion systems constructed in ultrastatic spacetimes and the Schwarzschild spacetime.
In these examples, the correspondence to the ADM mass and similarities to the Brown-York mass are worked out.
\end{abstract}

\tableofcontents

\section{Introduction} \label{secintro}
The theory of {\em{causal fermion systems}} is a recent approach to fundamental physics
where spacetime is no longer modelled by a Lorentzian manifold but may instead have a
more general, possibly discrete structure on a microscopic length scale (which can be thought of
as the Planck scale). In the setting of causal fermion systems, the physical equations are formulated
via a variational principle, the {\em{causal action principle}}
(for the general context see the reviews~\cite{nrstg, dice2014}, the textbooks~\cite{cfs, intro}
or the website~\cite{cfsweblink}).
{\em {Causal variational principles}} were introduced in~\cite{continuum}
as a mathematical generalization of the causal action principle.
From the mathematical perspective, causal fermion systems and causal variational principles
are of interest because they provide a setting for describing and analyzing non-smooth geometries
and singular spaces (see~\cite{topology, lqg} for more details on different aspects).

In general terms, given a manifold~$\G$ together with a non-negative function~$\L : \G \times \G \rightarrow \R^+_0$,
in a causal variational principle one minimizes the action~$\Sact$ defined by
\[ %\label{Sactintro} 
\Sact (\mu) = \int_\G d\mu(x) \int_\G d\mu(y)\: \L(x,y) \]
under variations of the measure~$\mu$ on~$\G$, keeping the total volume~$\mu(\G)$ fixed
(for the precise mathematical setup see Section~\ref{seccvp} below).
The support of the measure~$\mu$ denoted by
\beq \label{Ndef}
N := \supp \mu \subset \G
\eeq
has the interpretation as the underlying {\em{space}} or {\em{spacetime}}.

In order to clarify the nature of the interaction described by the causal variational principle, it is an important task
to investigate whether quantities which are defined geometrically and which have been extensively studied in the context of classical general relativity and differential geometry
can be formulated and analyzed in the broader setting of causal variational
principles. In~\cite{pmt} a notion of {\em{total mass}}~$\Mass^\text{\rm{tot}}$ was introduced for a class of
causal variational principles which include the causal action principle for static causal fermion systems.
To this end, one considers two critical measures~$\mu$ (describing the vacuum) and~$\tilde{\mu}$
(describing the gravitating system) and takes the limit of double integrals
\begin{align}
\Mass^\text{\rm{tot}} = &\lim_{\Omega \nearrow N, \; \tilde{\Omega} \nearrow \tilde{N}} \notag \\
&\; \bigg(
\int_{\tilde{\Omega}} d\tilde{\mu}(\x) \int_{N\setminus \Omega} d\mu(\y)\: \L_\kappa(\x,\y)
- \int_{\Omega} d\mu(\x) \int_{\tilde{N} \setminus \tilde{\Omega}} d\tilde{\mu}(\y)\: \L_\kappa(\x,\y) \bigg) \:,
\label{massintro}
\end{align}
where~$\Omega \nearrow N$ and~$\tilde{\Omega} \nearrow \tilde{N}$ denote exhaustions
under the
\beq \label{volconstraintintro}
\text{\em{volume constraint}} \qquad
\mu(\Omega) = \tilde{\mu}(\tilde{\Omega}) < \infty \:.
\eeq
The limit in~\eqref{massintro} is well-defined provided that the measures~$\mu$ and~$\tilde{\mu}$
are {\em{asymptotically close}} (for the precise definition and more details see Section~\ref{sectotmass}).
Moreover, a {\em{positive mass theorem}} was proven which states that the total mass is non-negative,
provided that a suitable local energy condition holds. The proof of this theorem
was inspired by and bears some similarity with
Witten's spinor proof of the positive mass theorem~\cite{witten, parker+taubes}.
Moreover, in~\cite{pmt} it was shown how the ADM mass is recovered in a limiting case in which the measure~$\tilde{\mu}$
is constructed from a static asymptotically flat Lorentzian spacetime.

In the present paper we shall improve and generalize the concepts, methods and results in~\cite{pmt}
in various ways. A key ingredient is a novel {\em{positivity argument for a surface layer integral}} (see Section~\ref{secposnonlin}). In the simplest case, our positivity argument yields that,
again for critical measures~$\mu$ and~$\tilde{\mu}$ and any subsets~$\tilde{\Omega} \subset N$
and~$\Omega \subset N$ satisfying the volume constraint~\eqref{volconstraintintro},
the following combination of double integrals is non-negative,
\beq \label{pososiintro}
\begin{split}
0 &\leq \Mass_{\tilde{\mu}, \mu} \big( \tilde{\Omega}, \Omega \big) := 2 \int_{\tilde{\Omega}} d\tilde{\mu}(\x) \int_{N \setminus \Omega} d\mu(\y)\: \L(\x,\y) \\
& \qquad - \int_{\tilde{\Omega}} d\tilde{\mu}(\x) \int_{\tilde{N} \setminus \tilde{\Omega}} d\tilde{\mu}(\y)\: \L(\x,\y)
- \int_\Omega \mu(\x) \int_{N \setminus \Omega} d\mu(\y)\: \L(\x,\y) \:.
\end{split} 
\eeq
Asymptotically as~$\tilde{\Omega} \nearrow \tilde{N}$ and~$\Omega \nearrow N$,
this goes over to~\eqref{massintro}, giving a new and much simpler proof of the
above-mentioned positive mass theorem (see Section~\ref{secnewproof}).
It is remarkable that this new proof does not require a local energy condition.
Instead, it is a general consequence of the assumption that the vacuum measure~$\mu$
is a minimizer of the causal variational principle.

Another main concern of the present paper is the proper {\em{treatment of
the volume constraint}}~\eqref{volconstraintintro}. Our motivation comes from the fact that this
volume constraint is not fully convincing because it
has no counterpart in general relativity, where the ADM mass is defined purely in terms of the asymptotic
geometry at infinity. This raises the important question of if and how the volume constraint can
be removed. In order to address these questions, we introduce the concept of
{\em{alignment}}. The idea is to remove the freedom in identifying~$N$ and~$\tilde{N}$
by imposing a condition which can be evaluated locally in a neighborhood of each point~$\bzeta \in N$
(see~\eqref{asyalign} in Definition~\ref{defasyalign}).
It turns out that the resulting identification of~$\Omega$ with~$\tilde{\Omega}$
satisfies the volume inequality~$\tilde{\mu}(\tilde{\Omega}) \geq \mu(\Omega)$,
provided that the so-called {\em{local volume condition}} holds (see Definition~\ref{deflvc}).
In this way, we obtain a positive mass theorem without volume constraint (see Theorem~\ref{pmt}).
As a further generalization, we define an {\em{equivariant}} positive mass theorem,
where we minimize over a group~${\mathcal{G}}$ of diffeomorphisms acting on~$\G$
which describe symmetries of the Lagrangian
(see Definition~\ref{defmassiso} and Theorem~\ref{pmtequi}).

Starting from the above concepts and results, we also succeed in introducing a {\em{quasilocal mass}}.
Given a subset~$\tilde{\Omega} \subset \tilde{N}$, it is defined by 
\[ \Mass(\tilde{\Omega}) := \inf_{\Phi, \Omega} \Mass_{\tilde{\mu}, \Phi_* \mu}(\tilde{\Omega}, \Omega) \:, \]
where we take the infimum over symmetry transformations~$\Phi \in {\mathcal{G}}$ and
subsets~$\Omega \subset \Phi(N)$ which must satisfy certain admissibility conditions
(see Definition~\ref{defquasilocal}). This quasilocal mass is shown to be non-negative
(Theorem~\ref{thmquasipos}). Moreover, it bounds the total mass from below (see Theorem~\ref{thmlower}).
In the example of causal fermion systems constructed in ultrastatic spacetimes,
our quasilocal mass has surprising similarities with the Brown-York mass~\cite{brown-york, shi-tam, liu-yau}
(see Theorem~\ref{thmquasilocal}).
Finally, we propose a notion of {\em{synthetic scalar curvature}} (see Definition~\ref{defsynthetic}).
Again in the example of ultrastatic spacetimes, this definition reduces, up to a constant,
to the scalar curvature of the spatial metric (see Theorem~\ref{thmsynthetic}).
Our findings and results are illustrated by detailed computations and examples
in ultrastatic spacetimes and the Schwarzschild spacetime.

A major advantage of describing the concept of mass in the setting of causal variational principles
is that the {\em{regularity and smoothness assumptions can be weakened}} drastically.
Indeed, the definition of mass~\eqref{massintro} does not require the spaces~$N, \tilde{N} \subset \G$
to be smooth or even topological manifolds. Instead, all we need are certain boundedness and decay
assumptions on the measures~$\mu$ and~$\tilde{\mu}$ near infinity
(for details see Section~\ref{sectotmass}).
Clearly, in order to compare our notions to the classical concepts of total and quasilocal mass,
we need to go back to the smooth setting. Moreover, local notions like the alignment, the local volume condition
and scalar curvature only make sense in smooth spaces.
For clarity in presentation, we begin as general as possible and specify the regularity and smoothness
assumptions on the way whenever needed.
More precisely, starting from Section~\ref{secnovol} we assume that the vacuum~$N$ is smooth
and translation invariant (see~\eqref{Nsmooth}) and that the gravitating spacetime~$\tilde{N}$
has a smooth manifold structure in the asymptotic end.
Beginning from Section~\ref{seclvc}, we need to assume that~$\tilde{N}$ has a smooth manifold structure
everywhere. An exception is the bound of the total mass in terms of the quasilocal mass
(see Theorem~\ref{thmlower}), were it suffices to assume smoothness in the exterior region~$\tilde{N}
\setminus \tilde{U}$.

The paper is organized as follows. After the necessary preliminaries on causal variational principles
and the total mass (Section~\ref{secprelim}), the general positivity statement for a nonlinear
surface layer integral is introduced in various versions (Section~\ref{secposnonlin}).
As an application, we give the new proof of the positive mass theorem (Section~\ref{secnewproof}).
Then the volume constraint is removed by working with asymptotic alignment (Section~\ref{secnovol}).
An equivariant version of the positive mass theorem is stated and proven, where we
minimize over a group of isometries of the Lagrangian (Section~\ref{secequi}).
We proceed by defining a quasilocal mass (Section~\ref{secquasilocal})
and studying its relation to the total mass (Section~\ref{secbounds}).
Moreover, a notion of {\em{synthetic scalar curvature}} is introduced (Section~\ref{secsynthetic}).
Finally, in Section~\ref{secultra} it is worked out that, for causal fermion systems constructed
in ultrastatic spacetimes, our synthetic scalar curvature reduces to the scalar curvature of the Riemannian
metric. Moreover, we work out an interesting similarity of our quasilocal mass with the Brown-York mass.
We conclude the paper with detailed computations in the example
of causal fermion systems in the Schwarzschild spacetime (Appendix~\ref{appA}).

\section{Preliminaries} \label{secprelim}
This section provides the necessary background on causal variational principles and the
definition of the total mass. More details can be found in~\cite{pmt}.
\subsection{Causal Variational Principles in the Non-Compact Setting} \label{seccvp}
We consider causal variational principles in the non-compact setting as
introduced in~\cite[Section~2]{jet}. Thus we let~$\G$ be a (possibly non-compact)
smooth manifold of dimension~$m \geq 1$
and~$\mu$ a (positive) Borel measure on~$\G$ (the {\em{universal measure}}).
Moreover, we are given a non-negative function~$\L : \G \times \G \rightarrow \R^+_0$
(the {\em{Lagrangian}}) with the following properties:
\begin{itemize}[leftmargin=2em]
\item[(i)] $\L$ is symmetric: $\L(\x,\y) = \L(\y,\x)$ for all~$\x,\y \in \G$.\label{Cond1}
\item[(ii)] $\L$ is lower semi-continuous, i.e.\ for all sequences~$\x_n \rightarrow \x$ and~$\y_{n'} \rightarrow \y$,
\label{Cond2}%
\[ \L(\x,\y) \leq \liminf_{n,n' \rightarrow \infty} \L(\x_n, \y_{n'})\:. \]
\end{itemize}
The {\em{causal variational principle}} is to minimize the action
\beq \label{Sact} 
\Sact (\mu) = \int_\G d\mu(\x) \int_\G d\mu(\y)\: \L(\x,\y) 
\eeq
under variations of the measure~$\mu$, keeping the total volume~$\mu(\G)$ fixed
({\em{volume constraint}}).

If the total volume~$\mu(\G)$ is finite, one minimizes~\eqref{Sact}
over all regular Borel measures with the same total volume.
If the total volume~$\mu(\G)$ is infinite, however, it is not obvious how to implement the volume constraint,
making it necessary to proceed as follows.
We make the following additional assumptions:
\begin{itemize}[leftmargin=2em]
\item[(iii)] The measure~$\mu$ is {\em{locally finite}}
(meaning that any~$\x \in \G$ has an open neighborhood~$U$ with~$\mu(U)< \infty$)
and {\em{regular}} (meaning that the measure of a set can be recovered by approximation from inside
with compact and from outside with open sets). \label{Cond3}
\item[(iv)] The function~$\L(\x,.)$ is $\mu$-integrable for all~$\x \in \G$, giving
a lower semi-continuous and bounded function on~$\G$. \label{Cond4}
\end{itemize}
Given a regular Borel measure~$\mu$ on~$\G$, we vary over all
regular Borel measures~$\tilde{\mu}$ with
\beq \label{totvol}
\big| \tilde{\mu} - \mu \big|(\G) < \infty \qquad \text{and} \qquad
\big( \tilde{\mu} - \mu \big) (\G) = 0
\eeq
(where~$|.|$ denotes the total variation of a measure).
We then consider the difference of the actions defined by
\beq \label{integrals}
\begin{split}
\big( &\Sact(\tilde{\mu}) - \Sact(\mu) \big) := \int_\G d(\tilde{\mu} - \mu)(\x) \int_\G d\mu(\y)\: \L(\x,\y) \\
&\quad + \int_\G d\mu(\x) \int_\G d(\tilde{\mu} - \mu)(\y)\: \L(\x,\y) 
+ \int_\G d(\tilde{\mu} - \mu)(\x) \int_\G d(\tilde{\mu} - \mu)(\y)\: \L(\x,\y) \:.
\end{split}
\eeq
The measure~$\mu$ is said to be a {\em{minimizer}} of the causal action
with respect to {\em{variations of finite volume}}
if this difference is non-negative for all~$\tilde{\mu}$ satisfying~\eqref{totvol},
\beq \label{Sdiffpos}
\big( \Sact(\tilde{\mu}) - \Sact(\mu) \big) \geq 0 \:.
\eeq
These variations of the causal action are well-defined.
The existence theory for minimizers is developed in~\cite{noncompact}.
It is shown in~\cite[Lemma~2.3]{jet} that a minimizer
satisfies the {\em{Euler-Lagrange (EL) equations}}
which state that for a suitable value of the parameter~$\s>0$,
the lower semi-continuous function~$\ell : \G \rightarrow \R_0^+$ defined by
\beq \label{ldef}
\ell(\x) := \int_\G \L(\x,\y)\: d\mu(\y) - \s
\eeq
is minimal and vanishes on the support~\eqref{Ndef} of the measure,
\beq \label{EL}
\ell|_N \equiv \inf_\G \ell = 0 \:.
\eeq
For further details we refer to~\cite[Section~2]{jet}.

\subsection{The Restricted Euler-Lagrange Equations and Jets} \label{secwEL}
\hspace*{0.5cm}
The EL equations~\eqref{EL} are nonlocal in the sense that
they make a statement on~$\ell$ even for points~$\x \in \G$ which
are far away from~$N$.
It turns out that for the applications in this paper, it is preferable to
evaluate the EL equations locally in a neighborhood of~$N$.
This leads to the {\em{restricted EL equations}} introduced in~\cite[Section~4]{jet}.
We here give a slightly less general version of these equations which
is sufficient for our purposes. In order to explain how the restricted EL equations come about,
we begin with the simplified situation that the function~$\ell$ is smooth.
In this case, the minimality of~$\ell$ implies that the derivative of~$\ell$
vanishes on~$N$, i.e.\
\beq \label{ELrestricted}
\ell|_N \equiv 0 \qquad \text{and} \qquad D \ell|_N \equiv 0
\eeq
(where~$D \ell(p) : T_p \G \rightarrow \R$ is the derivative).
In order to combine these two equations in a compact form,
it is convenient to consider a pair~$\u := (a, \bu)$
consisting of a real-valued function~$a$ on~$N$ and a vector field~$\bu$
on~$T\G$ along~$N$, and to denote the combination of 
multiplication and directional derivative by
\beq \label{Djet}
\nabla_{\u} \ell(\x) := a(\x)\, \ell(\x) + \big(D_\bu \ell \big)(\x) \:.
\eeq
Then the equations~\eqref{ELrestricted} imply that~$\nabla_{\u} \ell(\x)$
vanishes for all~$\x \in N$.
The pair~$\u=(a,\bu)$ is referred to as a {\em{jet}}.

In the general lower-continuous setting, one must be careful because
the directional derivative~$D_\bu \ell$ in~\eqref{Djet} need not exist.
Our method for dealing with this problem is to restrict attention to vector fields
for which the directional derivative is well-defined.
Moreover, we must specify the regularity assumptions on~$a$ and~$u$.
To begin with, we always assume that~$a$ and~$u$ are {\em{smooth}} in the sense that they
have a smooth extension to the manifold~$\G$. Thus the jet~$\u$ should be
an element of the jet space
\[ %\label{Jdef}
\J := \big\{ \u = (a,\bu) \text{ with } a \in C^\infty(N, \R) \text{ and } \bu \in \Gamma(N, T\G) \big\} \:, \]
where~$C^\infty(N, \R)$ and~$\Gamma(N,T\G)$ denote the space of real-valued functions and vector fields
on~$N$, respectively, which admit smooth extensions to~$\G$.

Clearly, the fact that a jet~$\u$ is smooth does not imply that the functions~$\ell$
or~$\L$ are differentiable in the direction of~$\u$. This must be ensured by additional
conditions which are satisfied by suitable subspaces of~$\J$
which we now introduce.
First, we let~$\Gdiff$ be those vector fields for which the
directional derivative of the function~$\ell$ exists,
\[ %\beq \label{Gdiffdef}
\Gdiff = \big\{ \bu \in C^\infty(N, T\G) \;\big|\; \text{$D_{\bu} \ell(\x)$ exists for all~$\x \in N$} \big\} \:. \]
This gives rise to the jet space
\[ %\label{Jdiffdef}
\Jdiff := C^\infty(N, \R) \oplus \Gdiff \;\subset\; \J \:. \]
For the jets in~$\Jdiff$, the combination of multiplication and directional derivative
in~\eqref{Djet} is well-defined. 
We choose a linear subspace~$\Jtest \subset \Jdiff$ with the property
that its scalar and vector components are both vector spaces,
\[ %\label{Gammatest}
\Jtest = \Ctest(N, \R) \oplus \Gtest \;\subseteq\; \Jdiff \:, \]
and the scalar component is nowhere trivial in the sense that
\[ %\label{Cnontriv}
\text{for all~$\x \in N$ there is~$a \in \Ctest(N, \R)$ with~$a(\x) \neq 0$}\:. \]
Then the {\em{restricted EL equations}} read (for details cf.~\cite[(eq.~(4.10)]{jet})
\beq \label{ELtest}
\nabla_{\u} \ell|_N = 0 \qquad \text{for all~$\u \in \Jtest$}\:.
\eeq
The purpose of introducing~$\Jtest$ is that it gives the freedom to restrict attention to the portion of
information in the EL equations which is relevant for the application in mind.
For example, if one is interested only in the macroscopic dynamics, one can choose~$\Jtest$
to be composed of jets pointing in directions where the 
microscopic fluctuations of~$\ell$ are disregarded.

We finally point out that the restricted EL equations~\eqref{ELtest}
do not hold only for minimizers, but also for critical points of
the causal action. For brevity, we also refer to a measure which satisfies the restricted EL equations~\eqref{ELtest}
as a {\em{critical measure}}.

\subsection{The Linearized Field Equations and Inner Solutions}
In words, the {\em{linearized field equations}}
describe variations of the measure~$\mu$ which preserve the EL equations.
In order to make this statement mathematically precise,
we consider variations where we multiply~$\mu$ by a weight function~$f_\tau$ and then
take the push-forward with respect to a mapping~$F_\tau$ from~$N$ to~$\G$.
More precisely, we consider the ansatz
\begin{align} \label{rhoFf}
\tilde{\mu}_\tau = (F_\tau)_* \big( f_\tau \, \mu \big) \:,
\end{align}
where~$f_\tau \in C^\infty(N, \R^+)$ and~$F_\tau \in C^\infty(N, \G)$ are smooth mappings,
and~$(F_\tau)_*\mu$ denotes the push-forward (defined 
for a subset~$\Omega \subset \G$ by~$((F_\tau)_*\mu)(\Omega)
= \mu ( F_\tau^{-1} (\Omega))$; see for example~\cite[Section~3.6]{bogachev}).
Demanding that the family of measures~\eqref{rhoFf} is critical for all~$\tau$
implies that the jet~$\v$ defined by
\[ %\label{vvary}
\v(x) := \frac{d}{d\tau} \big( f_\tau(x), F_\tau(x) \big) \Big|_{\tau=0} \]
satisfies the linearized field equations
\beq \label{eqlinlip}
\la \u, \Delta \v \ra|_N = 0 \qquad \text{for all~$\u \in \Jtest$} \:,
\eeq
where
\[ %\label{Deldef}
\la \u, \Delta \v \ra(\x) := \nabla_{\u} \bigg( \int_N \big( \nabla_{1, \v} + \nabla_{2, \v} \big) \L(\x,\y)\: d\mu(\y) - \nabla_\v \,\s \bigg) \:, \]
where $\nabla_1$ and~$\nabla_2$ refer to derivatives acting on the first and second argument of the
Lagrangian, respectively.
Here we do not enter the details but refer instead to the general derivation in~\cite[Section~3.3]{perturb}
or to the simplified presentation in the smooth setting in the textbook~\cite[Chapter~6]{intro}.
We denote the vector space of all solutions of the linearized field equations by~$\Jlin$.

A specific class of linearized solutions are described by vector fields on~$N$.
In order to make sense of the notion of a vector field, we need to assume that~$N$
has the following smoothness property (we restrict attention to the three-dimensional case throughout).
\begin{Def} \label{defsms}
Space~$N:= \supp \mu$ has a {\bf{smooth manifold structure}} if
the following conditions hold:
\bitem
\item[\rm{(i)}] $N$ is a three-dimensional, smooth, oriented and connected submanifold of~$\G$.
\item[\rm{(ii)}] In a chart~$(\x,U)$ of~$N$, the measure~$\mu$ is absolutely continuous with respect
to the Lebesgue measure with a smooth, strictly positive weight function,
\[ %\label{hdef}
d\mu = h(\x)\: d^3\x \qquad \text{with} \quad h \in C^\infty(N, \R^+) \:. \]
\eitem
\end{Def} \noindent
Let~$\bv \in \Gamma(N, TN)$ be a vector field. Then, under the above assumptions,
its {\em{divergence}} $\div \bv \in C^\infty(N, \R)$ can be defined by the relation
\[ \int_N \div \bv(\x)\: \eta(\x)\: d\mu(\x) = -\int_N D_\bv \eta(\x)\: d\mu(\x) \:, \]
to be satisfied by all test functions~$\eta \in C^\infty_0(N, \R)$.
In a local chart~$(\x,U)$, the divergence is computed by
\[ %\label{divdef}
\div \bv = \frac{1}{h}\: \partial_\alpha \big( h\, \bv^\alpha \big) \]
(where, using the Einstein summation convention, we sum over~$\alpha=1,2,3$).
\begin{Def} \label{definner} An {\bf{inner solution}} is a jet~$\v$ of the form
\beq \label{vinner}
\v = (\div \bv, \bv) \qquad \text{with} \qquad \bv \in \Gamma(N, TN) \:.
\eeq
\end{Def}
The name ``inner {\em{solution}}'' stems from the fact that a jet~$\v$ of the form~\eqref{vinner}
satisfies the linearized field equations~\eqref{eqlinlip}.
The reason is that, applying the Gauss divergence theorem, integrating its jet derivative
of a compactly supported function gives zero, i.e.\ for for every~$f \in C^1_0(N, \R)$
\[ \int_N \nabla_\v f\: d\mu = 
\int_N \big( \div \bv\: f + D_\bv f \big)\: d\mu 
= \int_N \div \big( f \bv \big)\: d\mu = 0 \:. \]
Integrating by parts formally, one finds that
\begin{align*}
\la \u, \Delta \v \ra_N &= \nabla_\u \bigg( \int_N \big(\nabla_{1,\v} + \nabla_{2,\v} \big) \L(\x,\y)\: d\mu(\y)
- \nabla_\v \,\s \bigg) \\
&= \nabla_\u \bigg( \int_N \nabla_{1,\v} \L(\x,\y)\: d\mu(\y)
- \nabla_\v \,\s \bigg) \\
&= \nabla_\u \nabla_\v \ell(\x) = \nabla_\v \big( \nabla_\u \ell(\x) \big) 
- \nabla_{D_\bv \u} \ell(\x) = 0 \:.
\end{align*}
Here~$D_\bv \u$ denotes the partial derivative computed in given charts.
Here we do not need to introduce a connection and work with covariant derivatives,
simply because the corresponding summand~$\nabla_{D_\bv \u} \ell(\x)$ vanishes by the restricted EL equations.
Moreover, the function~$\nabla_\u \ell$ vanishes identically
on~$N$ in view of the restricted EL equations. Therefore, it is differentiable in the direction
of every vector field on~$N$, and this directional derivative is zero.

This formal computation can be made rigorous by imposing suitable
regularity and decay assumptions of the vector field~$\bv$ near infinity.
In order to avoid excessive overlap with previous works, we here omit the details and refer
instead for example to~\cite[Section~3]{fockbosonic} or~\cite[Section~2.1.3]{pmt}.

\subsection{The Total Mass as a Surface Layer Integral} \label{sectotmass}
In~\cite{pmt} the total mass was defined for causal variational principles. Moreover, it was shown that
in the example of asymptotically flat, static causal fermion system, this total mass gives back the
familiar ADM mass. We now recall the basic definitions.
We consider two measures: A measure~$\mu$ which describes the
vacuum spacetime, and another measure~$\tilde{\mu}$ which typically describes a gravitating spacetime.
In order to compare the measures~$\mu$ and~$\tilde{\mu}$, we introduce the functions
\[ %\label{nkleindef}
\left\{ \begin{array}{rl} 
n \,:\, N \rightarrow \R^+_0 \cup \{\infty\} \:,\qquad
n(\x) &\!\!\!\!= \displaystyle \int_{\tilde{N}} \L(\x,\y)\: d\tilde{\mu}(\y) \\[1em]
\tilde{n} \,:\, \tilde{N} \rightarrow \R^+_0 \cup \{\infty\} \:,\qquad
\tilde{n}(\x) &\!\!\!\!= \displaystyle \int_N \L(\x,\y)\: d\mu(\y) \:.
\end{array} \right. \]
For simplicity, throughout this paper we shall restrict attention to one asymptotic end (but
all our methods and results could be extended in a straightforward way
to several asymptotic ends).
Then we need to make the following assumption.
\begin{Def} \label{defasyclose}
The measures~$\tilde{\mu}$ and~$\mu$ are {\bf{asymptotically close}} if
they are both $\sigma$-finite with infinite total volume,
\[ %\label{nuinf}
\tilde{\mu}(\tilde{N}) = \mu(N) = \infty \:, \]
but
\[ \int_N \big| n(\x) - \s \big| \: d\mu(\x) < \infty \qquad \text{and} \qquad
\int_{\tilde{N}} \big| \tilde{n}(\x) - \s \big| \: d\tilde{\mu}(\x) < \infty 
 \:. \]
\end{Def} \noindent
We now state the most general definition of the total mass.
\begin{Def} \label{defmassgen}
Assume that~$\mu$ and~$\tilde{\mu}$ are asymptotically close.
Then the {\bf{total mass}}~$\Mass^\text{\rm{tot}}$ of~$\tilde{\mu}$ relative to~$\mu$ is defined by
\begin{align}
&\Mass^\text{\rm{tot}} := \lim_{\Omega \nearrow N} \;\lim_{\tilde{\Omega} \nearrow \tilde{N}} 
\bigg( -\s \Big( \tilde{\mu}(\tilde{\Omega}) - \mu(\Omega) \Big) \notag \\
& \qquad\qquad
+ \int_{\tilde{\Omega}} \!d\tilde{\mu}(\x) \int_{N\setminus \Omega} d\mu(\y)\: \L(\x,\y)
- \int_{\Omega} \!d\mu(\x) \int_{\tilde{N} \setminus \tilde{\Omega}}  d\tilde{\mu}(\y)\: \L(\x,\y) \bigg) \:,
\label{massgen}
\end{align}
where the notation~$\Omega \nearrow N$ means that we take an exhaustion of~$N$ by 
sets of finite $\mu$-measure.
\end{Def} \noindent
Restricting attention to sets~$\tilde{\Omega}$ and~$\Omega$ which satisfy
the volume constraint~\eqref{volconstraintintro}, one gets back the definition of the mass~\eqref{massintro}
stated in the introduction.

For what follows, it is important that, when evaluating the double integrals in~\eqref{massgen}
asymptotically near infinity, one may {\em{linearize}} in the sense that~$\tilde{\mu}$ may be described
in the asymptotic end by a first order perturbation of the vacuum measure~$\mu$.
The needed technical assumptions are subsumed in the following definition,
which is a simplified and weakened version of~\cite[Definition~2.3]{pmt}.
\begin{Def} \label{defasyflat}
The measure~$\tilde{\mu}$ is {\bf{asymptotically flat}} (with respect to~$\mu$)
if it is asymptotically close to~$\mu$
(see Definition~\ref{defasyclose}) and has the following additional properties:
There is a function~$f \in C^\infty(N, \R^+_0)$ and a mapping~$F \in C^\infty(N, \G)$ such that
\[ \tilde{\mu} = F_*\big( f \mu) \]
(where $F_* \mu$ is the push-forward measure defined by $(F_* \mu)(\tilde{\Omega})
= \mu(F^{-1}(\tilde{\Omega}))$). This transformation tends to the identity at infinity in the sense that
there is a jet~$\v \in \Jvary$ with
\begin{align}
\lim_{\Omega \nearrow N}
\int_\Omega \!d\mu(\x) \int_{N \setminus \Omega} \!\!\!\!\!\!\!d\mu(\y)\: \Big( f(\x)\: \L\big( F(\x),\y \big)
- \L(\x, \y) - \nabla_{1,\v} \L(\x,\y)\Big) &= 0 \label{osilin1} \\
\lim_{\Omega \nearrow N}
\int_\Omega \!d\mu(\x) \int_{N \setminus \Omega} \!\!\!\!\!\!\!d\mu(\y)\: \Big( \L\big( \x,F(\y) \big)\: f(\y)
- \L(\x, \y) - \nabla_{2,\v} \L(\x,\y)\Big) &= 0 \:. \label{osilin2}
\end{align}
\end{Def} \noindent
If this condition holds, the surface layer integral in~\eqref{massintro} can be linearized to obtain
\beq \label{Mtot}
\Mass^\text{\rm{tot}} = \lim_{\Omega \nearrow N} \Mass(\Omega, \v) \:,
\eeq
where the {\em{linearized quasilocal mass}}~$\Mass(\Omega, \v)$ is defined by
\beq \label{linosi}
\Mass(\Omega, \v) := \int_\Omega d\mu(\x) \int_{N \setminus \Omega} d\mu(\y) \:\big( \nabla_{1,\v} - \nabla_{2,\v}\big) \L(\x, \y) \:.
\eeq

The double integrals in~\eqref{massgen} and~\eqref{linosi} (and similarly in~\eqref{massintro}
and~\eqref{pososiintro} in the introduction) have the structure of a {\em{surface layer integral}},
as we now briefly explain (for more details see~\cite[Section~2.3]{noether} and~\cite[Section~4]{fockbosonic}).
The main point is that in the double integral~\eqref{linosi} the two variables~$\x$ and~$\y$
are integrated over~$\Omega$ and its complement~$N \setminus \Omega$, respectively.
In typical situations, the Lagrangian and its derivatives are very small if~$\x$ and~$\y$ are
far apart. Therefore, we only get a relevant contribution to the double integral if both~$\x$ and~$\y$ are
near the boundary of~$\Omega$. This picture can be made more quantitative if one assumes
that the Lagrangian is of {\em{short range}} in the following sense.
We let~$d \in C^0(N \times N, \R^+_0)$ be a distance function on~$N$. The assumption
of short range means that~$\L$ vanishes on distances larger than~$\delta$, i.e.
\beq \label{shortrange}
d(\x,\y) > \delta \quad \Longrightarrow \quad \L(\x,\y) = 0 = \nabla \L(\x,\y)
\eeq
Then the integrand in~\eqref{linosi} vanishes unless both arguments lie in a layer around the
boundary of~$\Omega$ of width~$\delta$, i.e.
\[ \x, \y \in B_\delta \big(\partial \Omega \big) \:. \]
For the purpose of this paper, we do not need to assume that the Lagrangian is of short range.
It suffices that it {\em{decays on the scale~$\delta$}}. Nevertheless, the reader who wants to avoid scaling arguments may assume the stronger assumption~\eqref{shortrange}.
We refer to~$\delta$ as the {\em{range}} of the Lagrangian.
The double integral in~\eqref{massgen} can be regarded as a nonlinear version of the linear
surface layer integral. Our above considerations again apply, provided that the measures~$\mu$ and~$\tilde{\mu}$
are close to each other near the boundary of~$\tilde{\Omega}$ and~$\Omega$.
This statement could be quantified in straightforward way a strict sense similar to~\eqref{shortrange};
we here omit the details for brevity.

We finally comment on the positive mass theorem as proven in~\cite[Section~5]{pmt}.
Inspired by Witten's spinor proof of the positive mass theorem, in this theorem one works with a
linear equation for the wave functions. This linear equation is obtained by linearizing the EL equations
with respect to a parameter~$\kappa$, being the Lagrange parameter of the so-called boundedness constraint.
This linearization makes it necessary to assume the existence of a whole family~$(\mu_\kappa)$
of minimizing measures. It is then shown under certain technical assumptions that if a suitable local energy condition holds,
then the total mass is non-negative.
We shall see that, in contrast to this result, the positive mass theorem that will be proved in Section~\ref{secnewproof} will be quite different in nature.
In particular, it does not require a local energy condition. Instead, positivity of
the total mass will be a direct consequence of the fact that the vacuum measure is a minimizer.

\section{A Positive Nonlinear Surface Layer Integral} \label{secposnonlin}
\subsection{A Positivity Argument under a Volume Constraint}
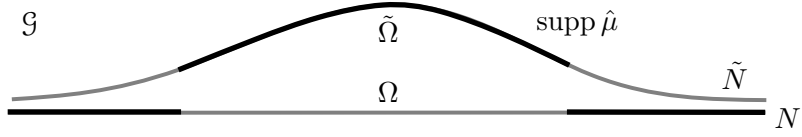
\begin{figure}
\psset{xunit=.5pt,yunit=.5pt,runit=.5pt}
\begin{pspicture}(573.96361344,85.55514358)
{
\newrgbcolor{curcolor}{0.50196081 0.50196081 0.50196081}
\pscustom[linewidth=2.99999995,linecolor=curcolor]
{
\newpath
\moveto(0.40141228,1.68943398)
\lineto(573.3260674,1.89844185)
}
}
{
\newrgbcolor{curcolor}{0.50196081 0.50196081 0.50196081}
\pscustom[linewidth=2.99999995,linecolor=curcolor]
{
\newpath
\moveto(3.14536441,11.870045)
\curveto(41.58976252,14.38101193)(80.03338583,16.89190327)(129.84447874,34.44641035)
\curveto(179.65557165,52.00087964)(240.8289222,84.59689728)(295.30771654,83.46354169)
\curveto(349.78651465,82.3301861)(397.56424063,47.4676009)(442.88882646,29.81199146)
\curveto(488.21341228,12.15638201)(531.08024693,11.70986862)(573.94795087,11.26331744)
}
}
{
\newrgbcolor{curcolor}{0 0 0}
\pscustom[linewidth=3.99998762,linecolor=curcolor]
{
\newpath
\moveto(0.00000378,2.32499933)
\lineto(131.17056,2.32499933)
}
}
{
\newrgbcolor{curcolor}{0 0 0}
\pscustom[linewidth=3.99998762,linecolor=curcolor]
{
\newpath
\moveto(129.24157984,34.15315681)
\curveto(149.65659213,42.35133004)(170.07161953,50.54946547)(187.51295244,57.30094894)
\curveto(204.95428535,64.0524324)(219.42133417,69.35701067)(235.30994268,73.89678264)
\curveto(251.1985474,78.43655083)(268.50737386,82.21117256)(283.77875528,83.25605697)
\curveto(299.05013669,84.30094516)(312.28252724,82.61592972)(327.82054299,78.31725004)
\curveto(343.35859654,74.01857035)(361.2012926,67.10653256)(377.75853354,59.89842925)
\curveto(394.31573669,52.69030327)(409.58650205,45.18658043)(424.85756976,37.68270642)
}
}
{
\newrgbcolor{curcolor}{0 0 0}
\pscustom[linewidth=3.99998762,linecolor=curcolor]
{
\newpath
\moveto(423.09627213,2.27023398)
\lineto(573.06195402,1.99999775)
}
\rput[bl](10,60){$\G$}
\rput[bl](580,-10){$N$}
\rput[bl](540,20){$\tilde{N}$}
\rput[bl](400,60){$\supp \hat{\mu}$}
\rput[bl](280,55){$\tilde{\Omega}$}
\rput[bl](280,10){$\Omega$}
}
\end{pspicture}
\caption{The measure~$\hat{\mu}$.}
\label{figpos1}
\end{figure}

We consider two measures: A measure~$\mu$ which describes the
vacuum spacetime, and another measure~$\tilde{\mu}$ which typically describes an interacting spacetime.
We assume that the vacuum measure is a {\em{minimizer
with respect to variations of finite volume}}
as defined in Section~\ref{seccvp}.
We choose subsets~$\Omega \subset N$
and~$\tilde{\Omega} \subset \tilde{N}$ having the same finite volume,
\beq \label{volconstraint}
\mu(\Omega) = \tilde{\mu}(\tilde{\Omega}) < \infty \:.
\eeq
In order to construct an admissible test measure~$\hat{\mu}$, we ``cut out'' $\Omega$ from~$\mu$
and ``glue in'' the set~$\tilde{\Omega}$, i.e.\
\[ \hat{\mu} := \chi_{\tilde{\Omega}}\, \tilde{\mu} + \chi_{N \setminus \Omega}\, \mu \]
(see Figure~\ref{figpos1}).
The measure~$\hat{\mu}$ differs from~$\mu$ only on a set of finite volume
and preserves the volume constraint (see~\eqref{totvol}).

Therefore, we obtain from~\eqref{Sdiffpos} and~\eqref{integrals} (with~$\tilde{\mu}$ replaced by~$\hat{\mu}$) that
\begin{align*}
0 &\leq \big( \Sact(\hat{\mu}) - \Sact(\mu) \big) \notag \\
&= 2 \int_\G d(\hat{\mu} - \mu)(\x) \int_N d\mu(\y)\: \L(\x,\y)
+ \int_\G d(\hat{\mu} - \mu)(\x) \int_\G d(\hat{\mu} - \mu)(\y)\: \L(\x,\y) \notag \\
&= 2 \int_{\tilde{\Omega}} d\tilde{\mu}(\x) \int_N d\mu(\y)\: \L(\x,\y)
- 2 \int_{\Omega} d\mu(\x) \int_N d\mu(\y)\: \L(\x,\y) \notag \\
&\quad\: + \int_{\tilde{\Omega}} d\tilde{\mu}(\x) \int_{\tilde{\Omega}} d\tilde{\mu}(\y)\: \L(\x,\y)
- 2 \int_{\tilde{\Omega}} d\tilde{\mu}(\x) \int_\Omega d\mu(\y)\: \L(\x,\y) \notag \\
&\quad\: + \int_\Omega d\mu(\x) \int_\Omega d\mu(\y)\: \L(\x,\y) \notag \\
&= 2 \int_{\tilde{\Omega}} d\tilde{\mu}(\x) \int_{N \setminus \Omega} d\mu(\y)\: \L(\x,\y)
- 2 \int_{\Omega} d\mu(\x) \int_{N \setminus \Omega} d\mu(\y)\: \L(\x,\y) \\
&\quad\: + \int_{\tilde{\Omega}} d\tilde{\mu}(\x) \int_{\tilde{\Omega}} d\tilde{\mu}(\y)\: \L(\x,\y)
- \int_\Omega d\mu(\x) \int_\Omega d\mu(\y)\: \L(\x,\y) \:.
\end{align*}
Our findings are summarized as follows.
\begin{Thm} {\bf{(Positivity argument under volume constraint}} \label{thm1}
Let~$\mu$ be a minimizer with respect to variations of finite volume and~$\tilde{\mu}$ a measure on~$\G$.
Moreover, let~$\Omega \subset N:= \supp \mu$ and~$\tilde{\Omega} \subset \tilde{N} := \supp \tilde{\mu}$
satisfying the volume constraint~\eqref{volconstraint}. Then
\begin{align}
0  & \leq 2 \int_{\tilde{\Omega}} d\tilde{\mu}(\x) \int_{N \setminus \Omega} d\mu(\y)\: \L(\x,\y)
- 2 \int_{\Omega} d\mu(\x) \int_{N \setminus \Omega} d\mu(\y)\: \L(\x,\y) \label{pososi1} \\
&\quad\: + \int_{\tilde{\Omega}} d\tilde{\mu}(\x) \int_{\tilde{\Omega}} d\tilde{\mu}(\y)\: \L(\x,\y)
- \int_\Omega d\mu(\x) \int_\Omega d\mu(\y)\: \L(\x,\y) \:. \label{pososi2}
\end{align}
\end{Thm} \noindent
The first summand in~\eqref{pososi1} coincides with the first summand in the nonlinear
surface layer integral as introduced in~\cite{fockbosonic}. Thus it is a nonlinear surface layer integral
with a somewhat different structure. This nonlinear surface layer integral it is not conserved, but instead it satisfies
an {\em{in}}equality. The second summand in~\eqref{pososi1} can be interpreted as the
surface area of~$\partial \Omega$. The two summands in~\eqref{pososi2}, on the other hand,
can be regarded as volume integrals over~$\tilde{\Omega}$ and~$\Omega$.

It is useful to rewrite this inequality in a more geometric way.
\begin{Def} {\bf{(Area and Volume)}} Given a measurable subset~$\Omega \subset N$, we define its
area~$A$ and volume~$V$ by
\begin{align*}
A &:= \int_\Omega d\mu(\x) \int_{N \setminus \Omega} d\mu(\y)\: \L(\x,\y) \\
V &:= \mu(\Omega) \:.
\end{align*}
For the measure~$\tilde{\mu}$ we use the same notation with additional tildes.
\end{Def} \noindent
We now let~$\mu$ be a minimizer with respect to variations of finite volume. It satisfies the EL equations
\[ \ell|_N \equiv \inf_N \ell = 0 \qquad \text{with} \qquad
\ell(\x) := \int_N \L(\x, \y)\: d\mu(\y) - \s \:. \]
Moreover, we assume that the measure~$\tilde{\mu}$ has the property that
\beq \label{tellconst}
\tilde{\ell}|_{\tilde{N}} \equiv 0
\eeq
(this assumption will be discussed at the end of this section).
Under these assumptions and using these notions,
we can reformulate Theorem~\ref{thm1} as follows.

\begin{Thm} \label{thm2} Let~$\mu$ be a minimizer with respect to variations of finite volume and~$\tilde{\mu}$ a measure on~$\G$.
Moreover, assume that~$\tilde{\mu}$ satisfies~\eqref{tellconst}.
Then for any measurable subsets~$\Omega \subset N$ and~$\tilde{\Omega} \subset \tilde{N}$
which satisfy the volume constraint
\beq \label{VVt}
V = \tilde{V} \:,
\eeq
the function~$\Mass(\tilde{\Omega}, \Omega)$ defined by
\beq \label{massOtOdef}
\Mass(\tilde{\Omega}, \Omega) := 
2 \int_{\tilde{\Omega}} d\tilde{\mu}(\x) \int_{N \setminus \Omega} d\mu(\y)\: \L(\x,\y)
\;-\; \tilde{A} - A
\eeq
is non-negative,
\beq \label{ineq2}
\Mass(\tilde{\Omega}, \Omega) \geq 0\:.
\eeq
\end{Thm} \noindent
In order to clarify the dependence on the measures~$\mu$ and~$\tilde{\mu}$,
we sometimes also denote the function~$\Mass(\tilde{\Omega}, \Omega)$ 
in~\eqref{massOtOdef} by~$\Mass_{\tilde{\mu}, \mu}(\tilde{\Omega}, \Omega)$
(as in~\eqref{pososiintro} in the introduction).
\Proof[Proof of Theorem~\ref{thm2}.] We rewrite the last integral in~\eqref{pososi2} as
\begin{align*}
&\int_\Omega d\mu(\x) \int_\Omega d\mu(\y)\: \L(\x,\y) \\
&= \int_\Omega d\mu(\x) \int_N d\mu(\y)\: \L(\x,\y) -
\int_\Omega d\mu(\x) \int_{N \setminus \Omega} d\mu(\y)\: \L(\x,\y) = \s V - A \:.
\end{align*}
Rewriting the first integral in~\eqref{pososi2} similarly and using the volume constraint~\eqref{VVt}
gives the result.
\QED

The assumption~\eqref{tellconst} can be understood as follows. We first point out that this condition
follows from the restricted EL equations~\eqref{ELtest}. Hence~\eqref{tellconst} holds
whenever~$\tilde{\mu}$ is a minimizing or critical measure.
But~\eqref{tellconst} is much weaker than the restricted EL equations, because
it is only the scalar component of these equations.
Indeed, this condition can be satisfied for any given measure~$\tilde{\mu}$ by changing its
weight (i.e.\ by varying in the class of measures with fixed support~$\supp \tilde{\mu}$).
We refer the interested reader to~\cite{support, static}.

We finally comment on the equality case in~\eqref{ineq2}.
\begin{Remark} {\bf{(Rigidity statements)}} {\em{
It is a natural question whether the equality~$\Mass(\tilde{\Omega}, \Omega)=0$
implies that~$\tilde{\mu}|_{\tilde{\Omega}}= \mu|_{\Omega}$.
The answer is yes, provided that the minimizer~$\mu$ is unique.
Such a uniqueness statement seems a sensible assumption. However, the complication arises
that uniqueness will hold only up to symmetry transformations of the Lagrangian.
For conceptual clarity, we decided to postpone such symmetry transformations
until later (see Definition~\ref{defiso} in Section~\ref{secequi}).
Consequently, for simplicity we decided to leave uniqueness and rigidity statements out of the present paper.
We merely remark at this point that, assuming that the measure~$\mu$ is unique up to
symmetry transformations, a rigidity statement should hold stating that if~$\Mass(\tilde{\Omega}, \Omega)$
vanishes, then the measures~$\tilde{\mu}|_{\tilde{\Omega}}$ and~$\mu|_{\Omega}$ coincide up to such symmetry transformations.
Similarly, if the total mass is zero, then~$\tilde{\mu}$ and~$\mu$ are related to each other
by symmetry transformations.
An analogous statement should also hold for the quasilocal mass as will be introduced
in Section~\ref{secquasilocal}.
}} \QEDrem
\end{Remark}

\subsection{Generalizations of the Positive Nonlinear Surface Layer Integral}
In this section, we shall get rid of the volume constraint~\eqref{VVt}. 
Let~$\mu$ and~$\tilde{\mu}$ again be measures as in the statement of Theorem~\ref{thm2}.
Moreover, we now need to specialize
our setting by assuming that the total volume of the minimizing measure~$\mu$ is infinite,
\beq \label{inftotvol}
\mu(N) = \infty \:.
\eeq
We choose subsets~$\Omega \subset N$ and~$\tilde{\Omega} \subset \tilde{N}$
of finite (but not necessarily the same) volume.
The idea for treating the case that the volumes~$V$ and~$\tilde{V}$ are different is to
compensate for this fact by adding to~\eqref{ineq2} the volume difference times the
``action per volume.'' In order to compute this ``action per volume'' heuristically, we choose
a subset~$U \subset N$ of finite volume and consider the variation~$(\mu_\tau)_{\tau \in [0,1]}$
with
\[ \mu_\tau := \mu + \tau \chi_U\: \mu \:. \]
Then the action changes by
\[ \frac{d}{d\tau} \big( \Sact(\mu_\tau) - \Sact(\mu) \big) \Big|_{\tau=0}
= 2 \int_U d\mu(\x) \int_N d\mu(\y)\: \L(\x, \y) = 2 \s\:\mu(U) \:, \]
suggesting that the ``action per volume'' is given simply by~$2\s$.
Taking into account the corresponding ``volume contribution'' gives
the following theorem, whose proof also makes the above heuristic argument precise.
\begin{Thm} \label{thm3}
Let~$\mu$ be a minimizer with respect to variations of finite volume which has infinite total volume~\eqref{inftotvol}.
Moreover, let~$\tilde{\mu}$ be a measure on~$\G$ which satisfies~\eqref{tellconst}.
Let~$\Omega \subset N$ and~$\tilde{\Omega} \subset \tilde{N}$ be subsets of finite volume.
Then the function~$\Mass(\tilde{\Omega}, \Omega)$ defined in~\eqref{massOtOdef}
satisfies the inequality
\[ \Mass(\tilde{\Omega}, \Omega) \geq  \s \,\big(\tilde{V} - V \big) \:. \]
\end{Thm}
\Proof
Given~$n \in \N$ we choose a subset~$U_n \subset N$
with~$n < \mu(U_n) < \infty$. We define the measure
\beq \label{hmu2}
\hat{\mu} := \chi_{\tilde{\Omega}}\, \tilde{\mu} + \chi_{N \setminus \Omega}\, \mu 
- \frac{\tilde{V} - V}{\mu(U_n)}\: \chi_{U_n}\, \mu\:.
\eeq
Obviously, the measures~$\hat{\mu}$ and~$\mu$ coincide outside
the set of finite volume~$\Omega \cup \tilde{\Omega} \cup U_n$. Moreover, with the last summand
in~\eqref{hmu2} we arranged the volume constraint. Hence
\begin{align*}
0 &\leq \big( \Sact(\hat{\mu}) - \Sact(\mu) \big) \notag \\
&= 2 \int_\G d(\hat{\mu} - \mu)(\x) \int_N d\mu(\y)\: \L(\x,\y)
+ \int_\G d(\hat{\mu} - \mu)(\x) \int_N d(\hat{\mu} - \mu)(\y)\: \L(\x,\y) \:.
\end{align*}
Substituting~\eqref{hmu2} and multiplying out, we can use transformations similar as
in the proof of Theorem~\ref{thm2} to obtain
\begin{align}
0 \leq \big( \Sact(\hat{\mu}) - \Sact(\mu) \big) 
&= 2 \int_{\tilde{\Omega}} d\tilde{\mu}(\x) \int_{N \setminus \Omega} d\mu(\y)\: \L(\x,\y)
+ \s \tilde{V} - \tilde{A} - \s V - A \\
&\quad\,-2\, \frac{\tilde{V} - V}{\mu(U_n)} \int_{U_n}\, d\mu(\x) 
\int_N d\mu(\y)\: \L(\x,\y) \label{t1} \\
&\quad\,+ \bigg( \frac{\tilde{V} - V}{\mu(U_n)} \bigg)^2
\int_{U_n}\, d\mu(\x) \int_{U_n} d\mu(\y)\: \L(\x,\y) \label{t2} \\
&\quad\,+2\, \frac{\tilde{V} - V}{\mu(U_n)} \int_{U_n}\, d\mu(\x) 
\int_\Omega d\mu(\y)\: \L(\x,\y) \label{t3} \\
&\quad\,-2\, \frac{\tilde{V} - V}{\mu(U_n)} \int_{U_n}\, d\mu(\x) 
\int_{\tilde{\Omega}} d\tilde{\mu}(\y)\: \L(\x,\y) \:. \label{t4}
\end{align}
Using the EL equations, \eqref{t1} simplifies to
\[ \eqref{t1} = -2 \s\, \big(\tilde{V} - V \big) \:. \]
The terms~\eqref{t2}--\eqref{t4}, on the other hand, tend to zero as~$n \rightarrow \infty$,
as one sees from the estimates
\begin{align*}
\big| \eqref{t2} \big| &\leq \bigg( \frac{\tilde{V} - V}{\mu(U_n)} \bigg)^2
\int_{U_n}\, d\mu(\x) \int_N d\mu(\y)\: \L(\x,\y) \leq
\s\: \frac{\big| \tilde{V} - V \big|^2}{n} \\
\big| \eqref{t3} \big| &\leq 2\, \frac{\big|\tilde{V} - V \big|}{\mu(U_n)}\:
\int_\Omega d\mu(\x) \int_N d\mu(\y)\: \L(\x,\y) \leq 2\:\s V\, \frac{\big|\tilde{V} - V \big|}{n} \\
\big| \eqref{t4} \big| &\leq 2\, \frac{\big| \tilde{V} - V \big|}{\mu(U_n) }
\int_{\tilde{\Omega}} d\tilde{\mu}(\y)\int_N\, d\mu(\x) \: \L(\x,\y)
\leq 2\, \frac{\big| \tilde{V} - V\big|}{n}\: \sup_{x \in \tilde{\Omega}} \big(\ell(\x) + \s \big) \:.
\end{align*}
According to our assumption~(iv) on page~\pageref{Cond4}, the function~$\ell$ is bounded on~$\G$.
Therefore, all the expressions on the right tend to zero as~$n$ tends to infinity.
\QED

We finally state a variant of the above theorem which does not require that the
interacting measure~$\tilde{\mu}$ has the property~\eqref{tellconst}.
\begin{Prp} \label{thm4} Let~$\mu$ be a minimizer with respect to variations of finite volume which has infinite total volume~\eqref{inftotvol}. Then for any measure~$\tilde{\mu}$ on~$\G$ and
all subsets~$\Omega \subset N$ and~$\tilde{\Omega} \subset \tilde{N}$ of finite volume,
the function~${\mathfrak{M}}(\tilde{\Omega}, \Omega)$ satisfies the lower bound
\beq \label{frakNin}
\Mass(\tilde{\Omega}, \Omega) \geq 2 \s \tilde{V} - \s V  - \int_{\tilde{\Omega}} d\tilde{\mu}(\x) \int_{\tilde{N}} d\tilde{\mu}(\y)\: \L(\x,\y) \:.
\eeq
\end{Prp}
\Proof Dropping the relation~\eqref{tellconst}, the first integral in~\eqref{pososi2} can be rewritten as
\begin{align*}
\int_{\tilde{\Omega}} d\tilde{\mu}(\x) \int_{\tilde{\Omega}} d\tilde{\mu}(\y)\: \L(\x,\y)
= \int_{\tilde{\Omega}} d\tilde{\mu}(\x) \int_{\tilde{N}} d\tilde{\mu}(\y)\: \L(\x,\y) - \tilde{A} \:.
\end{align*}
Hence the inequality~\eqref{ineq2} is modified to
\begin{align*}
2 \int_{\tilde{\Omega}} d\tilde{\mu}(\x) \int_{N \setminus \Omega} d\mu(\y)\: \L(\x,\y) \geq A + \s V
+ \tilde{A} - \int_{\tilde{\Omega}} d\tilde{\mu}(\x) \int_{\tilde{N}} d\tilde{\mu}(\y)\: \L(\x,\y) \:.
\end{align*}
Evaluating this inequality for~$\tilde{\mu} = \hat{\mu}$ according to~\eqref{hmu2}, we obtain
\begin{align*}
0 &\leq \big( \Sact(\hat{\mu}) - \Sact(\mu) \big) \\
&= 2 \int_{\tilde{\Omega}} d\tilde{\mu}(\x) \int_{N \setminus \Omega} d\mu(\y)\: \L(\x,\y)
-A - \s V - \tilde{A} + \int_{\tilde{\Omega}} d\tilde{\mu}(\x) \int_{\tilde{N}} d\tilde{\mu}(\y)\: \L(\x,\y)  \\
&\quad\,-2\, \frac{\tilde{V} - V}{\mu(U_n)} \int_{U_n}\, d\mu(\x) 
\int_N d\mu(\y)\: \L(\x,\y) + \bigg( \frac{\tilde{V} - V}{\mu(U_n)} \bigg)^2
\int_{U_n}\, d\mu(\x) \int_{U_n} d\mu(\y)\: \L(\x,\y) \\
&\quad\,+2\, \frac{\tilde{V} - V}{\mu(U_n)} \int_{U_n}\, d\mu(\x) 
\int_\Omega d\mu(\y)\: \L(\x,\y) -2\, \frac{\tilde{V} - V}{\mu(U_n)} \int_{U_n}\, d\mu(\x) 
\int_{\tilde{\Omega}} d\tilde{\mu}(\y)\: \L(\x,\y) \:.
\end{align*}
Now we can take the limit~$n \rightarrow \infty$ exactly as in the proof of Theorem~\ref{thm3}
to obtain the result.
\QED

\section{A New Proof of the Positive Mass Theorem with Volume Constraint} \label{secnewproof}
We let~$(\Omega_n)_{n \in \N}$ and~$(\tilde{\Omega}_n)_{n \in \N}$ be exhaustions
of~$N$ and~$\tilde{N}$, respectively.

\begin{Thm} \label{pmtconsrtaint}
Let~$\mu$ be a minimizer with respect to variations of finite volume which has infinite total volume~\eqref{inftotvol}.
Moreover, let~$\tilde{\mu}$ be a measure on~$\G$ which has the property~\eqref{tellconst}.
Assume that~$\tilde{\mu}$ is asymptotically flat (see Definition~\ref{defasyflat}).
Then the total mass (see Definition~\ref{defmassgen}) is non-negative,
\[ \Mass^\text{\rm{tot}} \geq 0 \:. \]
\end{Thm} 
\Proof We let~$\tilde{\Omega}_n \subset \tilde{N}$ and~$\Omega_n \subset N$ be exhaustions
by sets of finite volume. From Theorem~\ref{thm3} we know that for any~$n$,
\[ 0 \leq \Mass(\tilde{\Omega}_n, \Omega_n) = 
2 \int_{\tilde{\Omega}_n} d\tilde{\mu}(\x) \int_{N \setminus \Omega_n} d\mu(\y)\: \L(\x,\y)
\;-\; \tilde{A}_n - A_n - \s \,\big(\tilde{V}_n - V_n \big) \:. \]
Taking the limit~$n \rightarrow \infty$, we can use~\eqref{osilin1} and~\eqref{osilin2} to obtain
\begin{align*}
0 &\leq \liminf_{n \rightarrow \infty} \bigg( 2 \int_{\tilde{\Omega}_n} d\mu(\x) \int_{N \setminus \Omega_n} d\mu(\y)\: \nabla_{1,\v}  \L(\x,\y) \\
&\qquad\qquad\;\; - \int_{\Omega_n} d\mu(\x) \int_{N \setminus \Omega_n} d\mu(\y)\: \big( \nabla_{1,\v}
+ \nabla_{2,\v} \big) \L(\x,\y)
 \;-\;  \s \,\big(\tilde{V}_n - V_n \big) \bigg) \\
&= \liminf_{n \rightarrow \infty} \bigg( \int_{\tilde{\Omega}_n} d\mu(\x) \int_{N \setminus \Omega_n} d\mu(\y)\: \big( \nabla_{1,\v} - \nabla_{2,\v} \big)  \L(\x,\y) \;-\;  \s \,\big(\tilde{V}_n - V_n \big) \bigg) \\
&= \liminf_{n \rightarrow \infty} \bigg( 
\int_{\tilde{\Omega}} \!d\tilde{\mu}(\x) \int_{N\setminus \Omega} d\mu(\y)\: \L(\x,\y) \\
&\qquad\qquad\;\;
- \int_{\Omega} \!d\mu(\x) \int_{\tilde{N} \setminus \tilde{\Omega}} d\tilde{\mu}(\y)\: \L(\x,\y) \;-\;  \s \,\big(\tilde{V}_n - V_n \big) \bigg) = \Mass^\text{tot} \:.
\end{align*}
This gives the result.
\QED

\section{A Positive Mass Theorem without Volume Constraint} \label{secnovol}

\subsection{Rewriting Surface Layer Integrals as Surface Integrals}
For the proof of the positive mass theorem, it is essential that the nonlinear surface layer integrals
in~\eqref{massintro} and~\eqref{pososiintro} can be linearized near infinity,
giving a surface layer integral of the form~\eqref{linosi}
(for details see the proof of Theorem~\ref{pmtconsrtaint}).
From the computational point of view, it is desirable to rewrite such linear surface layer integrals
as standard surface integrals over the boundary of~$\Omega$.
We now give a general procedure for doing so. This procedure will also leads us to a method for
treating the volume constraint.

We again assume that~$\mu$ describes the vacuum. Moreover, from now on we assume
that the vacuum is smooth and three-dimensional and that the measure is translation invariant, i.e.\
\beq \label{Nsmooth}
N \simeq \R^3 \qquad \text{and} \qquad d\mu(\x) = d^3\x \:.
\eeq
Note that, if we assume again that the Lagrangian is of short range~\eqref{shortrange}, in the surface layer integral~\eqref{linosi}, the jet~$\v$ enters only in a $\delta$-neighborhood of~$\partial \Omega$. 
For the following construction, however, we need to extend~$\v$ smoothly to all of~$N$.
Then, making use of the fact that the integrand is anti-symmetric, we can rewrite~\eqref{linosi} as
\beq \label{linosi2}
\Mass(\Omega, \v) = \int_\Omega d\mu(\x) \int_N d\mu(\y) \:\big( \nabla_{1,\v} - \nabla_{2,\v} \big) \L(\x, \y) \:.
\eeq
Given a parameter~$s \in [0,1]$ we introduce the new variables~$(\bzeta, \bxi)$ by
\begin{align*}
\bzeta &= s \x + (1-s)\, \y \:,\hspace*{-3cm} &\bxi  &= \y - \x \:. \\
\x &= \bzeta - (1-s) \bxi \:,\hspace*{-3cm} & \y &= \bzeta + s \bxi \:.
\end{align*}
Note that~$\bzeta$ is a convex combination of~$\x$ and~$\y$. In this way, increasing the parameter~$s$
from zero to one, one can continuously deform these variables and interchange the roles of~$\x$ and~$\y$.
We now consider the integral
\[ \varphi(s, \bzeta) := \int_N \nabla_{1,\v} \L(\x,\y)\: d^3\bxi
= \int_N \nabla_{1,\v} \L\big( \bzeta - (1-s) \bxi, \bzeta + s \bxi \big)\: d^3\bxi \:. \]
Then
\begin{align*}
\varphi(1, \bzeta) &= \int_N \nabla_{1,\v} \L\big( \bzeta, \bzeta + \bxi \big)\: d^3\bxi
= \int_N \nabla_{1,\v} \L\big(\bzeta, \y \big)\: d^3\y \\
\varphi(0, \bzeta) &= \int_N \nabla_{1,\v} \L\big( \bzeta - \bxi, \bzeta \big)\: d^3\bxi
= \int_N \nabla_{1,\v} \L\big(\x, \bzeta \big)\: d^3\x
= \int_N \nabla_{2,\v} \L\big(\bzeta, \y \big)\: d^3\y \:,
\end{align*}
making it possible to write the inner integral of the surface layer integral~\eqref{linosi} as the difference
\beq \label{divform}
\int_N \big(\nabla_{1,\v} - \nabla_{2,\v} \big) \L(\bzeta,\y)\: d^3\y
= \varphi(1,\bzeta) - \varphi(0,\bzeta) \:.
\eeq
We next expand the right side in a Taylor series about~$s=\frac{1}{2}$.
It is a remarkable fact that each term of this Taylor series is a total divergence:
\begin{Thm} \label{thmdivA}
The inner integral of the surface layer integral~\eqref{linosi2} can be written as a divergence,
\beq \label{Aexp}
\int_N \big(\nabla_{1,\v} - \nabla_{2,\v} \big) \L(\bzeta,\y)\: d^3\y
= \sum_{\alpha=1}^3 \frac{\partial}{\partial \bzeta_\alpha} A^\alpha(\bzeta) \:,
\eeq
where the vector field~$A(\bzeta)$ has the formal power expansion
\begin{align}
A_\alpha(\bzeta) &:= \sum_{k=0}^\infty A_\alpha^{(k)}(\bzeta) \qquad \text{and} \label{Aser} \\
A_\alpha^{(k)}(\bzeta) &:= \frac{1}{4^k\,(2k+1)!}
\int_N \bxi_\alpha\: \bigg( \bxi\: \frac{\partial}{\partial \bzeta} \bigg)^{2k}
\nabla_{1,\v} \L\Big( \bzeta - \frac{\bxi}{2}, \bzeta + \frac{\bxi}{2} \Big)\: d^3\bxi \:. \label{Aalphadef}
\end{align}
\end{Thm} \noindent
The vector field~$A_\alpha(\zeta)$ will be referred to as the {\em{alignment vector field}}.

Before coming to the proof of this theorem, we make a two short remarks. 
We first point out that, similar to
the Taylor series of a smooth function, the formal power series~\eqref{Aser} will in general {\em{not}} converge.
Nevertheless, the summands for larger~$k$ will become small in the sense that they
involve higher scaling factors~$\delta/ \ell_{\macro}$. This will be explained in more detail in the paragraph
after Corollary~\ref{cormass}.

The second remark is to note that the method of this theorem works immediately extends to
other general surface layer integrals. More precisely, one may replace the
integrand~$(\nabla_{1,\v} - \nabla_{2,\v} ) \L(\bzeta,\y)$ in~\eqref{Aexp} by any other
function which is smooth and {\em{anti-symmetric}} in its two arguments~$\bzeta$ and~$y$.

\Proof[Proof of Theorem~\ref{thmdivA}]
We begin by computing the~$s$-derivatives of the function~$\varphi(s, \bzeta)$.
The first derivative can be calculated by
\begin{align*}
\frac{d}{ds} \varphi(s, \bzeta) &= \frac{d}{ds} \int_N \nabla_{1,\v} \L\big( \bzeta - (1-s) \bxi, \bzeta + s \bxi \big)\: d^3\bxi \\
&= \int_N \big( D_{1,\bxi} + D_{2,\bxi} \big) \nabla_{1,\v} \L\big( \bzeta - (1-s) \bxi, \bzeta + s \bxi \big)\: d^3\bxi \\
&= \int_N \sum_{\alpha=1}^3 \bxi^\alpha \: \frac{\partial}{\partial \bzeta^\alpha} \nabla_{1,\v} \L\big( \bzeta - (1-s) \bxi, \bzeta + s \bxi \big)\: d^3\bxi \:.
\end{align*}
Computing the higher derivatives iteratively gives
\[ \frac{d^p}{ds^p} \varphi(s, \bzeta) = 
\int_N \bigg( \bxi \: \frac{\partial}{\partial \bzeta} \bigg)^p \nabla_{1,\v} \L\big( \bzeta - (1-s) \bxi, \bzeta + s \bxi \big)\: d^3\bxi \:. \]

Using these formulas in~\eqref{divform} gives
\begin{align*}
& \int_N \big(\nabla_{1,\v} - \nabla_{2,\v} \big) \L(\bzeta,\y)\: d^3\y
= \sum_{\ell=0}^\infty \frac{1}{\ell!} \:\partial_s^\ell \varphi(s, \bzeta) \Big|_{s=\frac{1}{2}}\:
\bigg( \Big( \frac{1}{2} \Big)^\ell - \Big(-\frac{1}{2} \Big)^\ell \bigg) \\
&= \sum_{k=0}^\infty \frac{1}{(2k+1)!} \:\partial_s^{2k+1} \varphi(s, \bzeta) \Big|_{s=\frac{1}{2}}\: \frac{2}{2^{2k+1}}
= \sum_{k=0}^\infty \frac{1}{4^k\,(2k+1)!} \:\partial_s^{2k+1} \varphi(s, \bzeta) \Big|_{s=\frac{1}{2}} \\
&= \sum_{k=0}^\infty \frac{1}{4^k\,(2k+1)!}
\int_N \bigg( \bxi \: \frac{\partial}{\partial \bzeta} \bigg)^{2k+1}
\nabla_{1,\v} \L\Big( \bzeta - \frac{\bxi}{2}, \bzeta + \frac{\bxi}{2} \Big)\: d^3\bxi \:.
\end{align*}
This concludes the proof.
\QED

Using this expansion in~\eqref{linosi2} we can express the linearized quasilocal mass in as a standard surface integral.
\begin{Corollary} \label{cormass}
Assume that~$\Omega \subset N \simeq \R^3$ has a smooth boundary with outer normal~$\nu$.
Then the linearized quasilocal mass~\eqref{linosi} can be written as
\[ \Mass(\Omega, \v) = \int_{\partial \Omega}
\sum_{\alpha=1}^3  A^\alpha(\x)\: \nu_\alpha(\x)\: d\mu_{\partial \Omega}(\x) \:. \]
\end{Corollary}
\Proof We integrate~\eqref{Aexp} over~$\Omega$ and use~\eqref{linosi2},
\[ \Mass(\Omega, \v) = \int_{\Omega} \sum_{\alpha=1}^3 \frac{\partial}{\partial \bzeta_\alpha} A^\alpha(\bzeta)\: d^3\bzeta \:. \]
Applying the Gauss divergence theorem gives the result.
\QED

We finally explain in which sense the higher orders of the above expansion are small.
Each factor~$\bxi$ in~\eqref{Aalphadef} gives a scaling factor~$\delta$
(where~$\delta$ denotes again the range of the Lagrangian as introduced after~\eqref{shortrange}).
Each such factor comes with a $\bzeta$-derivative acting on the jet~$\v$.
We assume that each $\bzeta$-derivative gives a scaling factor~$\ell_{\macro}$, where~$\ell_{\macro}$
can be thought of as the length scale on which the macroscopic physical objects (like currents, fields,
curvature) change. Then
\beq \label{scalefact}
\text{$A^{(k)}$ involves the scaling factor} \quad \bigg( \frac{\delta}{\ell_{\macro}} \bigg)^{2k} \:.
\eeq
Working with corresponding error terms, we may truncate the formal power series~\eqref{Aser}.
Indeed, in this paper, it will be sufficient to consider the vector fields~$A^{(0)}$ and~$A^{(1)}$.
We note that the reader who prefers to avoid formal power expansions may replace~\eqref{Aser}
by a Taylor polynomial and estimate the remainder term using the scaling~\eqref{scalefact}.

We finally remark that the scalings could be avoided by imposing strict inequalities of the form
\[ \sup_{\x \in N} \|\partial^\kappa \v^i(\x) \| \leq \frac{c}{|\ell_\macro^{|\kappa|}|} \:, \]
to be satisfied for all multi-indices~$\kappa$ with~$|\kappa| \leq p$ and some maximal order
of differentiability~$p$ ($p=2$ seems sufficient for our purposes).
Here the index~$i$ refers to the components in a chart of~$\G$, for example in symmetric wave
charts for causal fermion systems as constructed in~\cite{gaugefix, banach}.
Since such inequalities are straightforward but lengthy to state, we do not enter the details
but prefer instead to work with scaling factors of the form~\eqref{scalefact}.

\subsection{Asymptotic Alignment}
The main difficulty in deriving a positive mass theorem without volume constraint is
to control the volume difference~$\tilde{V}-V$ in the positivity statement of Theorem~\ref{thm3}.
Here the basic problem is that, at the present stage, for given~$\tilde{\Omega}$ the set~$\Omega$
can be chosen arbitrarily. In order to remove this freedom, one needs to canonically identify~$\tilde{N}$
with~$N$ near infinity. We now explain how this can be done in the linearized setting
(the nonlinear setting will be treated in Section~\ref{seclvc}).
In the linearized description, the gravitating system is described by a jet~$\v$. The freedom in identifying~$\tilde{N}$ with~$N$
corresponds to the fact that~$\v$ is determined only up to inner solutions. Thus we have the freedom to
transform~$\v$ according to
\beq \label{vinner2}
\v \rightarrow \v + \u \:,
\eeq
where~$\u = (\div \bu, \bu)$ is an inner solution. This inner solution changes the linearized quasilocal mass,
as is worked out in the next lemma (a similar statement in the four-dimensional case is given 
in~\cite[Proposition~3.5]{fockbosonic}).
\begin{Lemma} For an inner solution~$\u = (\div \bu, \bu)$, the linearized quasilocal mass reduces to
the flux integral
\[ \Mass(\Omega, \u) = \s \int_{\partial \Omega} \sum_{\alpha=1}^3 \bu^\alpha(\x)\: \nu_\alpha(\x)\: d\mu_{\partial \Omega}(\x)\:. \]
\end{Lemma}
\Proof We compute the linearized quasilocal mass~\eqref{linosi} for the inner solution~$\v=\u$,
and apply the Gauss divergence theorem,
\begin{align*}
\Mass(\Omega, \u)
&= \int_{\Omega} \mu(\x) \int_{N \setminus \Omega} d\mu(\y) \:\sum_{\alpha=1}^3
\bigg( \frac{\partial}{\partial \x_\alpha} \Big( \bu^\alpha(\x)\: \L(\x,\y) \Big)
- \frac{\partial}{\partial \y_\alpha} \Big( \bu^\alpha(\y)\: \L(\x,\y) \Big) \bigg) \\
&= \int_{\partial \Omega} d\mu_{\partial \Omega}(\x) \int_{N \setminus \Omega} d\mu(\y) \:
\sum_{\alpha=1}^3 \nu_\alpha(\x) \,\bu^\alpha(\x)\: \L(\x,\y) \\
&\quad\: + \int_\Omega d\mu(\x) \int_{\partial \Omega} d\mu_{\partial \Omega}(\y) \: 
\sum_{\alpha=1}^3 \nu_\alpha(\y) \,\bu^\alpha(\y)\: \L(\x,\y) \\
&= \int_{\partial \Omega} d\mu_{\partial \Omega}(\x)\:\sum_{\alpha=1}^3
\nu_\alpha(\x)\, \bu^\alpha(\x) \int_N d\mu(\y) \: \L(\x,\y) \:.
\end{align*}
Carrying out the last integral with the help of~\eqref{ldef} and~\eqref{EL} gives the result.
\QED

This lemma shows that, in order to get a well-defined linearized quasilocal mass, we must fix the
freedom to add inner solutions~\eqref{vinner2}. To this end, it turns out to be useful to demand that the
vector field~$A^{(0)}(\bzeta)$ defined in~\eqref{Aalphadef} must vanish.
\begin{Def} \label{defasyalign}
The jet~$\v \in \Jlin$ is {\bf{asymptotically aligned}} if there is a compact set~$K \subset N$
such that
\beq \label{asyalign}
0 = A^{(0)}_\alpha(\bzeta) :=
\int_N \bxi_\alpha\:
\nabla_{1,\v} \L\Big( \bzeta - \frac{\bxi}{2}, \bzeta + \frac{\bxi}{2} \Big)\: d^3\bxi
\eeq
for all~$\alpha \in \{1,2,3\}$ and~$\bzeta \in N \setminus K$.
The corresponding total mass~\eqref{Mtot} is referred to as the 
{\bf{aligned total mass}}~$\Mass^\text{\rm{\tiny{aligned}}}$.
\end{Def}
Clearly, we need to verify that this condition can be satisfied and that it determines the inner solution
uniquely. In preparation, we compute the alignment vector field for an inner solution~$\v=\u$.
In the computations, we always assume that the Lagrangian on~$N$ is {\em{translation invariant}},
meaning that~$\L(\x,\y)$ depends only on the difference vector~$\y-\x$. We use the short notation
with square brackets,
\beq \label{ti}
\L(\x,\y) = \L[\bxi] \qquad \text{for all~$\x, \y \in N \simeq \R^3$ and~$\bxi := \y-\x$}\:.
\eeq
Moreover, we assume that the Lagrangian in the vacuum is spherically symmetric, i.e.\
\beq \label{spherical}
\L[\bxi] = \L \big[ \bxi' \big] \qquad \text{for all~$\bxi, \bxi' \in \R^3$ with~$|\bxi|=|\bxi|$} \:. 
\eeq

\begin{Lemma} \label{lemmaA0inner} Assume that the Lagrangian in the vacuum is
translation invariant~\eqref{ti} and spherically symmetric~\eqref{spherical}.
Let~$\bu=(a:=\div \bu, \bu)$ be an inner solution. Then the corresponding
alignment vector field is given by
\beq \label{Ainner}
A^{(0)}_\alpha(\bzeta) = \s\,\bu_\alpha(\x) + \frac{1}{24} \: \delta^2\, \s_2 \: \Delta_{\R^3} \bu_\alpha(\x)
- \frac{1}{12}\:\delta^2\, \s_2\: \partial_\alpha a(\x)
 + \O\Big(\frac{\delta^3}{\ell_\macro^3} \Big) \:,
\eeq
where~$\Delta_{\R^3}$ denotes the Laplacian in~$\R^3$,
$\delta$ is again the range of the Lagrangian (see below~\eqref{shortrange}) and
\beq \label{s2def}
\s_2 := \frac{1}{\delta^2} \int_N |\bxi|^2 \L[\bxi]\: d^3\bxi \:.
\eeq
\end{Lemma}
\Proof We denote the two arguments of the Lagrangian as usual by~$\x$ and~$\y$. Thus
\begin{align*}
\bzeta &= \frac{1}{2} \:(\y+\x) \:,&\hspace*{-2.2cm} \bxi &= \y-\x \\
\x &= \bzeta-\frac{\bxi}{2} \:,&\hspace*{-2.2cm} \y &= \bzeta + \frac{\bxi}{2} \:.
\end{align*}
Then the partial derivatives transform according to
\[ \frac{\partial}{\partial \bxi_\alpha}
= \frac{1}{2} \Big( \frac{\partial}{\partial \y_\alpha} - \frac{\partial}{\partial \x_\alpha} \Big) \:,\qquad
\frac{\partial}{\partial \bzeta_\alpha}
= \frac{\partial}{\partial \y_\alpha} + \frac{\partial}{\partial \x_\alpha} \:. \]
We also make use of the fact that the vacuum Lagrangian depends only on the difference vector~$\bxi$; we write
\[ \L(\x, \y) = \L[\bxi] \:. \]
This implies that
\[ \frac{\partial}{\partial \x_\alpha} \L(\x, \y) = -\frac{\partial}{\partial \bxi_\alpha} \L[\xi] \qquad \text{and} \qquad
\frac{\partial}{\partial \y_\alpha} \L(\x, \y) = \frac{\partial}{\partial \bxi_\alpha} \L[\xi] \:.\]
Using these relations, we obtain
\begin{align*}
A^{(0)}\alpha(\bzeta) &= \int_N \bxi_\alpha\: a(\x)\:  \L[\bxi]\: d^3\bxi
+ \int_N \bxi_\alpha\: \bu_\beta(\x)\: \frac{\partial}{\partial \x_\beta} \L[\bxi]\: d^3\bxi \\
&= \int_N \bxi_\alpha\: a(\x)\:  \L[\bxi]\: d^3\bxi
- \int_N \bxi_\alpha\: \bu_\beta(\x)\: \frac{\partial}{\partial \bxi_\beta} \L[\bxi]\: d^3\bxi \:.
\end{align*}
Integrating by parts gives
\begin{align*}
A^{(0)}\alpha(\bzeta) &= \int_N \Big( \bxi_\alpha\: a(\x) + \bu_\alpha(\x) \Big)\: \L[\bxi]\: d^3\bxi
+ \int_N \bxi_\alpha\: \frac{\partial \bu_\beta(\x)}{\partial \bxi_\beta}\:  \L[\bxi]\: d^3\bxi \\
&= \int_N \Big( \bxi_\alpha\: a(\x) + \bu_\alpha(\x) \Big)\: \L[\bxi]\: d^3\bxi
- \frac{1}{2} \int_N \bxi_\alpha\: \div \bu(\x)\:  \L[\bxi]\: d^3\bxi \\
&= \int_N \Big( \bu_\alpha(\x)  + \bxi_\alpha\: \frac{a(\x)}{2} \Big)\: \L[\bxi]\: d^3\bxi \:.
\end{align*}
We next expand~$a$ and~$\u_\alpha$ in a Taylor series about~$\bzeta$. This gives
\begin{align}
A^{(0)}\alpha(\bzeta) &= 
\int_N \Big( \bu_\alpha(\bzeta) - \frac{1}{2}\: \bxi^\beta \partial_\beta
\bu_\alpha(\x) + \frac{1}{8}\: \bxi^\beta \bxi^\gamma\: \partial_{\beta \gamma}
\bu_\alpha(\x) \Big)\: \L[\bxi]\: d^3\bxi \notag \\
&\quad\:+ \int_N \Big( \frac{1}{2}\: \bxi_\alpha\: a(\bzeta) - \frac{1}{4}\:
\bxi_\alpha \bxi^\beta \: \partial_\beta a(\bzeta) \Big)\: \L[\bxi]\: d^3\bxi + \O \Big( \frac{\delta^3}{\ell_\macro^3}
\Big) \:. \label{Aexp2}
\end{align}
Now we can use spherical symmetry of the unperturbed Lagrangian to obtain the formulas
\beq \label{linLcomp}
\begin{split}
\int_N \bxi_\alpha\: \L[\bxi]\: d^3\y &= 0 \\\
\int_N \bxi_\alpha \bxi_\beta \: \L[\bxi]\: d^3\y &= \frac{1}{3}\: \delta_{\alpha \beta} \int_N |\bxi|^2 \L[\bxi]\: d^3\bxi 
\overset{\eqref{s2def}}{=} \frac{1}{3}\: \delta^2\: \s_2\, \delta_{\alpha \beta} \:.
\end{split}
\eeq
Using these relations in~\eqref{Aexp2} gives the result.
\QED

This lemma immediately shows that asymptotic alignment can be arranged by a unique inner solution:
As explained after~\eqref{scalefact}, the first summand in~\eqref{Ainner} dominates
all the other terms. Then the inhomogeneous equation~$A^{(0)}[\u] = -A^{(0)}[\v]$ reduces to the local equation
\[ \u(x) = -\frac{1}{s}\: A^{(0)}[\v] \:, \]
which clearly has a unique solution. The errors can be treated with an iteration argument.

We remark that, in the present asymptotic expansion, the errors could be quantified in terms of
decay properties of the derivatives of~$\u$ like~$\|D^\kappa \u(\x)\| = \O(|\x|^{-|\kappa|})$.
Then the errors could be made arbitrarily small by choosing the compact set~$K$ sufficiently large.
The iteration argument could be carried out with the help of Banach's fixed point theorem.
For brevity, we do not enter the details of these standard constructions.

\subsection{Local Alignments and the Local Volume Condition} \label{seclvc}
We now return to the nonlinear setting with two measures~$\tilde{\mu}$ and~$\mu$.
Linearizing near infinity and choosing asymptotic alignment, we get a canonical identification of~$\tilde{N}$
and~$N$ in the asymptotic end. This also gives a unique identification of~$\tilde{\Omega}$ with~$\Omega$.
The next question is how to control the difference of inner volumes~$\tilde{V}-V$.
In order to analyze this question, we assume that~$\tilde{\mu}$ can be obtained from~$\mu$ by
a {\em{smooth deformation}}~$(f_\tau, F_\tau)$ with~$\tau \in [0,1]$ and~$f_\tau \in C^\infty(N, \R)$
and~$F_\tau \in C^\infty(N, \G)$, i.e.\
\beq \label{mtmrel}
(f_0, F_0) = \big(1_N, \text{\rm{id}}_N \big)
\qquad \text{and} \qquad \tilde{\mu} = (F_1)_* \big( f_1\, \mu)
\eeq
(where~$1_N$ is the constant function one and~$\text{\rm{id}}_N$ is the identity map on~$N$).
Before entering the construction, we point out that the smoothness of~$F_\tau$ 
poses restrictions for the topology of~$\tilde{N}$.
Indeed, in the typical situation that~$F_\tau$ is a diffeomorphism to its image, the topology of~$\tilde{N}$
must be trivial (interesting physical situations like a topology change or an event horizon will be
considered in Section~\ref{secbounds}).

Given~$\tau \in [0,1]$, we denote the corresponding space by~$\tilde{N}_\tau := \supp \tilde{\mu}_\tau$.
The infinitesimal change of the measure is described by the jet
\[ %\label{vtau}
\v_\tau := \frac{d}{d\tau} \big( f_\tau, F_\tau \big) \in \Jvary_\tau \:. \]
Given~$\bzeta \in \tilde{N}_\tau$, we assume that there is a canonical family of coordinate systems
around~$\bzeta$, with are centered in the sense that~$\x^\alpha(\bzeta)=0$ and
are unique up to linear transformations
\beq \label{lintrans}
\x^\alpha \rightarrow A^\alpha_\beta\: \x^\beta
\eeq
(with a real $3 \times 3$-matrix~$A$). Clearly, in the vacuum~$\tau=0$, one
takes the linear coordinates of~$N \simeq \R^3$. If~$\tilde{N}_\tau$ comes from a causal fermion system
constructed in a Riemannian or Lorentzian spacetime (as in the examples in Section~\ref{secultra}
and Appendix~\ref{appA}), then one can choose the coordinates obtained
from the exponential map.
Alternatively, for causal fermion systems a distinguished family of coordinates
is obtained by restricting the so-called symmetric wave charts~\cite{gaugefix, banach},
being charts on~$\G$, to~$N$.
\Felix{Wie geht das genau?}%
More generally, for causal variational principles one can work with
the Gaussian-type coordinates as constructed in~\cite{gauss}.
Here we do not need to specify how precisely the coordinate systems are chosen. Instead, we simply
assume that we are given canonical centered charts around~$\bzeta$, which are unique up to linear transformations~\eqref{lintrans}.

Working in these coordinates, similar to~\eqref{asyalign} we impose the {\em{local alignment condition}}
\beq \label{lac}
\int_N \bxi_\alpha\: \nabla_{1,\v_\tau} \L\Big( \bzeta - \frac{\bxi}{2}, \bzeta + \frac{\bxi}{2} \Big) \:
d\tilde{\mu}_\tau (\bxi) = 0\:.
\eeq
Exactly as explained for asymptotic alignment, this condition can be satisfied
by adding to~$\v_\tau$ a unique inner solution~$\u_\tau$. For clarity, we denote the
jet constructed in this way from coordinates around a point~$\x \in \tilde{N}_\tau$ by~$\v_{\tau, \x}$.

Having carried out this construction for every~$\x \in \tilde{N}_\tau$, we obtain a unique global
vector field~$\bv_\tau$ by setting
\[ \hat{\bv}_\tau(\x) = \bv_{\tau, \x}(\x) \:. \]
We assume that this vector field is again smooth.
We choose the scalar component~$\hat{a}_\tau$ such that the corresponding
jet~$\hat{\v}_\tau = (\hat{a}_\tau, \hat{\bv}_\tau)$ describes the variation of measures. More precisely, we choose
\[ \hat{a}_\tau(\x) = a_{\tau, \y}(\x) + \div \hat{\bv}_\tau(\x) - \div \bv_{\tau, \y}(\x) \]
for an arbitrary base point~$\y \in \tilde{N}_\tau$. We refer to~$\hat{\v}_\tau$ as the {\em{locally aligned jet}}.
We note for clarity that the locally aligned jet will in general {\em{not}} satisfy the local alignment condition
(i.e., the equation obtained by replacing~$\v_\tau$ in~\eqref{lac} by~$\hat{\v}_\tau$ will be violated
by an error term). But it is asymptotically aligned as in Definition~\ref{defasyalign}
(because the error term decays at infinity).

Performing this construction for every~$\tau \in [0,1]$, we obtain a family of jets~$(\hat{\v}_\tau)_{\tau \in [0,1]}$
which define a unique flow~$(\hat{f}_\tau, \hat{F}_\tau)_{\tau \in [0,1]}$ via
\[ \frac{d}{d\tau} \big( \hat{f}_\tau, \hat{F}_\tau \big) = \hat{\v}_\tau \:. \]
By construction, the resulting flow of measures coincides with the original flow, i.e.\
\[ (\hat{F}_\tau)_* \big( \hat{f}_\tau\, \mu) = (F_\tau)_* \big( f_\tau\, \mu) \qquad \text{for all~$\tau \in [0,1]$} \]
(but the flows differ by an inner solution for every~$\tau$).

We now demand that this flow increases the volume:
\begin{Def} \label{deflvc}
The smooth deformation~$(f_\tau, F_\tau)$ satisfies the {\bf{local volume condition}} if the
locally aligned jets~$\hat{\v}_\tau = (\hat{a}_\tau, \hat{\bv}_\tau)$ have a non-negative scalar component, i.e.\
\beq \label{lvc}
\hat{a}_\tau(\x) \geq 0 \qquad \text{for all~$\tau \in [0,1]$ and~$\x \in \tilde{N}_\tau$}\:.
\eeq
\end{Def}

We now explain what this condition means and how it can be verified in the applications.
The smooth deformation is obtained typically by continuously bringing matter fields into the system.
For example, in the setting of causal fermion systems one can introduce an additional physical
wave function~$\psi$ and can continuously increase its amplitude.
The local volume condition states that the aligned jets
only increase the volume. This condition can be understood similar to a local energy condition.
Instead of stating it for the energy-momentum tensor (which does not exist at the present level of generality),
it is a statement on the volume change when introducing matter.
We point out that the local volume condition can be verified at a point~$\x_0$ by
performing constructions in a neighborhood~$U$ of this point, provided that this
neighborhood is much larger than the range~$\delta$ of the Lagrangian
(here we assume again that the Lagrangian is of short range as introduced before~\eqref{shortrange}).
More precisely, choosing Gaussian-type coordinates around every~$\x \in U$ and satisfying
the local alignment condition~\eqref{lac} at~$\x$, we obtain a vector field~$\hat{\bv}$,
and computing its divergence at~$\x_0$ allows us to compute~$\hat{a}(\x_0)$ in~\eqref{lvc}.
In this way, similar to the usual local energy conditions, the local volume condition
can be verified locally.

We finally remark that, under additional assumptions on the smooth deformation~$(f_\tau, F_\tau)$,
the local volume condition can be stated in an alternative form which may be of advantage for computations.
We first state this alternative formulation and explain if afterward.
\begin{Prp} \label{prperror}
Assume that all the measures~$\tilde{\mu}_\tau := (F_\tau)_*(f_\tau \mu)$ are critical
in the sense that they satisfy the restricted EL equations~\eqref{ELrestricted}.
Then the local volume condition in Definition~\ref{deflvc} can be stated equivalently by demanding that
the divergence of the vector field~$A_\tau(\bzeta)$ in~\eqref{Aser} is non-negative,
\[ \div A_\tau(\bzeta) \geq 0 \qquad \text{for all~$\tau \in [0,1]$ and~$\bzeta \in \tilde{N}_\tau$}\:. \]
\end{Prp}
\Proof For any~$\tau \in [0,1]$ and~$\x_\tau \in \tilde{N}_\tau$, the restricted EL equations imply that
\[ 0 = \nabla_{\hat{\v}_\tau} \ell(\x_\tau) = \int_{\tilde{N}_\tau} \nabla_{1,\hat{\v}} \L(\x_\tau,\y)\: d\tilde{\mu}_\tau(\y)
- \big( \nabla_{\hat{\v}} \s \big)(\x) \]
with~$\x_\tau := F_\tau(\x)$.
On the other hand, differentiating the scalar component of the restricted EL equations with respect to~$\tau$
gives
\[ 0 = \frac{d}{d\tau} \ell\big( F_\tau(x) \big) = 
\int_{\tilde{N}_\tau} \big( \nabla_{1,\hat{\v}} + \nabla_{2, \hat{\v}} \big) \L(\x_\tau,\y)\: d\tilde{\mu}_\tau(\y)
- \big( \nabla_{\hat{\v}} \s \big)(\x) \:. \]
Using these equations, the infinitesimal volume change~\eqref{lvc} can be written as
\begin{align*}
\hat{a}_\tau(\x) &= \big( \nabla_{\hat{\v}} \s \big)(\x) 
= \int_{\tilde{N}_\tau} \big( \nabla_{1,\hat{\v}_\tau} - \nabla_{2,\hat{\v}_\tau} \big) \L(\x,\y)\: d\tilde{\mu}_\tau(\y)
= \div A_\tau(\x) \:,
\end{align*}
where in the last step we applied Theorem~\ref{thmdivA}.
\QED

This result can be used in computations as follows. First, in order to arrange that the measures~$\tilde{\mu}_\tau$
are all critical, one can solve the inhomogeneous linearized field equations for any~$\tau$
for example using the methods developed in~\cite{static}.
Once this has been accomplished, one can compute the vector field~$A_\tau(\bzeta)$
in~\eqref{Aser} term by term by evaluating the integrals~\eqref{Aalphadef}.
As explained after~\eqref{scalefact}, this gives an expansion in powers of the range~$\delta$
of the Lagrangian, making it possible to truncate the series.
We now illustrate this procedure for the contribution~$\sim \delta^2$
(as we shall see in the examples of Section~\ref{secultra} and Appendix~\ref{appA}, this is
indeed the contribution which gives the correspondence to the ADM mass and the Brown-York mass).
To this end, we expand according to
\beq \label{divAexp}
\div A_\tau(\bzeta) = \div A^{(0)}_\tau(\bzeta) + \div A^{(1)}_\tau(\bzeta) + \O \big( \delta^4 \big) \:.
\eeq
In order to get rid of the first summand, we arrange that the local alignment condition~\eqref{lac}
holds, i.e.\
\beq \label{A0zero}
0 = A^{(0)}_\tau(\bzeta) :=
\int_N \bxi_\alpha\: \nabla_{1,\hat{\v}_\tau} \L\Big( \bzeta - \frac{\bxi}{2}, \bzeta + \frac{\bxi}{2} \Big) \:
d\tilde{\mu}_\tau (\bxi) = 0 \:.
\eeq
Then the local volume condition can be stated as
\beq \label{A1div}
\div A^{(1)}_\tau(\bzeta) \geq 0 + \O \big( \delta^4 \big) \qquad \text{for all~$\tau \in [0,1]$
and~$\bzeta \in \hat{N}_\tau$}\:.
\eeq
In the above computation it is very convenient that in~\eqref{A0zero} we may allow for an error term
of the order
\beq \label{A0error}
A^{(0)}_\tau(\bzeta) = \O \big( \delta^2 \big) \:.
\eeq
Indeed, this error determines the inner solutions up to error terms of order~$\O(\delta^2)$.
This error of the inner solutions gives rise to an error in~$A^{(1)}$ which is of the order~$\O(\delta^4)$
and can thus be absorbed into the error term in~\eqref{A1div}.
Therefore, the condition~\eqref{A1div} is unaffected by the error in~\eqref{A0error}
(note that this is not apparent when analyzing the error terms in the expansion~\eqref{divAexp}).

\subsection{Statement and Proof of the Positive Mass Theorem} 
We are now in the position for stating and proving the positive mass theorem without volume constraint.

\begin{Thm} {\bf{(Positive mass theorem for aligned total mass)}} \label{pmt}
Let~$\mu$ be a minimizer with respect to variations of finite volume which has infinite total volume~\eqref{inftotvol}.
Moreover, let~$\tilde{\mu}$ be a measure on~$\G$ which has the property~\eqref{tellconst}
and is asymptotically flat (see Definition~\ref{defasyflat}).
Moreover, assume that~$\tilde{\mu}$ can be obtained from~$\mu$
by a smooth deformation~$(f_\tau, F_\tau)$ which satisfies the local volume condition
(see Definition~\ref{deflvc}). Then the aligned total mass is non-negative,
\[ \Mass^\text{\rm{\tiny{aligned}}} \geq 0 \:. \]
\end{Thm}
\Proof The resulting variation~$(\hat{f}_\tau, \hat{F}_\tau)$ increases the volume and is asymptotically aligned.
Therefore, applying Theorem~\ref{thm3} choosing~$\tilde{\Omega} = \hat{F}_1(\Omega)$, we obtain
\begin{align*}
\Mass = \lim_{\Omega \nearrow N} \Mass \big( \tilde{\Omega} , \Omega \big)
\geq \s \big( \tilde{V} - V \big) \geq 0 \:.
\end{align*}
This concludes the proof.
\QED

Computing this aligned total mass in the Schwarzschild geometry, one finds that this mass
coincides, up to a prefactor, with the ADM mass~$M$.
\begin{Thm} \label{pmtschwarz}
Identifying the asymptotic ends with linear alignment, the total mass in the Schwarzschild geometry
is given by
\beq \label{Massaligned}
\Mass = \frac{\pi}{9}\: \delta^2 \, \s_2\: M \:.
\eeq
\end{Thm} \noindent
The proof is given in Appendix~\ref{appA} (see page~\pageref{proofpmtschwarz}).

\section{An Equivariant Positive Mass Theorem} \label{secequi}
In the context of causal fermion systems, the two causal fermion systems describing the vacuum and
the gravitating system are defined on two different Hilbert spaces. Before defining the mass, one needs to unitarily
identify these Hilbert spaces. The fact that this identification is not canonical, gives rise to
the freedom to unitarily transform these measures (this point, which is of central importance
in getting the connection to quantum field theory, is described in detail in~\cite[Section~3]{fockfermionic}).
We now generalize this unitary freedom to the setting of causal variational principles, where it corresponds
to the freedom in performing isometries of the Lagrangian. This freedom will also be important
in our definitions of the quasilocal mass and synthetic scalar curvature
in Sections~\ref{secquasilocal} and~\ref{secsynthetic}.

The following notion has been introduced similarly in~\cite[Section~3.1]{noether}.
\begin{Def} \label{defiso}
A diffeomorphism~$\Phi \in C^\infty(\G, \G)$ describes a {\bf{symmetry of the Lagrangian}} if
\[ \L\big(\Phi(\x), \Phi(\y) \big) = \L(\x,\y) \qquad \text{for all~$\x,\y \in \G$}\:. \]
\end{Def}
Such diffeomorphisms clearly form a group, denoted by~${\mathcal{G}}$, the {\em{group of symmetries of the Lagrangian}}.
Clearly, transforming the vacuum measure by a diffeomorphism~$\Phi \in {\mathcal{G}}$ does not
leaves the causal action and the Euler-Lagrange equations unchanged. In particular,
\[ \int_\G \L\big( \Phi(\x),\y \big)\: d\big(\Phi_*\mu\big)(\y) = \int_\G \L(\x,\y)\: d\mu(\y)
\qquad \text{for all~$\x \in \G$}\:. \]
In order to clarify the dependence on the measures, following the notation~\eqref{pososiintro}
in the introduction we also denote the aligned total mass (see Definition~\ref{defasyalign}) by~$\Mass^\text{\rm{\tiny{aligned}}}_{\tilde{\mu}, \mu}$.

\begin{Def} \label{defmassiso} The measure~$\tilde{\mu}$ is {\bf{asymptotically deformation-admissible}}
w.r.t.~$\mu$ if it is asymptotically flat w.r.t.~$\mu$ and if~$\tilde{\mu}$ can be obtained from~$\mu$
by a smooth deformation~$(f_\tau, F_\tau)$ which satisfies the local volume condition
(see Definition~\ref{deflvc}). 

The {\bf{equivariant aligned total mass}} is defined by
\begin{align*}
\Mass^\text{\rm{\tiny{equi}}} &:=
\inf \Big\{ \Mass^\text{\rm{\tiny{aligned}}}_{\tilde{\mu}, \Phi_*\mu} \;\Big|\;
\text{$\Phi \in {\mathcal{G}}$ with} \\
&\qquad\qquad \text{$\tilde{\mu}$ asymptotically deformation-admissible w.r.t.~$\Phi_* \mu$} \Big\} \:.
\end{align*}
\end{Def}

\begin{Thm} \label{pmtequi} Under the assumptions of Theorem~\ref{pmt}, also the
equivariant aligned total mass is non-negative,
\[ \Mass^\text{\rm{\tiny{equi}}} \geq 0 \:. \]
\end{Thm}
\Proof Theorem~\ref{pmt} implies that~$\Mass^\text{\rm{\tiny{aligned}}}_{\tilde{\mu}, \Phi_*\mu} \geq 0$
for all~$\tilde{\mu}$ which are asymptotically deformation-admissible w.r.t.~$\Phi_* \mu$.
Therefore, the infimum of all these aligned masses is again non-negative.
\QED

\section{A Positive Quasilocal Mass} \label{secquasilocal}
Given an open subset~$\tilde{\Omega} \subset \tilde{N}$ of the gravitating system,
the goal of this section is to introduce its quasilocal mass~$\Mass(\tilde{\Omega})$.
This quasilocal mass will be non-negative, and it will bound the total mass from below.
In preparation, we need to think about how to ``localize'' the above notions to the subset~$\tilde{\Omega}$.
Ideally, the quasilocal mass should depend only on~$\tilde{\mu}$ restricted to~$\tilde{\Omega}$,
but not on the ``ambient space'' $\tilde{N} \setminus \tilde{\Omega}$.
This requirement is indeed satisfied for the functional~$\Mass(\tilde{\Omega}, \Omega)$
in~\eqref{massOtOdef}, as becomes obvious when rewriting it as
\begin{align}
\Mass(\tilde{\Omega}, \Omega) &= 
2 \int_{\tilde{\Omega}} d\tilde{\mu}(\x) \int_{N \setminus \Omega} d\mu(\y)\: \L(\x,\y) \notag \\
&\quad\: + \int_{\tilde{\Omega}} d\tilde{\mu}(\x) \int_{\tilde{\Omega}} d\tilde{\mu}(\y)\: \L(\x,\y) 
-A - \s \tilde{V} \label{Mnew}
\end{align}
Likewise, this requirement is also satisfied for the functional in Proposition~\ref{thm4},
which we now denote and rewrite as follows,
\begin{align}
{\mathfrak{N}}(\tilde{\Omega}, \Omega) &:= 
\Mass(\tilde{\Omega}, \Omega) \;-\; 2 \s \tilde{V} + \s V  + \int_{\tilde{\Omega}} d\tilde{\mu}(\x) \int_{\tilde{N}} 
d\tilde{\mu}(\y)\: \L(\x,\y) \label{frakNdef} \\
&\;= 2 \int_{\tilde{\Omega}} d\tilde{\mu}(\x) \int_{N \setminus \Omega} d\mu(\y)\: \L(\x,\y) \notag \\
&\quad\: + \int_{\tilde{\Omega}} d\tilde{\mu}(\x) \int_{\tilde{\Omega}} d\tilde{\mu}(\y)\: \L(\x,\y) 
\:-\: A  - 2 \s \tilde{V} + \s V \label{Nnew}
\end{align}
(note that in the derivation of~\eqref{Mnew} we used~\eqref{tellconst}, whereas in~\eqref{Nnew}
we did not).
However, in order to satisfy the local alignment condition~\eqref{lac} for~$\bzeta \in \tilde{\Omega}$,
we need to integrate over all~$\bxi$ for which the integrand is non-zero.
This will in general make it necessary to evaluate the Lagrangian for
arguments~$\bzeta \pm \bxi/2$ which lie outside~$\tilde{\Omega}$.
This is the reason why our notion of quasilocal mass will not depend only on~$\tilde{\Omega}$,
but on a neighborhood~$\tilde{K}$ of~$\tilde{\Omega}$.
If we assumed that the Lagrangian were of short range~\eqref{shortrange}, then this
neighborhood could be chosen as the $\delta$-neighborhood~$\tilde{K}= \overline{B_\delta(\tilde{\Omega})}$.
More generally, the ``width of the boundary layer $\tilde{K} \setminus \tilde{\Omega}$ should be
large on the scale~$\delta$.'' In order to make this statement mathematically precise, we
adapt the notion of {\em{compact range}} introduced in~\cite[Def.~3.3]{noncompact}
to our setting.
\begin{Def} \label{defrealize} The set~$\tilde{\Omega}$ can be realized by
a {\bf{local smooth deformation of~$\Omega$ inside~$K$}} if 
there is a pair of mappings~$(f, F) : [0,1] \times N \rightarrow \R^+_0 \times \G$ with the following properties:
\bitem
\item[{\rm{(i)}}] Both mappings~$f$ and~$F$ are measurable.
Moreover, their restrictions to~$K$ are smooth, i.e.\
\[ f|_{[0,1] \times K} \in C^\infty\big( [0,1] \times K, \R+_0 \big) \:,\qquad
F|_{[0,1] \times K} \in C^\infty\big( [0,1] \times K, \G \big) \:. \]
\item[{\rm{(ii)}}] The mappings~$f$ and~$F$ have the correct boundary values at~$\tau=0$ and~$\tau=1$
inside~$K$; i.e., weakening the conditions~\eqref{mtmrel},
\begin{gather*}
f_0 = 1_N \:, \quad F_0 = \text{\rm{id}}_N \\
F_1(N) \subset \tilde{N} \qquad \text{and} \qquad (F_1)_* \big(f_1\, \mu|_K \big) = \tilde{\mu}|_{F_1(K)} \:.
\end{gather*}
\item[{\rm{(iii)}}] Setting~$N_\tau = F_\tau(N)$, $K_\tau := F_\tau(K)$ and~$\Omega_\tau := F_\tau(\Omega)$,
for any~$\tau \in [0,1]$ and~$\bzeta \in \Omega_\tau$ we again choose the distinguished coordinates
centered at~$\bzeta$ unique up to linear transformations~\eqref{lintrans}.
We assume that, in these coordinates, for every~$\y \in N_\tau \setminus \Omega_\tau$
and every~$\x_0 \in N$ with~$F_\tau(\x_0) = 2 \bzeta - \y$, there is~$\varepsilon>0$ such that
\[ \L\big( F_{\tau'}(\x_0),\y \big) = 0 \qquad \text{for all~$\tau' \in [0,1]$ with~$|\tau'-\tau|<\varepsilon$} \:. \]
\eitem
\end{Def} \noindent
The last condition ensures that, for any~$\bzeta \in \Omega_\tau$,
the integrand in~\eqref{lac} is well-defined and vanishes whenever one of the arguments~$\bzeta \pm \bxi/2$
lies outside~$K_\tau$. Therefore, the local alignment condition~\eqref{lac} is well-defined
for any~$\bzeta \in \Omega_\tau$.
We again consider the resulting flow~$(\hat{f}_\tau, \hat{F}_\tau)$ and choose~$\Omega \subset N$
such that
\[ \tilde{\Omega} = \hat{F}_1(\Omega) \:. \]
We assume that the local volume condition (see Definition~\ref{deflvc}) is satisfied on~$\Omega$, i.e.\
\beq \label{lvcloc}
\hat{a}_\tau(\x) \geq 0 \qquad \text{for all~$\tau \in [0,1]$ and~$\x \in \hat{F}_\tau(\Omega)$}\:.
\eeq
Then the volume difference~$\tilde{V} - V$ is positive. Therefore, Theorem~\ref{thm3} yields that
\[ \Mass \big(\tilde{\Omega}, \Omega \big) \geq 0 \:. \]
Following the procedure for the equivariant mass, we define the quasilocal mass by taking the
infimum over the isometries of the Lagrangian.

\begin{Def} {\bf{(Quasilocal mass)}} \label{defquasilocal}
Let~$\mu$ be a minimizer with respect to variations of finite volume which has infinite total volume~\eqref{inftotvol}.
Moreover, let~$\tilde{\mu}$ be a measure on~$\G$ which has the property~\eqref{tellconst}.
The subset~$\tilde{\Omega} \subset \tilde{N}$ is {\bf{deformation-admissible}} w.r.t.~$\mu$
for~$\Omega$ if~$\tilde{\Omega}$ can be realized by a local deformation of~$\Omega$ inside~$K$
(see Definition~\ref{defrealize}), and if the resulting deformation~$(\hat{f}, \hat{F})$
satisfies the local volume condition~\eqref{lvcloc}.
Given a subset~$\tilde{\Omega} \subset \tilde{N}$, we define the {\bf{quasilocal mass}} by
\beq \label{quasiloc}
\begin{split}
\Mass(\tilde{\Omega}) := \inf \Big\{ \Mass_{\tilde{\mu}, \Phi_* \mu} \big(\tilde{\Omega}, \Omega \big) \;\Big|&\;
\text{$\Phi \in {\mathcal{G}}, \Omega \subset \Phi(N)$ with $\tilde{\Omega}$ is} \\
&\;\; \text{deformation-admissible w.r.t.~$\Phi_* \mu$ for~$\Omega$} \Big\}
\end{split}
\eeq
(where for the function~$\Mass(.,.)$ we used again the notation with subscripts~\eqref{pososiintro}).
\end{Def}

\begin{Thm} \label{thmquasipos} The quasilocal mass is non-negative,
\[ \Mass(\tilde{\Omega}) \geq 0 \:. \]
\end{Thm}
\Proof Follows exactly as the proofs of Theorems~\ref{pmt} and~\ref{pmtequi}.
\QED

\section{Bounding the Total Mass by the Quasilocal Mass} \label{secbounds}
\begin{figure}
\psset{xunit=.5pt,yunit=.5pt,runit=.5pt}
\begin{pspicture}(573.96361344,85.55514358)
{
\newrgbcolor{curcolor}{0.50196081 0.50196081 0.50196081}
\pscustom[linewidth=2.99999995,linecolor=curcolor]
{
\newpath
\moveto(204.75994205,63.39018358)
\curveto(211.09111559,66.7183335)(218.2772674,68.40326957)(225.42795969,68.23622957)
\curveto(237.0172989,67.96550201)(247.91514709,62.99219933)(258.39525921,58.03708752)
\curveto(268.87536756,53.08193791)(279.66796346,47.95943083)(291.24473953,47.35584027)
\curveto(303.10220598,46.73762295)(314.63131465,50.92163555)(325.70359181,55.20958516)
\curveto(336.77583118,59.49757256)(348.10311307,64.0196261)(359.97623433,64.12310957)
\curveto(368.62069417,64.19832216)(377.27350677,61.8500261)(384.69373984,57.41467492)
}
}
{
\newrgbcolor{curcolor}{0.50196081 0.50196081 0.50196081}
\pscustom[linestyle=none,fillstyle=solid,fillcolor=curcolor]
{
\newpath
\moveto(223.19366551,70.12834043)
\lineto(247.0120252,77.73830579)
\lineto(276.3481474,82.79625854)
\lineto(295.33951748,83.80784909)
\lineto(320.12243906,80.77307744)
\lineto(338.14711937,76.26690169)
\lineto(367.29933354,65.13942453)
\lineto(349.09070362,62.74838201)
\lineto(333.82488945,57.94959146)
\lineto(313.96093984,50.7932072)
\lineto(295.66035402,46.93079461)
\lineto(279.4748863,49.68966075)
\lineto(248.39146583,63.48406705)
\closepath
}
}
{
\newrgbcolor{curcolor}{0.50196081 0.50196081 0.50196081}
\pscustom[linewidth=0.99999995,linecolor=curcolor]
{
\newpath
\moveto(223.19366551,70.12834043)
\lineto(247.0120252,77.73830579)
\lineto(276.3481474,82.79625854)
\lineto(295.33951748,83.80784909)
\lineto(320.12243906,80.77307744)
\lineto(338.14711937,76.26690169)
\lineto(367.29933354,65.13942453)
\lineto(349.09070362,62.74838201)
\lineto(333.82488945,57.94959146)
\lineto(313.96093984,50.7932072)
\lineto(295.66035402,46.93079461)
\lineto(279.4748863,49.68966075)
\lineto(248.39146583,63.48406705)
\closepath
}
}
{
\newrgbcolor{curcolor}{0.50196081 0.50196081 0.50196081}
\pscustom[linewidth=2.99999995,linecolor=curcolor]
{
\newpath
\moveto(-0.00058583,11.2156198)
\curveto(39.4924611,13.94470327)(78.98471433,16.67374894)(129.3201411,34.33733319)
\curveto(179.65557165,52.00087964)(240.8289222,84.59689728)(295.30771654,83.46354169)
\curveto(349.78651465,82.3301861)(397.56424063,47.4676009)(442.88882646,29.81199146)
\curveto(488.21341228,12.15638201)(531.08024693,11.70986862)(573.94795087,11.26331744)
}
}
{
\newrgbcolor{curcolor}{1 1 1}
\pscustom[linestyle=none,fillstyle=solid,fillcolor=curcolor]
{
\newpath
\moveto(270.55449827,67.62240027)
\lineto(276.25618772,69.92147177)
\lineto(283.61320819,71.39287587)
\lineto(289.59078803,71.57678768)
\lineto(296.57995087,70.65715681)
\lineto(303.1093115,68.45005256)
\lineto(309.91457764,64.40370169)
\lineto(305.1324926,62.28856468)
\lineto(300.62631307,60.7252009)
\lineto(293.26929638,59.98947807)
\lineto(285.17657575,60.81715681)
\lineto(276.89992819,63.20819933)
\lineto(272.57767559,65.78315366)
\closepath
}
}
{
\newrgbcolor{curcolor}{1 1 1}
\pscustom[linewidth=0,linecolor=curcolor]
{
\newpath
\moveto(270.55449827,67.62240027)
\lineto(276.25618772,69.92147177)
\lineto(283.61320819,71.39287587)
\lineto(289.59078803,71.57678768)
\lineto(296.57995087,70.65715681)
\lineto(303.1093115,68.45005256)
\lineto(309.91457764,64.40370169)
\lineto(305.1324926,62.28856468)
\lineto(300.62631307,60.7252009)
\lineto(293.26929638,59.98947807)
\lineto(285.17657575,60.81715681)
\lineto(276.89992819,63.20819933)
\lineto(272.57767559,65.78315366)
\closepath
}
}
{
\newrgbcolor{curcolor}{0.50196081 0.50196081 0.50196081}
\pscustom[linewidth=2.99999995,linecolor=curcolor]
{
\newpath
\moveto(0.40141228,1.68943398)
\lineto(573.3260674,1.89844185)
}
}
{
\newrgbcolor{curcolor}{0 0 0}
\pscustom[linewidth=3.99998762,linecolor=curcolor]
{
\newpath
\moveto(0.00000378,2.32499933)
\lineto(64.04116157,2.30610169)
}
}
{
\newrgbcolor{curcolor}{0 0 0}
\pscustom[linewidth=3.99998762,linecolor=curcolor]
{
\newpath
\moveto(502.68405165,1.79393791)
\lineto(573.06195402,1.99992216)
}
}
{
\newrgbcolor{curcolor}{0 0 0}
\pscustom[linewidth=3.99998762,linecolor=curcolor]
{
\newpath
\moveto(205.91960315,1.79393791)
\lineto(386.67398551,1.79393791)
}
}
{
\newrgbcolor{curcolor}{0 0 0}
\pscustom[linewidth=3.99998762,linecolor=curcolor]
{
\newpath
\moveto(65.48597858,17.92434963)
\curveto(74.17883528,19.57377325)(82.84240063,21.37759057)(91.47120945,23.33462994)
\curveto(100.57853669,25.40021735)(109.64714268,27.63648821)(118.67066646,30.04189294)
\curveto(127.76160567,33.11827719)(136.77642898,36.41954333)(145.70411906,39.94172286)
\curveto(154.33832126,43.34809766)(162.88960063,46.96045672)(171.4883811,50.45527246)
\curveto(182.65832126,54.99508979)(193.90880693,59.33675404)(205.23275906,63.47749861)
}
}
{
\newrgbcolor{curcolor}{0 0 0}
\pscustom[linewidth=3.99998762,linecolor=curcolor]
{
\newpath
\moveto(384.38482016,57.48855259)
\curveto(393.51018709,53.19507731)(402.60176504,48.82976078)(411.65838236,44.39308676)
\curveto(422.48170205,39.09097653)(433.27225323,33.67937338)(444.49321701,29.28083479)
\curveto(454.24821543,25.45693558)(464.29983118,22.41135448)(474.48724535,19.96210724)
\curveto(483.78306898,17.72723479)(493.19775874,15.98676235)(502.67821228,14.75051668)
}
}
{
\newrgbcolor{curcolor}{0 0 0}
\pscustom[linewidth=0.99999871,linecolor=curcolor]
{
\newpath
\moveto(264.62847496,72.13794169)
\curveto(271.45451717,64.8100009)(281.24868661,60.33050484)(291.2564674,59.95924185)
\curveto(301.26267213,59.58809224)(311.36024693,63.32827492)(318.71014299,70.12834043)
}
}
{
\newrgbcolor{curcolor}{0 0 0}
\pscustom[linewidth=0.99999871,linecolor=curcolor]
{
\newpath
\moveto(270.31623307,67.15941571)
\curveto(276.82601953,70.57133886)(284.34194646,72.03397445)(291.65615622,71.31226988)
\curveto(298.41733039,70.64513413)(304.98584693,68.12053823)(310.45338709,64.0876198)
}
\rput[bl](10,60){$\G$}
\rput[bl](580,-10){$N$}
\rput[bl](540,20){$\tilde{N}$}
\rput[bl](400,60){$\supp \hat{\mu}$}
\rput[bl](332,33){$\tilde{U}$}
\rput[bl](100,45){$\tilde{K} \setminus \tilde{U}$}
\rput[bl](280,10){$U$}
}
\end{pspicture}
\caption{The measure~$\hat{\mu}$.}
\label{figpos2}
\end{figure}
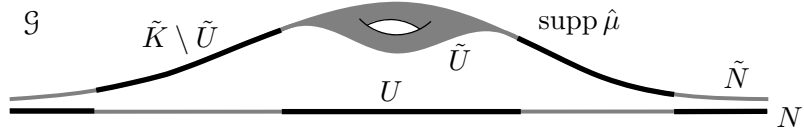

In this section we prove that the total mass is at least as large as the quasilocal mass.
\begin{Thm} \label{thmlower}
Let~$\mu$ be a minimizer with respect to variations of finite volume which has infinite total volume~\eqref{inftotvol}.
Moreover, let~$\tilde{\mu}$ be a measure on~$\G$ which has the property~\eqref{tellconst}
and is asymptotically flat (see Definition~\ref{defasyflat}).
Given sets~$U \subset N$ and~$\tilde{U} \subset \tilde{N}$ we assume that~$\tilde{N} \setminus \tilde{\Omega}$
can be realized by a local smooth deformation of~$N \setminus \Omega$ inside a set~$N \setminus K$
(see Definition~\ref{defrealize}) which satisfies the local volume condition
(see Definition~\ref{deflvc}). Then the {\bf{quasilocal mass}} of the subset~$\tilde{U} \subset \tilde{N}$
bounds the equivariant total mass by
\[ \Mass^\text{\rm{\tiny{equi}}} \geq \Mass \big(\tilde{U} \big) \:. \]
\end{Thm}
\Proof We choose a large relatively compact set~$\tilde{K} \subset \tilde{N}$ and apply Theorem~\ref{thm3} for~$\tilde{\Omega}$ replaced by~$\tilde{K} \setminus \tilde{\Omega}$.
(see Figure~\ref{figpos2}).
\QED
We point out that this theorem also applies in cases when~$\tilde{N}$ has a {\em{non-trivial
topology}}. All we need is that the exterior region~$\tilde{N} \setminus \tilde{U}$
is diffeomorphic to~$\R^3$ minus a ball. In other words, the non-trivial topology must be
included in the inner region~$\tilde{U}$, as is indicated by the handle in Figure~\ref{figpos2}.
Our theorem also applies in the case that the inner region~$\tilde{U}$ contains an event
horizon. In this case, the resulting inequalities are reminiscent and should be closely related
to the Riemannian Penrose inequality~\cite{huisken+ilmanen, bray}.
We plan to investigate this connection in a separate publication.

\section{A Synthetic Definition of Scalar Curvature} \label{secsynthetic}
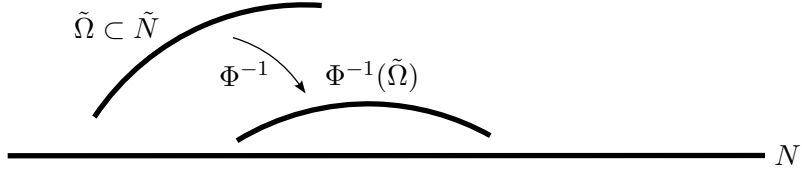
\begin{figure}
\psset{xunit=.5pt,yunit=.5pt,runit=.5pt}
\begin{pspicture}(572.9261119,118.26056293)
{
\newrgbcolor{curcolor}{0 0 0}
\pscustom[linewidth=3.99998762,linecolor=curcolor]
{
\newpath
\moveto(0.00072945,2.0000086)
\lineto(572.92538457,2.20901647)
}
}
{
\newrgbcolor{curcolor}{0 0 0}
\pscustom[linewidth=3.99998762,linecolor=curcolor]
{
\newpath
\moveto(173.54762154,13.41394523)
\curveto(204.88200643,32.58019404)(241.87073461,42.37132475)(278.58391635,41.21766176)
\curveto(308.77240517,40.26903814)(338.70452485,31.98423814)(365.08441776,17.27556412)
}
}
{
\newrgbcolor{curcolor}{0 0 0}
\pscustom[linewidth=3.99998762,linecolor=curcolor]
{
\newpath
\moveto(65.62493908,31.3241446)
\curveto(85.92352934,61.93714843)(115.30879109,86.44303844)(149.06979263,100.91312574)
\curveto(176.83076079,112.81158743)(207.45978172,117.95286907)(237.58423833,115.77090776)
}
}
{
\newrgbcolor{curcolor}{0 0 0}
\pscustom[linewidth=0.99999871,linecolor=curcolor]
{
\newpath
\moveto(170.45831433,91.4188571)
\curveto(181.6067263,88.27918844)(192.15056504,83.00726907)(201.3513411,75.97233616)
\curveto(208.75650142,70.3103393)(215.29212472,63.51329364)(220.65949228,55.89187631)
}
}
{
\newrgbcolor{curcolor}{0 0 0}
\pscustom[linestyle=none,fillstyle=solid,fillcolor=curcolor]
{
\newpath
\moveto(222.58760301,59.47577712)
\lineto(225.37521408,49.19576627)
\lineto(216.6355052,55.28402441)
\curveto(219.41699186,55.01684793)(221.89455258,56.761665)(222.58760301,59.47577712)
\closepath
}
\rput[bl](50,90){$\tilde{\Omega} \subset \tilde{N}$}
\rput[bl](160,55){$\Phi^{-1}$}
\rput[bl](240,50){$\Phi^{-1}(\tilde{\Omega})$}
\rput[bl](580,-5){$N$}
}
\end{pspicture}
\caption{Isometric transformation of~$\tilde{\Omega}$.}
\label{figscalar}
\end{figure}

In this section we explore the inequality in Theorem~\ref{thm3} in the case that~$\tilde{\Omega}$
and~$\Omega$ are small neighborhoods of given points~$\tilde{\x} \in \tilde{N}$ and~$\x \in N$.
In this case, the linearized description applies not only near the boundary of~$\tilde{\Omega}$
but in all of~$\tilde{\Omega}$. Therefore, the quasilocal mass inside the curly brackets in~\eqref{quasiloc}
simplifies to
\begin{align}
&\Mass_{\tilde{\mu}, \Phi_* \mu} \big(\tilde{\Omega}, \Phi \Omega \big)
= \int_\Omega d\mu(\x) \int_{N \setminus \Omega} d\mu(\y) \:\big( \nabla_{1,\v} - \nabla_{2,\v} \big) \L(\x, \y)
\notag \\
&= \int_\Omega d\mu(\x) \int_N d\mu(\y) \:\big( \nabla_{1,\v} - \nabla_{2,\v} \big) \L(\x, \y) 
= \int_\Omega \big( \nabla_\v \s - \Delta \v \big)(\x) \:d\mu(\x)\:, \label{masslin}
\end{align}
where the jet~$\v$ depends on the choice of~$\Phi$ (see Figure~\ref{figscalar}).
In order to define scalar curvature, we take the integrand in~\eqref{masslin}
and minimize it under variations of~$\Phi$. Following the procedure for the
quasilocal mass, we again assume linear alignment.
This leads us to the following definition.
\begin{Def} \label{defsynthetic}
We say that~$\Phi$ and the identification of~$\tilde{\Omega}$ with~$\Omega$ is
{\bf{deformation-admissible}} if the quasilocal mass can be described linearly by~\eqref{masslin}
and if the local alignment condition
\beq \label{localign}
\int_N \bxi_\alpha\:
\nabla_{1,\v} \L\Big( \bzeta - \frac{\bxi}{2}, \bzeta + \frac{\bxi}{2} \Big)\: d^3\bxi = 0
\eeq
holds. The {\bf{synthetic scalar curvature}}~$\text{\rm{scal}}$ at~$\x \in \Omega$ is defined by
\beq \label{scaldef}
\text{\rm{scal}}(\x) := \inf \big\{ \big( \nabla_\v \s - \Delta \v \big)(\x) \:\big|\:
\text{$\Phi$ and~$\Omega$ are deformation-admissible} \big\} \:.
\eeq
\end{Def} \noindent
Comparing this definition with that of the quasilocal mass (see Definition~\ref{defquasilocal}),
it is obvious in view of~\eqref{masslin} that
\[ \Mass(\tilde{\Omega}) \geq \int_\Omega \text{\rm{scal}}(\x)\: d\mu(\x) \:. \]
In this sense, our synthetic definition of scalar curvature fits together with
our notion of quasilocal mass.

\section{Example: Ultrastatic Spacetimes} \label{secultra}
We now consider the example of ultrastatic spacetimes.
In this case, the geometry is determined by the Riemannian metric on the Cauchy surfaces.
Our main goal is to show that the synthetic definition of scalar curvature introduced in
Section~\ref{secsynthetic} agrees in a well-defined limiting case with the scalar curvature of the
Riemannian metric. Moreover, we will work out similarities of the quasilocal mass with the
Brown-York mass.

\subsection{Construction of the Lagrangian}
In order to make the example as simple as possible, we keep the background on causal
fermion systems to a minimum. In particular, we will only introduce those objects which are
essential for the computations, namely the Lagrangian and its jet
derivatives~$\nabla_{1,\v} \L(\x,\y)$ and~$\nabla_{2,\v} \L(\x,\y)$.
However, we do not explain what the space~$\G$ is, because this would make it necessary
to introduce the so-called local correlation operators.
The reader who wants to get a deeper understanding of the connection to causal fermion systems
may find it helpful to consult the introduction to static causal fermion systems in~\cite[Section~3]{pmt},
the survey article~\cite{nrstg} or the textbooks~\cite{cfs, intro}.

We let~$(\scrN, g)$ be a three-dimensional orientable, complete and asymptotically flat Riemannian manifold.
We let~$\scrM := \R \times \scrN$ be the corresponding ultrastatic spacetime with the line element
\beq \label{ultraline}
ds^2 = dt^2 - g_{\alpha \beta}(\x)\: d\x^\alpha\, d\x^\beta \:.
\eeq
The completeness of~$\scrN$ implies that this spacetime is globally hyperbolic.
Next, being three-dimensional, the manifold~$\scrN$ is spin.
Let $\Dir_\scrN$ denote the intrinsic Dirac operator on~$\scrN$.
Using standard elliptic theory (see~\cite[Proposition~8.2.7]{taylor2} and~\cite{chernoff}),
the operator~$\Dir_\scrN$ with domain~$C^\infty_0(\scrN, S\scrN)$ is
essentially self-adjoint on the Hilbert space~$L^2(\scrN, S\scrN)$. Thus its closure, which we again denote
by~$\Dir_\scrN$, is a self-adjoint operator with domain~$\D(\Dir_\scrN)$. The spectral theorem yields
\[ %\label{ultrast_Dirac-op_spectraldecomp}
\Dir_\scrN = \int_\R \lambda\: dF_\lambda \:, \]
where $F_\lambda$ denotes the spectral-measure of $\Dir_\scrN$.

The Dirac operator in spacetime can be written in block matrix notation as
\[ %\label{Dirultra}
\Dir = \begin{pmatrix} i \partial_t & -\Dir_\scrN \\ \Dir_\scrN & -i \partial_t \end{pmatrix} . \]
Since the Dirac operator is time independent, we can separate the time dependence with a plane
wave ansatz,
\[ \psi(t,x) = e^{-i \omega t}\: \chi(x)\:. \]
The sign of~$\omega$ gives a natural decomposition of the solution space into two
subspaces. This is often referred to as ``frequency splitting,'' and the subspaces
are called the solutions of {\em{positive}} and {\em{negative frequency}}, respectively.

We now construct corresponding causal fermion systems.
To this end, we choose a parameter~$\varepsilon>0$, referred to as the {\em{regularization length}}
(this length should be thought of as being very small; it can be identified with the Planck length).
Given a parameter~$m$ (the {\em{rest mass}}) and setting
\beq \label{omegadef}
\omega(\lambda) := -\sqrt{k^2 + \lambda^2} \:,
\eeq
the {\em{regularized kernel of the fermionic projector}}~$P^\varepsilon(x,y)$ is defined by
\[ P^\varepsilon\big( (t,\x), (t', \y) \big)
:= - \int_\R \begin{pmatrix} \omega(\lambda) + m & -\lambda \\ \lambda & -\omega(\lambda) + m \end{pmatrix}\:
e^{-i \omega(k)\, (t-t')}\:e^{\varepsilon \omega} \: dF_\lambda \big(\x, \y\big) \:. \]
By direct computation, one verifies that this kernel satisfies the Dirac equation
\[ (\Dir_x - m) P(x,y) = 0 \:. \]
Due to the minus sign in~\eqref{omegadef}, this kernel is composed of all negative-frequency
solutions of the Dirac equation. The factor~$\exp(\varepsilon \omega)$ in the integrand can be
understood as a convergence-generating factor. In position space, it gives rise to a smoothing
of the kernel on the regularization scale.

We next compute the Lagrangian of the causal action principle (for details see for example~\cite[Section~1.1]{cfs}
or~\cite[Section~2.2]{pmt}).
To this end, we form the {\em{closed chain}}
\[ A_{xy} := P(x,y)\, P(y,x) \:. \]
Denoting its eigenvalues by~$\lambda^{xy}_1, \cdots, \lambda^{xy}_4 \in \C$, the Lagrangian in spacetime is
defined by
\[ \L(x,y) = \frac{1}{8} \sum_{i,j=1}^{4} \Big( \big|\lambda^{xy}_i \big| - \big|\lambda^{xy}_j \big| \Big)^2 \:. \]
The {\em{static Lagrangian}} is obtained by integrating over one time variable
(for more details on this construction see~\cite[Section~3]{pmt})
\beq \label{Ltime}
\L(\x, \y) := \int_{-\infty}^\infty \L \big((0,\x), (t,\y) \big) \: dt\:.
\eeq
We remark that this Lagrangian decays on the Compton scale~$m^{-1}$.
Therefore, choosing~$\delta \gg m^{-1}$, the assumption of short range as introduced
before~\eqref{shortrange} holds approximately, meaning that~$\L(\x,\y)$ and its derivative
is very small if~$d(\x,\y) > \delta$. Nevertheless, one should keep in mind that
the strict assumption~\eqref{shortrange} is a mathematical idealization.

Due to the microlocal nature of the construction, one may expect that the static Lagrangian
should depend only the geodesic distance~$d(\x,\y)$. This is indeed the case, up to errors of the
following form,
\beq \label{Lgeodesic}
\L(\x,\y) = \L\big[ d(\x,\y) \big] \;\bigg( 1 + \O \Big( \frac{\varepsilon}{\ell_{\macro}} \Big)
+ \O \Big( \frac{1}{m\,\ell_{\macro}} \Big) \bigg) \:.
\eeq
In what follows, we shall not use this formula (details can be found in~\cite[Appendix~A]{jacobson},
also based on~\cite[Section~5]{lqg}).
The reason is that, for getting the connection to the Brown-York mass and
scalar curvature, it will suffice to compute the surface layer integrals in regions of spacetime
which can be described by linearized gravity. This simplifies the computations considerably,
as we now explain in detail.

\subsection{Linearized Gravity} \label{secultralin}
In linearized gravity, one assumes that, in a suitable coordinate chart, the metric~$g_{\alpha \beta}$
differs from the Euclidean metric~$\delta_{\alpha \beta}$ by a small tensor denoted by~$h_{\alpha \beta}$,
\beq \label{glinab}
g_{\alpha \beta}(\x) = \delta_{\alpha \beta} + h_{\alpha \beta}(\x) \:.
\eeq
Here ``small'' means that we only take into account the tensor~$h_{\alpha \beta}$ linearly
(in other words, we work with error terms~$\O((h_{\alpha \beta})^2)$ throughout).
In this formalism, tensor indices are raised and lowered with respect to the Euclidean metric
(for details in the more general Lorentzian setting see for example~\cite[\S105]{landau2}).
Therefore, we do not need to distinguish between upper and lower indices.
For notational simplicity, we write all indices as lower indices
(this convention is useful mainly in view of keeping track of the signs in
the computations in Lorentzian signature in Appendix~\ref{appA}).
Using the Einstein summation convention, we sum over all double indices
(this is unproblematic because we only consider Euclidean coordinate transformations).

In the resulting formalism, one has
the freedom to perform infinitesimal coordinate transformations
\beq \label{infcoordultra}
\x'^\alpha = \x^\alpha + \bzeta^\alpha(\x) \:.
\eeq
This transforms the linearized metric according to
\beq \label{infmetricultra}
h'_{\alpha \beta} = h_{\alpha \beta} - \partial_\alpha \bzeta_\beta - \partial_\beta \bzeta_\alpha \:.
\eeq
The trace of the metric denoted by
\[ h := h_{\alpha \alpha} = \delta^{\alpha \beta}\, h_{\alpha \beta} \]
transforms according to
\[ %\label{htrans}
h' = h - 2 \partial_\alpha \bzeta_\alpha \:. \]
It follows that the volume form behaves as follows,
\beq \label{volform}
\sqrt{\det g'} = \sqrt{ 1 + h'} = 1 + \frac{h'}{2} = 1+ \frac{h}{2} - \partial_\alpha \bzeta_\alpha 
= \sqrt{\det g} - \partial_\alpha \bzeta_\alpha \:.
\eeq
Moreover, the Lagrangian transforms according to
\beq \label{delLagultra}
\delta \L(\x,\y) := \L\big(\x', \y' \big) - \L(\x,\y)
= -\Big( \bzeta_\alpha(\x)\: \frac{\partial}{\partial \x_\alpha} + \bzeta_\alpha(\y)\: \frac{\partial}{\partial \y_\alpha}
\Big) \L(\x, \y) \:.
\eeq
We remark that the relative sign of the two terms can be verified for example by integrating over~$y$ and using that
\begin{align*}
&\int_N  \bigg( \partial_\alpha \bzeta_\alpha(\y) + \bzeta_\alpha(\y)\: \frac{\partial}{\partial \y_\alpha}
\bigg) \L(\x, \y)\: d^3\y
= \int_N  \frac{\partial}{\partial \y_\alpha} \big( \bzeta_\alpha(\y)\: \L(\x, \y) \big)\: d^3\y = 0 \:,
\end{align*}
being consistent with the fact that the spacetime integral is diffeomorphism invariant.

\subsection{Description with the Diffeomorphism Jet and Inner Solutions} \label{secultrajets}
By translational symmetry, the unperturbed static Lagrangian in Minkowski space depends only on the
difference vector~$\bxi := \y-\x$. We again use the short notation~$\L[\bxi]$
(due to rotational symmetry, it actually depends only on the norm of the vector~$\bxi$, but we prefer not to
make this explicit in our notation). We denote the linear perturbation of the Lagrangian by~$\delta \L(\x,\y)$, i.e.\
\[ \L(\x,\y) = \L[\bxi] + \delta \L(\x,\y) + \O \big( (h_{\alpha \beta})^2 \big) \:. \]
The first question is how to compute~$\delta \L(\x,\y)$. 
In view of~\eqref{Lgeodesic}, the first idea is to compute the linear perturbation of the geodesic distance.
A direct calculation gives (for details see for example~\cite[eq.~B.6]{pmt}))
\beq \label{delLheuristic}
\delta \L(\x, \y) = \frac{1}{2} \int_0^1
h_{\alpha \beta}|_{\tau \y + (1-\tau) \x}\:d\tau \: \bxi_\alpha\, \frac{\partial}{\partial \bxi_\beta} \:
\L[\bxi] \:.
\eeq
This formula indeed holds, up to error terms which will be specified below.
Before doing so, we make a few explanatory remarks on this formula.
The integral can be understood as an integration of the perturbation along the
unperturbed geodesic joining~$\x$ and~$\y$ (which here simply is the straight line segment).
A simple way of understanding the above formula for~$\delta \L(\x, \y)$
is to verify that it describes the correct behavior under infinitesimal coordinate transformations. Indeed,
for the perturbed metric~$h_{\alpha \beta} =-\partial_\alpha \bzeta_\beta -\partial_\beta \bzeta_\alpha$
(see~\eqref{infmetricultra}) one gets
\begin{align}
\delta \L(\x, \y) &= -\frac{1}{2} \int_0^1 \big(\partial_\alpha \bzeta_\beta + \partial_\beta \bzeta_\alpha
\big) \big|_{\tau \y + (1-\tau) \x}\:d\tau \: \bxi_\alpha\, \frac{\partial}{\partial \bxi_\beta} \L[\bxi] \label{szero} \\
&= -\int_0^1 \bxi_\alpha \partial_\alpha \bzeta_\beta|_{\tau \y + (1-\tau) \x}\:d\tau \: \frac{\partial}{\partial \bxi_\beta} \L[\bxi] \label{s0} \\
&= -\int_0^1 \frac{d}{d\tau} \bzeta_\beta|_{\tau \y + (1-\tau) \x}\:d\tau \: \frac{\partial}{\partial \bxi_\beta} \L[\bxi] \label{s1} \\
&= -\big( \bzeta_\beta(\y) - \bzeta_\beta(\x) \big) \, \frac{\partial}{\partial \bxi_\beta} \L[\bxi]
= -\Big( \bzeta_\alpha(\x)\: \frac{\partial}{\partial \x_\alpha} + \bzeta_\alpha(\y)\: \frac{\partial}{\partial \y_\alpha}
\Big) \L[\xi] \:, \label{s2}
\end{align}
giving agreement with~\eqref{delLagultra}.
For clarity, we point out that in~\eqref{s0} we used that the function~$\L[\bxi]$ depends only
on the norm~$|\bxi|$. Therefore, its gradient points in the direction~$\xi$, implying that
\[ \bxi_\alpha\, \frac{\partial}{\partial \bxi_\beta} \L[\bxi] =
\bxi_\beta\, \frac{\partial}{\partial \bxi_\alpha} \L[\bxi] \:. \]
In~\eqref{s1}, on the other hand, we rewrote the derivative in the direction~$\bxi$ as a
$\tau$-derivatives (this formula is immediately verified by carrying out the $\tau$-derivative with the chain rule
and comparing with~\eqref{s0}). Finally, in~\eqref{s2} we integrated the $\tau$-derivative by parts,
giving boundary terms at~$\y$ and~$\x$.

The formula~\eqref{delLheuristic} can be derived rigorously using the method of integration along
characteristics (for the general context see for example~\cite{friedlander1})
or, more specifically for the Dirac equation in Minkowski space,
the so-called light-cone expansion of the kernel of the
fermionic projector (see~\cite[Appendix~B]{firstorder} and~\cite[Section~2.3 and~\S4.5.2]{cfs}
as well as~\cite[Appendix~F]{cfs} and~\cite{reghadamard} for the issues related to the regularization).
This procedure also gives rise to correction terms involving the curvature tensor.
All these correction terms will be negligible in our computations, because, following the
considerations after Proposition~\ref{prperror}, they can be
absorbed into the error term~\eqref{A0error}.

Our next step is to write the formula~\eqref{delLheuristic} in the jet formalism.
Since both arguments~$\x$ and~$\y$ are perturbed, we have
\[ \delta \L(\x,\y) = \big(D_{1,\bv} + D_{2,\bv} \big) \L(\x,\y) \:, \]
where~$\bv$ is a vector field on~$\G$ along~$N$.
In the jet formalism, the change of the volume form is described by the scalar component of
the jet. According to~\eqref{volform}, we need to choose the scalar component as~$h/2$,
leading us to the formula
\beq
\begin{split} \label{D12L}
\big( \nabla_{1,\v} + \nabla_{2,\v} \big) \L(\x,\y)
&= \frac{1}{2} \:\big( h(\x) + h(\y) \big)\: \L[\bxi]  \\
&\quad\: +\frac{1}{2} \int_0^1
h_{\alpha \beta}|_{\tau \y + (1-\tau) \x}\:d\tau \: \bxi_\alpha\, \frac{\partial}{\partial \bxi_\beta} \:
\L[\bxi] \:.
\end{split}
\eeq
The shortcoming of the considerations so far is that we can only compute the
combination of derivatives~$(\nabla_{1,\v} + \nabla_{2,\v}) \L(\x,\y)$, but it remains unclear what
a single derivative like~$\nabla_{1,\v} \L(\x,\y)$ should be.
Such derivative terms, however, are essential for computing anti-symmetric derivatives
as in~\eqref{linosi} or single derivatives as in~\eqref{Aalphadef}.
At this point, we need to use the detailed form of the light-cone expansion involving unbounded
integrals as developed in~\cite[Appendix~F]{pfp}
and, more generally and more systematically, in~\cite[Appendix~C]{nonlocal}.
For simplicity, we shall not derive these formulas, but merely state and briefly explain them.
The resulting formulas for the jet derivatives are
\begin{align}
\nabla_{1,\v} \L(\x, \y) &= \frac{h(\x)}{2}\: \L[\bxi] - \frac{1}{4} \int_{-\infty}^\infty \epsilon(\tau) \: d\tau \:
h_{\alpha \beta}|_{\tau \y + (1-\tau) \x}\: \bxi_\alpha\, \frac{\partial}{\partial x_\beta} \L[\bxi] \label{vxultra} \\
\nabla_{2,\v} \L(\x, \y) &= \frac{h(\y)}{2}\: \L[\bxi] + \frac{1}{4} \int_{-\infty}^\infty \epsilon(1-\tau) \: d\tau \:
h_{\alpha \beta}|_{\tau \y + (1-\tau) \x}\: \bxi_\alpha\, \frac{\partial}{\partial y_\beta} \L[\bxi] \label{vyultra}
\end{align}
(here~$\epsilon$ is the {\em{sign function}}~$\epsilon(\tau)=1$ if~$\tau>0$ and~$\epsilon(\tau)=-1$ otherwise).
Similar to~\eqref{D12L}, these formulas again involve line integrals, but this time along the
whole straight line through the points~$\x$ and~$\y$. These unbounded integrals are nonlocal
in the sense that, even if~$\x$ and~$\y$ are close together, the integral extends all the way to infinity.
If we add the formulas~\eqref{vxultra} and~\eqref{vyultra}, the unbounded integrals combine
to a bounded integral, giving back the previous formula~\eqref{D12L}.
A simple way of understanding the form of~\eqref{vxultra} (and similarly~\eqref{vyultra})
is by considering an infinitesimal coordinate transformation.
In this case, we obtain similar to~\eqref{szero} 
\begin{align}
\nabla_{1,\v} \L(\x, \y) &= -\div \bzeta(\x)\: \L[\bxi] - \frac{1}{2} \int_{-\infty}^\infty \epsilon(\tau)
\:\bxi_\alpha \partial_\alpha \bzeta_\beta|_{\tau \y + (1-\tau) \x}\:d\tau \:
\frac{\partial}{\partial \bxi_\beta} \: \L[\bxi] \notag \\
&= -\div \bzeta(\x)\: \L[\bxi] - \frac{1}{2} \int_{-\infty}^\infty \epsilon(\tau) \frac{d}{d\tau} \bzeta_\beta|_{\tau \y + (1-\tau) \x}\:d\tau \: \frac{\partial}{\partial \bxi_\beta} \L[\bxi] \notag \\
&= -\div \bzeta(\x)\: \L[\bxi] - \bzeta_\alpha(\x) \: \frac{\partial}{\partial \x_\alpha} \L[\bxi] \:. \label{Linner}
\end{align}
The first summand describes the infinitesimal change of the volume form at~$\x$.
Comparing with~\eqref{delLagultra}, one sees that the
second summand describes the variation of the Lagrangian by the
infinitesimal coordinate change, again at the point~$\x$.
The fact that the jet~\eqref{vxultra} and~\eqref{vyultra} describes the behavior under
infinitesimal coordinate transformations or, in more mathematical language, under infinitesimal
diffeomorphisms, is the reason why we refer to~$\v$ as the {\em{diffeomorphism jet}}.
We remark that the formula~\eqref{Linner} can be expressed by saying that,
for infinitesimal coordinate transformations, the jet~$\v$ reduces to an inner solution~$\u$ with
\beq \label{zetainner}
\u = -\big( \div \bzeta, \bzeta \big) \:.
\eeq

We finally verify by direct computation that the EL equations hold when testing with~$\v$.
\begin{Lemma} \label{lemmaELultra}
The jet~$\v$ defined by~\eqref{vxultra} and~\eqref{vyultra} has the property
\begin{align*}
0 &= \nabla_\v \ell(\x) = \int_N \nabla_{1,\v} \L(\x,\y)\: d\mu(\y) - \big( \nabla_\v \s)(x) \:.
\end{align*}
\end{Lemma}
\Proof Since~$\s$ is chosen such that~$\ell$ vanishes, we know that
\begin{align*}
\nabla_\v \ell(\x) &= \int_N D_{1,\v} \L(\x,\y)\: d\mu(\y) \\
&= \frac{1}{4} \int_N  d\mu(\y) \int_{-\infty}^\infty \epsilon(\tau) \: d\tau \:
h_{\alpha \beta}|_{\tau \y + (1-\tau) \x}\: \bxi_\alpha\, \frac{\partial}{\partial \bxi_\beta} \L[\bxi] \:.
\end{align*}
Applying the transformations
\beq \label{fliptrafo1}
\x \rightarrow \x\:, \qquad \y \rightarrow 2 \x - \y \qquad \text{and} \qquad \tau \rightarrow -\tau \:,
\eeq
the arguments in the integrand transform according to
\beq \label{fliptrafo2}
\bxi \rightarrow -\bxi \:,\qquad \tau \y + (1-\tau)\, \x = \x + \tau \bxi \;\rightarrow\;
\x + \tau \bxi = \tau \y + (1-\tau)\, \x \:.
\eeq
Therefore, the integrand flips sign. We conclude that the integral vanishes by symmetry.
\QED

\subsection{Satisfying the Alignment Condition}
After these preparations, we can now analyze the local alignment condition~\eqref{lac}.
For linear gravity, it simplifies to
\beq \label{alignultra}
0 = A^{(0)}_\alpha(\bzeta) = \int_N \bxi_\alpha\: \nabla_{1,\v} \L(\x,\y)\: d^3\bxi = 0 \:.
\eeq
Using~\eqref{vxultra}, we obtain
\begin{align}
A^{(0)}_\alpha(\bzeta)
&= \frac{1}{2} \int_N  \bxi_\alpha\: h(\x)\: \L(\x,\y)\: d^3\bxi \\
&\quad\: + \frac{1}{4} \int_N d^3\bxi\: \bxi_\alpha\: 
\int_{-\infty}^\infty \epsilon(\tau) \: d\tau \:
h_{\beta \gamma}|_{\bzeta + (\tau -\frac{1}{2}) \bxi}\: \bxi_\beta\, \frac{\partial}{\partial \bxi_\gamma} \L(\x, \y) \:.
\label{lineunbounded}
\end{align}
In the special case of an infinitesimal coordinate transformation, the jet~$\v$ reduces to the
inner solution~$\u$ given in~\eqref{zetainner}. In this case, the vector field~$A^{(0)}$
was already computed in Lemma~\ref{lemmaA0inner}.
As explained after this lemma, we can arrange by a suitable choice of the inner solution
that the vector field~$A^{(0)}$ vanishes. With this in mind, from now on we may assume that
the alignment condition~\eqref{alignultra} is satisfied.

We point out that, due to the unbounded line integral in~\eqref{lineunbounded},
the above formula for~$A^{(0)}$ is nonlocal. For this reason, it does not seem possible to
satisfy the alignment condition~\eqref{alignultra} explicitly or in closed form.
Instead, we proceed by working out the consequences of the local alignment condition.
To this end, we use that, if the vector field~$A^{(0)}$ vanishes identically, then also its divergence is zero.
This divergence is indeed a local quantity:
\begin{Lemma} \label{lemmavolcond}
For the jet~$\v$ given by~\eqref{vxultra},
\[ \frac{\partial}{\partial \bzeta_\alpha} A^{(0)}_\alpha
= \frac{\s}{2}\: h(\bzeta) + \O\Big(\frac{\delta^2}{\ell_\macro^2} \Big) \:. \]
\end{Lemma}
\Proof A direct computation using~\eqref{vxultra} yields
\begin{align*}
\frac{\partial}{\partial \bzeta_\alpha} A^{(0)}_\alpha
&= \int_N \bigg( \bxi\: \frac{\partial}{\partial \bzeta} \bigg)\: \nabla_{1,\v} \L(\x,\y)\: d^3\bxi \\
&= \frac{1}{2} \int_N  \bigg( \bxi\: \frac{\partial}{\partial \bzeta} \bigg)\: h(\x)\: \L[\bxi]\: d^3\bxi \\
&\quad\: + \frac{1}{4} \int_N d^3\bxi\:\bigg( \bxi\: \frac{\partial}{\partial \bzeta} \bigg)\: 
\int_{-\infty}^\infty \epsilon(\tau) \: d\tau \:
h_{\beta \gamma}|_{\bzeta + (\tau -\frac{1}{2}) \bxi}\: \bxi_\beta\, \frac{\partial}{\partial \bxi_\gamma} \L[\bxi] \\
&= \frac{1}{4} \int_N d^3\bxi \int_{-\infty}^\infty \epsilon(\tau) \: d\tau \: \frac{d}{d\tau}
h_{\beta \gamma}|_{\bzeta + (\tau -\frac{1}{2}) \bxi}\: \bxi_\beta\, \frac{\partial}{\partial \bxi_\gamma} \L[\bxi]
+ \O\Big(\frac{\delta^2}{\ell_\macro^2} \Big) \\
&= -\frac{1}{2} \int_N d^3\bxi \:h_{\beta \gamma}|_{\bzeta -\frac{1}{2} \bxi}\: \bxi_\beta\, \frac{\partial}{\partial \bxi_\gamma} \L[\bxi] + \O\Big(\frac{\delta^2}{\ell_\macro^2} \Big)  \\
&= \frac{1}{2} \int_N d^3\bxi \:h|_{\bzeta -\frac{1}{2} \bxi}\: \L[\bxi] + \O\Big(\frac{\delta^2}{\ell_\macro^2} \Big) 
= \frac{\s}{2}\: h(\bzeta) + \O\Big(\frac{\delta^2}{\ell_\macro^2} \Big) \:,
\end{align*}
giving the result.
\QED
This lemma shows that, up to the specified error term, the alignment condition implies
that~$h$ vanishes. In other words, alignment gives rise to volume conservation.

\subsection{Similarity between the Quasilocal Mass and the Brown-York Mass}
We assume that the system is weakly gravitating near the boundary~$\partial \tilde{\Omega}$,
so that the quasilocal mass can be computed linearly via~\eqref{linosi}
with the jet~$\v$ as given by~\eqref{vxultra} and~\eqref{vyultra}.
Moreover, we assume that the local alignment condition~\ref{lac} is satisfied, i.e.\
\[ A^{(0)}(\bzeta) = 0 \qquad \text{for all~$\bzeta \in \partial \Omega$}\:. \]
Then, according to Corollary~\ref{cormass}, the quasilocal mass is given by
\[ \Mass \big(\tilde{\Omega} \big) = \int_\Omega A^{(1)}_\alpha(\x)\: \nu_\alpha(\x)\: d\mu_{\partial \Omega}(\x) 
\;\bigg( 1 + \O \Big( \frac{\delta}{\ell_\macro} \Big) \bigg) \:. \]
Therefore, our task is to get a connection between the vector field~$A^{(1)}(\x)$ 
for a boundary point~$\x \in \partial \Omega$ and
the mean curvatures of~$\tilde{\Omega}$ and~$\Omega$. We first compute~$A^{(1)}(\x)$ 
in an expansion in powers of the range~$\delta$.

\begin{Lemma} \label{lemmaA1ultra} For the jet~$\v$ given by~\eqref{vxultra} and~\eqref{vyultra},
\[ A_\alpha^{(1)}(\bzeta) = \frac{1}{144}\: \delta^2\, \s_2 \Big( 
2 \partial_\beta h_{\alpha \beta}(\bzeta) + \partial_\alpha h(\bzeta) \Big) +\O\Big(\frac{\delta^3}{\ell_\macro^3} \Big) \:. \]
\end{Lemma}
\Proof We make the straightforward computation using~\eqref{linLcomp}
\begin{align*}
& \int_N \bxi_\alpha\: \bigg( \bxi\: \frac{\partial}{\partial \bzeta} \bigg)^2 \nabla_{1,\v} \L(\x,\y)\: d^3\bxi \\
&= \frac{1}{2} \int_N \bxi_\alpha\: \bigg( \bxi\: \frac{\partial}{\partial \bzeta} \bigg)^2 h(\x)\, \L[\bxi]\: d^3\bxi \\
&\quad\: +\frac{1}{4} \int_N  d^3\bxi \:\bxi_\alpha\: \bigg( \bxi\: \frac{\partial}{\partial \bzeta} \bigg)^2
\int_{-\infty}^\infty \epsilon(\tau) \: d\tau \:
h_{\beta \gamma}|_{\bzeta + (\tau -\frac{1}{2}) \bxi}\: \bxi_\beta\, \frac{\partial}{\partial \bxi_\gamma} \L[\bxi] \\
&=\frac{1}{4} \int_N  d^3\bxi \:\bxi_\alpha\: \bigg( \bxi\: \frac{\partial}{\partial \bzeta} \bigg)^2
\int_{-\infty}^\infty \epsilon(\tau) \: d\tau \:
h_{\beta \gamma}|_{\bzeta + (\tau -\frac{1}{2}) \bxi}\: \bxi_\beta\, \frac{\partial}{\partial \bxi_\gamma} \L[\bxi]
+\O\Big(\frac{\delta^3}{\ell_\macro^3} \Big) \\
&=\frac{1}{4} \int_N  d^3\bxi \:\bxi_\alpha\: \bigg( \bxi\: \frac{\partial}{\partial \bzeta} \bigg)
\int_{-\infty}^\infty \epsilon(\tau) \: d\tau \: \frac{d}{d\tau}
h_{\beta \gamma}|_{\bzeta + (\tau -\frac{1}{2}) \bxi}\: \bxi_\beta\, \frac{\partial}{\partial \bxi_\gamma} \L[\bxi]
+\O\Big(\frac{\delta^3}{\ell_\macro^3} \Big) \\
&=-\frac{1}{2} \int_N  d^3\bxi \:\bxi_\alpha\: \bigg( \bxi\: \frac{\partial}{\partial \bzeta} \bigg)\:
h_{\beta \gamma}|_{\bzeta -\frac{1}{2} \bxi}\: \bxi_\beta\, \frac{\partial}{\partial \bxi_\gamma} \L[\bxi]
+\O\Big(\frac{\delta^3}{\ell_\macro^3} \Big) \\
&=-\frac{1}{2} \int_N  d^3\bxi \:\bxi_\alpha\: \bxi_\delta\: \partial_\delta
h_{\beta \gamma}|_{\bzeta -\frac{1}{2} \bxi}\: \bxi_\beta\, \frac{\partial}{\partial \bxi_\gamma} \L[\bxi]
+\O\Big(\frac{\delta^3}{\ell_\macro^3} \Big) \\
&= \frac{1}{2} \int_N  d^3\bxi \:\bxi_\delta\: \partial_\delta
h_{\beta \alpha}|_{\bzeta -\frac{1}{2} \bxi}\: \bxi_\beta\: \L[\bxi] \\
&\quad\: +\frac{1}{2} \int_N  d^3\bxi \:\bxi_\alpha\: \partial_\gamma
h_{\beta \gamma}|_{\bzeta -\frac{1}{2} \bxi}\: \bxi_\beta\: \L[\bxi] \\
&\quad\: +\frac{1}{2} \int_N  d^3\bxi \:\bxi_\alpha\: \bxi_\delta\: \partial_\delta
h|_{\bzeta -\frac{1}{2} \bxi}\: \L[\bxi]
+\O\Big(\frac{\delta^3}{\ell_\macro^3} \Big) \\
&= \frac{1}{6}\: \delta^2\, \s_2 \Big( 
 \partial_\beta h_{\beta \alpha}(\bzeta) + \partial_\gamma h_{\alpha \gamma}(\bzeta)
+ \partial_\alpha h(\bzeta) \Big) +\O\Big(\frac{\delta^3}{\ell_\macro^3} \Big) \\
&= \frac{1}{6}\: \delta^2\, \s_2 \Big( 
2 \partial_\beta h_{\alpha \beta}(\bzeta) + \partial_\alpha h(\bzeta) \Big) +\O\Big(\frac{\delta^3}{\ell_\macro^3} \Big) \:.
\end{align*}
Applying~\eqref{Aalphadef} gives the result.
\QED

Let~$(N, g)$ be the Riemannian manifold with metric~$g_{\alpha \beta}$ given by~\eqref{glinab}.
Given~$\Omega \subset N$, we denote the first fundamental form (i.e.\ the
induced Riemannian metric) on~$\partial \Omega$
by~$g|_{\partial \Omega}$. Moreover, we denote the outer normal on~$\partial \Omega$ by~$\nu$,
and the second fundamental form by~$k|_{\partial \Omega}$.
For the definition of the Brown-York mass (see~\cite{brown-york, shi-tam, liu-yau}),
one considers an isometric embedding of~$\partial \tilde{\Omega}$ into~$N$ and sets
\[ \Mass_{\text{BY}} := \frac{1}{8 \pi \kappa} \int_{\partial \Omega} \big( \tr \tilde{k}|_{\partial \Omega}
- \tr k|_{\partial \Omega} \big)\: d\mu_{\partial \Omega} \]
(where~$\kappa$ is Newton's constant).
In our linearized description near the boundary,
the isometry of the embedding means that~$\delta g|_{\partial \Omega}=0$.
On the other hand, Lemma~\ref{lemmavolcond} shows that local alignment yields
volume preservation up to an error term.
Putting all these properties together, our quasilocal mass indeed coincides with the
Brown-York mass, as is made precise in the next theorem.

\begin{Thm} \label{thmquasilocal}
Assume that the induced metric on~$\partial \Omega$ is preserved by the variation, i.e.\
\beq \label{metricpreserve}
\delta g|_{\partial \Omega} = 0 \:.
\eeq
Moreover, assume that the volume form is preserved approximately in the sense that
\beq \label{volpreserve}
h_{\beta \beta} = \O\Big(\frac{\delta^2}{\ell_\macro^2} \Big) \:.
\eeq
Then the flux of~$A^{(1)}$ through~$\partial \Omega$ is related to the
first variation of the integral of mean curvature by
\beq \label{A1boundary}
\int_{\partial \Omega} A^{(1)}_{\alpha}\: \nu_\alpha\: d\mu_{\partial \Omega}
= -\frac{1}{36}\: \delta^2\, \s_2 
\int_{\partial \Omega} \tr  \delta k|_{\partial \Omega}\: d\mu_{\partial \Omega}
+\O\Big(\frac{\delta^3}{\ell_\macro^3} \Big) \:.
\eeq
\end{Thm}
\Proof Given a point~$\x_0 \in \partial \Omega$, we choose specific local coordinates near~$\x_0$
corresponding to the unperturbed metric. To this end, we first choose a Gaussian coordinate system~$(\x_1, \x_2)$ on~$\partial \Omega$. Thus,
denoting tensor indices on~$\partial \Omega$ by~$i,j \in \{1,2\}$, the first fundamental form can be written as
\[ g_{ij}^{(0)}(\x_0) = \delta_{ij} \qquad \text{and} \qquad \partial_k g_{ij}^{(0)}(\x_0) = 0 \]
(here the superscript~$(0)$ clarifies that we consider the unperturbed metric).
Next, we extend these coordinates to coordinates on~$N$ by choosing~$\x_3$ as the
arc-length parametrization of geodesics which leave~$\partial \Omega$ orthogonally
(in the literature, these coordinates are sometimes referred to as Fermi coordinates or 
a tubular neighborhood). In these coordinates, the unperturbed metric takes the form
\[ g^{(0)}_{\alpha \beta} = \begin{pmatrix} g^{(0)}_{11} & g^{(0)}_{12}  & 0 \\[0.2em] g^{(0)}_{21}  & g^{(0)}_{22}  & 0 \\[0.2em]
0 & 0 & 1 \end{pmatrix} \:. \]

In these coordinates, the variation of the metric~$\delta g$ looks as follows.
The fact that the first fundamental form is preserved means that
\[ \delta g_{ij}|_{(\partial \Omega) \cap U} = 0 \:. \]
Moreover, since the volume form is approximately preserved, we know
that~$\delta g_{\alpha \alpha} = \O(\delta^2/\ell_\macro^2)$, so that
\[ \delta g_{\alpha \beta} |_{(\partial \Omega) \cap U}
= \begin{pmatrix} 0 & 0 & \delta g_{13} \\[0.2em] 0 & 0 & \delta g_{23} \\[0.2em] \delta g_{31} & \delta g_{32} & \O(\delta^2
/\ell_\macro^2) \end{pmatrix} \:. \]

We next compute the normal vector field~$\nu$ and the second fundamental form.
Clearly, the unperturbed normal is given by
\[ \nu^{(0)}_\alpha = (0,0,1)\:. \]
Moreover, since~$\delta g_{33}$ is approximately zero, we know that
\[ \delta \nu^{(0)}_3 = \O\Big(\frac{\delta^2}{\ell_\macro^2} \Big) \:. \]
Let~$e_\alpha = \partial_{\x_\alpha}$ be the basis of~$T_{\x_0} \partial \Omega$.
Then the second fundamental form and its first variation take the form
\begin{align*}
k_{ij}(\x_0) &= -g \big( \nu, \nabla_i e_j \big) = -\Gamma^k_{ij} \, \nu_k \\
\delta k_{ij}(\x_0) &= - \big( \delta \Gamma^k_{ij} \big) \, \nu_k
+ \Gamma^k_{ij} \, \delta \nu_k
= -\delta \Gamma^3_{ij} + \Gamma^\gamma_{ij} \, \delta \nu_\gamma
= -\delta \Gamma^3_{ij} + \O\Big(\frac{\delta^2}{\ell_\macro^2} \Big) \\
&= -\frac{1}{2}\:\big( \partial_i \delta g_{3 j} + \partial_j \delta g_{3 i}
- \partial_3 \delta g_{ij} \big)  + \O\Big(\frac{\delta^2}{\ell_\macro^2} \Big) \\
\tr \delta  k(\x_0) &= \delta k_{i i}(\x_0)
= - \partial_i \delta g_{3 i} + \frac{1}{2}\: \partial_3 \delta g_{ii}  + \O\Big(\frac{\delta^2}{\ell_\macro^2} \Big) \:.
\end{align*}
In general coordinates, the last identity can be written as
\beq \label{delk}
\tr \delta k = -\big( \nabla_i \delta g^i_\alpha \big)\: \nu^\alpha + \frac{1}{2}\: D_\nu \delta g^i_i
 +\O\Big(\frac{\delta^2}{\ell_\macro^2} \Big) \:.
 \eeq

On the other hand, applying Lemma~\ref{lemmaA1ultra}, in the chosen coordinates around~$\x_0$
we have
\begin{align*}
A_\alpha^{(1)}(\x_0) \:\nu^\alpha &= \frac{1}{72}\: \delta^2\, \s_2\,
 \partial_\beta \delta g_{3 \beta} +\O\Big(\frac{\delta^3}{\ell_\macro^3} \Big) \\
&= \frac{1}{72}\: \delta^2\, \s_2\,
\big( \partial_i \delta g_{3i} + \partial_3 \delta g_{33} \big) +\O\Big(\frac{\delta^3}{\ell_\macro^3} \Big) \\
&= \frac{1}{72}\: \delta^2\, \s_2\,
\big( \partial_i \delta g_{3i} - \partial_3 \delta g_{ii} \big) +\O\Big(\frac{\delta^3}{\ell_\macro^3} \Big) \:.
\end{align*}
Thus in general coordinates,
\beq \label{A1flux}
A_\alpha^{(1)} \:\nu^\alpha = 
\frac{1}{72}\: \delta^2\, \s_2\,
\Big( \big( \nabla_i \delta g^i_\alpha \big) \nu^\alpha  - D_\nu \delta g^i_i \Big) +\O\Big(\frac{\delta^3}{\ell_\macro^3} \Big) \:.
\eeq

The identities~\eqref{delk} and~\eqref{A1flux} are not multiples of each other because of the
different relative prefactors of the two summands. Therefore, we must make use of the fact
that, integrating over~$\partial \Omega$, gives an additional identity.
Indeed, 
\[ \int_{\partial \Omega} \big( \nabla_i \delta g^i_\alpha \big)\: \nu^\alpha \: d\mu_{\partial \Omega} \\
= \int_{\partial \Omega} \nabla_i \big( \delta g^i_\alpha \: \nu^\alpha \big) \: d\mu_{\partial \Omega}
- \int_{\partial \Omega} \delta g^i_\alpha \: \nabla_i \nu^\alpha \: d\mu_{\partial \Omega} \:. \]
The first term vanishes by the Gauss divergence theorem. The last summand, on the other hand,
also vanishes because~$\nabla_i \nu^\alpha$ is non-zero only if~$\alpha \in \{1,2\}$, in which case~$\delta g^i_\alpha$ vanishes. We conclude that
\[ \int_{\partial \Omega} \big( \nabla_i \delta g^i_\alpha \big)\: \nu^\alpha \: d\mu_{\partial \Omega} = 0 \:. \]

Using this identity in~\eqref{delk} and~\eqref{A1flux} gives the result.
\QED

We finally explain this result. The formula~\eqref{A1boundary} shows that, under the assumptions
of the theorem, the quasilocal mass coincides with the Brown-York mass
(up to an irrelevant prefactor and the error term).
This result gives a connection between the quasilocal mass
for causal variational principles and corresponding notions of differential geometry.
This connection is surprising, because the positivity proof used here is completely different
from the proof for the Brown-York mass in~\cite{shi-tam}.
However, we point out that, despite these similarities, there are also major differences
between our quasilocal mass and the Brown-York mass. In order to explain this point,
we need to discuss the assumptions of Theorem~\ref{thmquasilocal}.
The preservation of the first fundamental form~\eqref{metricpreserve}
is the linearized statement of the isometric embedding used in the Brown-York mass.
Likewise, the approximate preservation of the volume form~\eqref{volpreserve}
is a consequence of the alignment condition (see Lemma~\ref{lemmavolcond}).
However, it is not clear how the isometry property~\eqref{metricpreserve}
comes about in the setting of
causal variational principles. In order to clarify this point, it seems necessary to take into account
that the quasilocal mass involves a minimization over the symmetry transformations~$\Phi \in {\mathcal{G}}$.
It might be that, as a consequence of this minimization process, the
identification of~$\tilde{\Omega}$ with~$\Omega$ is an isometry of the boundary.
However, verifying this conjecture goes beyond the scope of the present work.

\subsection{Computing the Synthetic Scalar Curvature}
We now compute synthetic scalar curvature. We again consider the linearized description.
Following~\eqref{masslin}, %Definition~\ref{defsynthetic},
the expression in~\eqref{scaldef} can be written as
\[ \big( \nabla_\v \s - \Delta \v \big)(\x)
= \int_N \big( \nabla_{1,\v} - \nabla_{2,\v} \big) \L(\x,\y)\: d\mu(\y) \:. \]
Applying Theorem~\ref{thmdivA}, we can rewrite this quantity as the divergence of the 
vector field~$A$,
\[ \big( \nabla_\v \s - \Delta \v \big)(\x) = \div A(\x)\:. \]
As a consequence of the local alignment condition~\ref{localign}, the vector field~$A^{(0)}$ vanishes.
Therefore, using the scaling~\eqref{scalefact}, we obtain
\[ \big( \nabla_\v \s - \Delta \v \big)(\x) = \div A^{(1)}(\x) + \O\Big(\frac{\delta^3}{\ell_\macro^3} \Big) \:. \]
This divergence is computed in the following lemma.

\begin{Lemma}
\beq \label{scalrel}
\big( \nabla_\v \s - \Delta \v \big)(\x) = \frac{1}{3}\: \delta^2\, \s_2\: \partial_{\alpha \beta} h_{\alpha \beta}(\x) 
+ \frac{1}{6}\: \delta^2\, \s_2\: \Delta_{\R^3} h(\x)
+\O\Big(\frac{\delta^3}{\ell_\macro^3} \Big) \:.
\eeq
\end{Lemma}
\Proof
We compute the divergence of~$A^{(1)}$ according to
\begin{align*}
& \frac{\partial}{\partial \bzeta_\alpha} \int_N \bxi_\alpha\: \bigg( \bxi\: \frac{\partial}{\partial \bzeta} \bigg)^2 \nabla_{1,\v} \L(\x,\y)\: d^3\bxi \\
&= \int_N \bigg( \bxi\: \frac{\partial}{\partial \bzeta} \bigg)^3 \nabla_{1,\v} \L(\x,\y)\: d^3\bxi \\
&= \frac{1}{2} \int_N \bigg( \bxi\: \frac{\partial}{\partial \bzeta} \bigg)^3 h(\x)\, \L[\bxi]\: d^3\bxi \\
&\quad\: +\frac{1}{4} \int_N  d^3\bxi \:\bigg( \bxi\: \frac{\partial}{\partial \bzeta} \bigg)^3
\int_{-\infty}^\infty \epsilon(\tau) \: d\tau \:
h_{\alpha \beta}|_{\bzeta + (\tau -\frac{1}{2}) \bxi}\: \bxi_\alpha\, \frac{\partial}{\partial \bxi_\beta} \L[\bxi] \\
&= \frac{1}{4} \int_N  d^3\bxi \:\bigg( \bxi\: \frac{\partial}{\partial \bzeta} \bigg)^3
\int_{-\infty}^\infty \epsilon(\tau) \: d\tau \:
h_{\alpha \beta}|_{\bzeta + (\tau -\frac{1}{2}) \bxi}\: \bxi_\alpha\, \frac{\partial}{\partial \bxi_\beta} \L[\bxi]
+\O\Big(\frac{\delta^3}{\ell_\macro^3} \Big) \\
&= \frac{1}{4} \int_N  d^3\bxi \:\bigg( \bxi\: \frac{\partial}{\partial \bzeta} \bigg)^2
\int_{-\infty}^\infty \epsilon(\tau) \: d\tau \: \frac{d}{d\tau}
h_{\alpha \beta}|_{\bzeta + (\tau -\frac{1}{2}) \bxi}\: \bxi_\alpha\, \frac{\partial}{\partial \bxi_\beta} \L[\bxi]
+\O\Big(\frac{\delta^3}{\ell_\macro^3} \Big) \\
&= -\frac{1}{2} \int_N  d^3\bxi \:\bigg( \bxi\: \frac{\partial}{\partial \bzeta} \bigg)^2
h_{\alpha \beta}|_{\bzeta -\frac{1}{2} \bxi}\: \bxi_\alpha\, \frac{\partial}{\partial \bxi_\beta} \L[\bxi]
+\O\Big(\frac{\delta^3}{\ell_\macro^3} \Big) \\
&= -\frac{1}{2} \int_N  d^3\bxi \:\bxi_\gamma \bxi_\delta\:
 \partial_{\gamma \delta}
 h_{\alpha \beta}|_{\bzeta -\frac{1}{2} \bxi}\: \bxi_\alpha\, \frac{\partial}{\partial \bxi_\beta} \L[\bxi]
+\O\Big(\frac{\delta^3}{\ell_\macro^3} \Big) \\
&= \int_N  d^3\bxi \:\bxi_\gamma\:
 \partial_{\gamma \beta} h_{\alpha \beta}|_{\bzeta -\frac{1}{2} \bxi}\: \bxi_\alpha\: \L[\bxi] \\
&\quad\: +\frac{1}{2} \int_N  d^3\bxi \:\bxi_\gamma \bxi_\delta\:
 \partial_{\gamma \delta} h|_{\bzeta -\frac{1}{2} \bxi}\: \L[\bxi]
+\O\Big(\frac{\delta^3}{\ell_\macro^3} \Big) \\
&= \frac{1}{3}\: \delta^2\, \s_2\: \partial_{\alpha \beta} h_{\alpha \beta}(\bzeta) + \frac{1}{6}\: \delta^2\, \s_2\: \Delta_{\R^3} h(\bzeta)
+\O\Big(\frac{\delta^3}{\ell_\macro^3} \Big) \:.
\end{align*}
This gives the result.
\QED

In order to relate the obtained expression to the scalar curvature of the spatial metric,
we first note that, according to Lemma~\ref{lemmavolcond}, linear alignment gives rise to
volume preservation. Therefore, taking into account the error term, we can rewrite~\eqref{scalrel} as
\begin{align*}
\big( \nabla_\v \s - \Delta \v \big)(\x) &= \frac{1}{3}\: \delta^2\, \s_2\: 
\Big( \partial_{\alpha \beta} h_{\alpha \beta}(\x) - \Delta_{\R^3} h(\x) \Big)
+\O\Big(\frac{\delta^3}{\ell_\macro^3} \Big) \\
&= \frac{1}{3}\: \delta^2\, \s_2\: \text{scal}_g(\x) +\O\Big(\frac{\delta^3}{\ell_\macro^3} \Big) \:,
\end{align*}
where~$\text{scal}_g$ is the linearized scalar curvature of the Riemannian metric~$g$~\eqref{glinab}
(see for example~\cite[Theorem~1.174~(e)]{besse}).
Since this relation does not depend on the choice of~$\Phi$ and~$\Omega$, in~\eqref{scaldef}
we can take the infimum to obtain the following result.
\begin{Thm} \label{thmsynthetic}
For an ultrastatic spacetime with linearized gravity, the synthetic scalar curvature
(see Definition~\ref{defsynthetic}) agrees, up to a constant and an error term,
with the scalar curvature of the Riemannian metric,
\[ \text{\rm{scal}}(\x) = \frac{1}{3}\: \delta^2\, \s_2\: \text{\rm{scal}}_g(\x)
+\O\Big(\frac{\delta^3}{\ell_\macro^3} \Big) \:. \]
\end{Thm}

\appendix
\section{Example: Schwarzschild Spacetime} \label{appA}
This appendix is devoted to a detailed computation of the total mass and the
asymptotic alignment in the prime example of the Schwarzschild geometry.

\subsection{The Static Lagrangian in the Schwarzschild Geometry}
We begin with the Schwarzschild metric in Schwarzschild coordinates,
\[ ds^2 = \Big( 1 - \frac{2M}{r} \Big)\: dt^2 - \Big( 1 - \frac{2M}{r} \Big)^{-1}\: dr^2 - r^2\, d\vartheta^2
- r^2 \sin^2 \vartheta \:d\varphi^2\:. \]
The volume form is given by
\[ %\label{rhodef}
d\rho = \sqrt{|\det g|}\: d^4x = dt\, r^2\,dr\,\sin \vartheta\, d\vartheta\, d\varphi \:. \]
We write it as
\[ d\rho = dt\, d\mu \qquad \text{with} \qquad d\mu := r^2\,dr\,\sin \vartheta\, d\vartheta\, d\varphi \:. \]
We also write~$x=(t,\x)$, where~$\x=(r,\vartheta, \varphi)$ are the spatial coordinates.

The static causal fermion system in the Schwarzschild geometry is obtained following the
general construction in~\cite{nrstg, grossmann}; for details see~\cite[Section~6.1]{pmt}.
The {\em{static Lagrangian}}~$\L(\x,\y)$ is obtained again by integrating over one time variable~\eqref{Ltime}.
We point out that here we do {\em{not}} show that~$\tilde{\mu}$ and~$\mu$ are asymptotically 
close and that~$\tilde{\mu}$ is asymptotically flat (see Definitions~\ref{defasyclose} and~\ref{defasyflat}).
These technical questions will be analyzed elsewhere.
Instead, we here restrict attention to the linearized description near infinity.
Our goal is to compute jets, alignments and the total mass explicitly and in detail.

\subsection{The Euler-Lagrange Equations in Linearized Gravity}
Denoting the radial coordinates by
\[ %\label{Rrdef}
R := |\x| \qquad \text{and} \qquad r := |\y| \:, \]
the function~$\ell$ defined in~\eqref{ldef} takes the form
\[ \ell(\x) + \s = \int_{2M}^\infty r^2\, dr \int_0^{2 \pi} d\varphi \int_0^\pi \sin\vartheta\: d\vartheta\:
\L(\x,\y) \:. \]
The EL equations~\eqref{EL} state that, for a suitable choice of the parameter~$\s>0$, this function
must be positive and must vanish for all~$x \in N$.

Since we are only interested in the asymptotic behavior near infinity, it suffices to consider
{\em{linearized gravity}}.
Following the presentation in~\cite[\S105]{landau2}, we write the metric as
\beq \label{gijlin}
g_{jk}(\x) = \eta_{jk} + h_{jk}(\x) \:,
\eeq
where~$\eta_{jk} =  \text{diag} (1, -1, -1,-1)$ is the Minkowski metric. 
We only consider the static setting; this is why~$h_{jk}$ depends only on the spatial coordinates~$\x$.
We always raise and lower the metric with respect to the Minkowski metric.
In order to describe the Schwarzschild metric in this formalism, we work in space with
Cartesian coordinates~$\x \in \R^3$. Then, introducing the abbreviation
\beq \label{Vdef}
V(\x) := -\frac{2M}{|\x|} \:,
\eeq
the linearized metric is given by
\beq \label{linschwarz}
h_{00}(x) = V(\x)\:,\quad
h_{\alpha \beta}(x) = V(\x)\: \hat{\x}_\alpha\, \hat{\x}_\beta \:,
\eeq
where~$\hat{\x}$ denotes the unit vector in the direction of~$\x$, i.e.\
\[ \hat{\x} := \frac{\x}{|\x|}\:. \]
The tensor~$h_{ij}$ is trace-free but not divergence-free, as one sees directly from the computation
(see also~\cite[\S105]{landau2})
\begin{align}
h &= h_{00} - h_{\alpha \alpha} = 0 \label{hsch} \\
 \partial_\alpha h_{\alpha \beta} &= -\partial_\alpha \Big( \frac{2M}{R^3}\, \x_\alpha\, \x_\beta \Big)
= -\frac{2M}{R^3}\: \x_\beta\: \big(-3+3+1) \notag \\
&= -\frac{2M}{R^2}\: \hat{\x}_\beta
= \partial_\beta \Big( \frac{2M}{R} \Big) = -\partial_\beta V(\x) \label{divspace} \\
 \partial^j h_{jk} &= -\sum_{\alpha=1}^3 \partial_\alpha h_{\alpha k} = \partial_k V(\x) \:. \label{divhsch}
\end{align}

Our next step is to compute the first variation of the Lagrangian and to work out the EL equations.
Here we make use of the light-cone expansion of the kernel of the
fermionic projector as developed in~\cite{firstorder, light}; see also~\cite[Section~2.3 and~\S4.5.2]{cfs}.
More specifically, the Lagrangian was computed in the linearized Schwarzschild geometry
in~\cite[Appendix~B]{pmt}. It was shown
that any homogeneous, spherically symmetric and static kernel~$\L(x,y)$ is perturbed linearly according to
(see~\cite[eq.~(B.6)]{pmt})
\beq \label{delL}
\delta \L(x,y) = \frac{1}{2} \int_0^1 d\tau \;h^i_{\;k} 
\big|_{\tau y + (1-\tau)\, x}\: \xi^k\: \frac{\partial}{\partial y^i} \L(x,y) \:.
\eeq
Exactly as explained for ultrastatic spacetimes in Section~\ref{secultrajets},
this formula holds up to corrections involving the curvature tensor.
All these correction terms will be negligible in our computations, because, again following the
considerations after Proposition~\ref{prperror}, they can be
absorbed into the error term~\eqref{A0error}.

The variation of the static Lagrangian~\eqref{Ltime} can be computed integrating by parts,
\begin{align}
\delta \L(\x, \y) &= \int_{-\infty}^\infty \delta \L \big((0,\x), (t,\y) \big) \: dt \notag \\
&= \frac{1}{2} \int_{-\infty}^\infty dt \int_0^1 d\tau \;h^i_{\;k} 
\big|_{\tau y + (1-\tau)\, x}\: \xi^k\: \frac{\partial}{\partial y^i} \L \big((0,\x), (t,\y) \big) \notag \\
&= -\frac{1}{2} \int_0^1 d\tau \;h_{00}
\big|_{\tau y + (1-\tau)\, x}\: \L(\x,\y) \label{dL1} \\
&\quad\:-\frac{1}{2} \int_0^1 d\tau \;h_{\alpha \beta}
\big|_{\tau y + (1-\tau)\, x}\: \bxi_\alpha\: \frac{\partial}{\partial \y_\beta} \L[\bxi] \:. \label{dL2}
\end{align}
Using that~$h_{ij}$ is trace-free~\eqref{hsch}, this formula can be written more compactly
as\footnote{To avoid confusion, we note that the sign difference compared to the integral in~\eqref{D12L}
comes about because in the ultrastatic setting, the metric~$h_{\alpha \beta}$ is Riemannian,
whereas here it is the spatial part of the Lorentzian metric (cf.~\eqref{ultraline}, \eqref{glinab} and~\eqref{gijlin}).}
\begin{align}
\delta \L(\x, \y) &= -\frac{1}{2} \int_0^1 d\tau \;h_{\alpha \beta}
\big|_{\tau y + (1-\tau)\, x}\: \frac{\partial}{\partial \y_\beta}\big( \bxi_\alpha\, \L[\bxi] \big) \:. \label{delL2}
\end{align}

After these preparation, we can now compute the first variation of the function~$\ell$
introduced in~\eqref{ldef}.
\begin{Lemma} \label{lemmaELlin0}
\[ \delta \ell(\x) = \frac{1}{2} \int_N d^3\y \bigg( \int_0^1 
V|_{\tau \y + (1-\tau)\, \x}\: d\tau - V(\y) \bigg)\, \L[\bxi] \:. \]
\end{Lemma}
\Proof Starting from the variation of the static Lagrangian~\eqref{delL2}, we obtain
\begin{align}
\delta \ell(\x) &= \int_N \delta \L (\x,\y)\:d^3y = -\frac{1}{2} \int_N d^3y \int_0^1 d\tau \;h_{\alpha \beta}
\big|_{\tau y + (1-\tau)\, x}\: \frac{\partial}{\partial \y_\beta} \big( \bxi_\alpha\, \L[\bxi] \big) 
\label{delLeq} \\
&= \frac{1}{2} \int_N d^3y \int_0^1 d\tau \;\frac{\partial}{\partial \y_\beta} h_{\alpha \beta}
\big|_{\tau y + (1-\tau)\, x}\: \bxi_\alpha\: \L[\bxi] \notag \\
&=\frac{1}{2} \int_0^1 d\tau \int_N d^3\y\, \tau \,\big(\partial_\beta h_{\alpha \beta}\big)
\big|_{\tau y + (1-\tau)\, x}\: \bxi_\alpha\, \L[\bxi] \notag \\
&\!\!\!\overset{\eqref{divspace}}{=}-\frac{1}{2} \int_0^1 d\tau \int_N d^3\y\, \tau \,\bxi_\alpha\,
 \partial_\alpha V|_{\tau \y + (1-\tau)\, \x}\, \L[\bxi] \notag \\
&=-\frac{1}{2} \int_N d^3\y \int_0^1 d\tau \:\tau \: \frac{d}{d\tau}
V|_{\tau \y + (1-\tau)\, \x}\, \L[\bxi] \:. \notag
\end{align}
Integrating by parts in~$\tau$ gives the result.
\QED

We next expand the obtained formula for~$\delta \ell$ asymptotically near infinity
and explain our findings.
\begin{align}
&\delta \ell(\x) \notag \\
&= - \frac{1}{2} \int_N d^3\y \:\bigg( \int_0^1 
\Big( \big(\tau - 1)\, \bxi_\alpha \partial_\alpha V|_\x + \frac{1}{2}\: \big(\tau^2 - 1 \big)\: \bxi_\alpha \bxi_\beta \partial_{\alpha \beta}
V|_\x \Big) \bigg)\, \L[\bxi] + \O \Big( \frac{M \delta^3}{|\x|^4} \Big) \notag \\
&= \frac{1}{2} \int_N d^3\y\: 
\Big( \frac{1}{2}\: \bxi_\alpha \partial_\alpha V|_\x + \frac{1}{3}\: \bxi_\alpha \bxi_\beta \partial_{\alpha \beta}
V|_\x \Big) \, \L[\bxi] + \O \Big( \frac{M \delta^3}{|\x|^4} \Big) \label{Lex}
\end{align}
(where~$\delta$ is again the range of the Lagrangian as introduced after~\eqref{shortrange}).
Carrying out the $\y$-integration with the help of~\eqref{linLcomp}, we obtain
\begin{align*}
\delta \ell(\x)
&= \frac{1}{18} \: \delta^2\: \s_2\: \Delta_{\R^3} V|_{\x} + \O \Big( \frac{M \delta^3}{|\x|^4} \Big)
\end{align*}
Now we can make use of the fact that the potential~$V$ in~\eqref{Vdef} is harmonic to conclude that
\beq \label{ellzero}
\delta \ell(\x) = \O \Big( \frac{M \delta^3}{|\x|^4} \Big) \:.
\eeq
This formula shows that the EL equations are indeed satisfied asymptotically at infinity.
This is all we need in order to ensure that the total mass is well-defined.
The critical reader might object that the EL equations should hold exactly,
not only asymptotically at infinity. In order to understand how the error terms comes about,
we first note that the error term in~\eqref{ellzero} includes the leading contributions of curvature, which scale like
the Riemann tensor,
\[ \text{Riem}(\x) \sim \partial^2 g(\x) \sim \frac{M}{|\x|^3} + \O \Big( \frac{1}{|\x|^4} \Big) \:. \]
In order to describe also the contributions of order~$\O(|\x|^{-4})$ in the EL equations,
one would have to take into account how the curvature tensor enters the Lagrangian.
This goes beyond the scope of the present analysis, where we 
took the simplified ansatz for the linearized Lagrangian~\eqref{delL} as our starting point.

\subsection{Description with Jets}
We next reformulate the previous findings in the jet formalism. To this end, we introduce
the jet derivatives
\begin{align}
D_{1,\bv} \L(\x, \y) &:= \frac{1}{4} \int_{-\infty}^\infty \epsilon(\tau)\: d\tau \;h_{\alpha \beta}
\big|_{\tau y + (1-\tau)\, x}\: \frac{\partial}{\partial \x_\beta} \big( \bxi_\alpha\, \L(\x,\y) \big) \label{vx} \\
D_{2,\bv} \L(\x, \y) &:= -\frac{1}{4} \int_{-\infty}^\infty \epsilon(1-\tau)\: d\tau \;h_{\alpha \beta}
\big|_{\tau y + (1-\tau)\, x}\: \frac{\partial}{\partial \y_\beta} \big( \bxi_\alpha\, \L(\x,\y) \big) \:. \label{vy}
\end{align}
This jet is constructed such that
\[ \big( D_{1,\bv} + D_{2,\bv} \big) \L(\x, \y) = \delta \L(\x,\y) \]
with the variation of the static Lagrangian as computed in~\eqref{delL2}.
Similar as explained after~\eqref{Linner} for ultrastatic spacetimes, we refer to~$\v$
as the {\em{diffeomorphism jet}}.
Before going on, we point out that this diffeomorphism jet is quite different from
the diffeomorphism jet in an ultrastatic spacetime if for the metric~$h_{\alpha \beta}$
in~\eqref{vxultra} and~\eqref{vyultra} is chosen as the spatial part of the Schwarzschild metric.
In other words, by comparing the diffeomorphism term in the Schwarzschild spacetime with
that in ultrastatic spacetimes, one sees that the metric component~$g_{00}$ of the Schwarzschild
metric has an important effect on the static Lagrangian. This is quite different from mathematical relativity,
where the ADM mass of the Schwarzschild spacetime is described purely in terms of the induced
Riemannian metric.

Let us verify that the EL equations are satisfied in the vacuum if we test with
the jet~$\v=(0, \bv)$.
\begin{Lemma} \label{lemmaELlin}
The diffeomorphism jet~\eqref{vx} satisfies for all~$\x \in N$ the equation
\[ \int_N D_{1,\bv} \L(\x, \y)\: d^3\y = 0 \:. \]
\end{Lemma}
\Proof By direct computation, we obtain
\begin{align*}
\int_N D_{1,\bv} \L(\x, \y)\: d^3\y
&=-\frac{1}{4} \int_N d^3\y \int_{-\infty}^\infty \epsilon(\tau)\: d\tau \;h_{\alpha \beta}
\big|_{\tau y + (1-\tau)\, x}\: \frac{\partial}{\partial \y_\beta} \big( \bxi_\alpha\, \L[\bxi] \big) \\
&=\frac{1}{4} \int_N d^3\y \int_{-\infty}^\infty \epsilon(\tau)\: d\tau \;\tau\,\partial_\beta h_{\alpha \beta}
\big|_{\tau y + (1-\tau)\, x}\: \bxi_\alpha\, \L[\bxi] \\
&\!\!\!\overset{\eqref{divspace}}{=} -\frac{1}{4} \int_N d^3\y \int_{-\infty}^\infty \epsilon(\tau)\: d\tau \;\tau\,
\frac{d}{d\tau} V|_{\tau \y + (1-\tau)\, \x}\, \L[\bxi] \\
&= \frac{1}{4} \int_N d^3\y \int_{-\infty}^\infty \epsilon(\tau)\: d\tau \;V|_{\tau \y + (1-\tau)\, \x}\, \L[\bxi] \:.
\end{align*}
Applying again the transformations~\eqref{fliptrafo1} and~\eqref{fliptrafo2},
the integrand flips sign. Therefore, the integral vanishes by symmetry.
\QED

Expressing the result of Lemma~\ref{lemmaELlin0} with jets gives a connection to the
linearized field operator.
\begin{Lemma} \label{lemmaELlin2}
The diffeomorphism jet~$\bv$ defined by~\eqref{vx} and~\eqref{vy} satisfies for all~$\x \in N$ the equation
\beq \label{intNex}
\int_N \big( D_{1,\bv} + D_{2,\bv} \big) \L(\x, \y)\: d^3\y
=\frac{1}{2} \int_N d^3\y \:\bigg( \int_0^1 
V|_{\tau \y + (1-\tau)\, \x}\: d\tau - V(\y) \bigg)\, \L[\bxi] \:.
\eeq
\end{Lemma}
\Proof Noting that
\[ \int_N \big( D_{1,\bv} + D_{2,\bv} \big) \L(\x, \y)\: d^3\y \\
=-\frac{1}{2} \int_0^1 d\tau \int_N d^3\y\, h_{\alpha \beta}
\big|_{\tau y + (1-\tau)\, x}\: \frac{\partial}{\partial \y_\beta}\big( \bxi_\alpha\, \L[\bxi] \big) \]
and comparing with~\eqref{delLeq}, we can proceed as in the proof of Lemma~\ref{lemmaELlin0}.
\QED
In order to better understand this result, it is useful to expand the potential on the right side of~\eqref{intNex}
in power of~$\bxi$. Proceeding as in~\eqref{Lex}, we obtain
\[ \int_N \big( D_{1,\bv} + D_{2,\bv} \big) \L(\x, \y)\: d^3\y = \frac{1}{18} \: \delta^2\: \s_2\: \Delta_{\R^3} V|_{\x} + \O \Big( \frac{M \delta^3}{|\x|^4} \Big) \:. \]
The right side vanishes because~$V$ is harmonic. In this way, we get a direct connection between
the linearized field equations and Newton's law of gravitation.

\subsection{Computation of the Total Mass}
We now compute the surface layer integral
\[ %\label{MR0def}
{\Mass}(R_0) = \int_{R<R_0} d^3\x \int_{r>R_0} d^3\y \: \big( D_{1,\bv} - D_{2,\bv} \big)
\L(\x, \y) \:. \]
\begin{Prp} \label{prpmassdirect} For any radius~$R_0>0$,
\[ {\Mass}(R_0) = \frac{4 \pi}{9}\:\delta^2\, \s_2 \:M + \O \Big( \frac{M \delta^3}{R_0} \Big) \:. \]
\end{Prp}
\Proof Using the anti-symmetry of the integrand together with Lemma~\ref{lemmaELlin}, we obtain
\[ {\Mass}(R_0) = -\int_{R<R_0} d^3\x \int_N d^3\y \: \big( D_{1,\bv} + D_{2,\bv} \big)
\L(\x, \y) \:. \]
Applying Lemma~\ref{lemmaELlin2} gives
\[ %\label{massformula}
{\Mass}(R_0) = - \frac{1}{2} \int_{R<R_0} d^3\x \int_N d^3\y \:\bigg( \int_0^1 
V|_{\tau \y + (1-\tau)\, \x}\: d\tau - V(\y) \bigg)\, \L[\bxi] \:. \]
We now expand the integrand in powers of~$\bxi$. Proceeding as in~\eqref{Lex}, we obtain
\[ {\Mass}(R_0) = \frac{1}{2} \int_{R<R_0} d^3\x \int_N d^3\y\: \bigg(
\Big( \frac{1}{2}\: \bxi_\alpha \partial_\alpha V|_\x + \frac{1}{3}\: \bxi_\alpha \bxi_\beta \partial_{\alpha \beta}
V|_\x \Big) \, \L[\bxi] + \O \Big( \frac{M \delta^3}{|\x|^3} \Big) \bigg) \:. \]
Carrying out the $\y$-integration using the formulas in~\eqref{linLcomp}, we obtain
\beq \label{MRO}
{\Mass}(R_0) =
\frac{1}{18}\:\delta^2\, \s_2 \int_{R<R_0} \Delta_{\R^3} V|_\x \: d^3\x + \O \Big( \frac{M \delta^3}{R_0} \Big) \:.
\eeq
Using~\eqref{Vdef}, we get
\[ \Delta_{\R^3} V|_\x = -2M\: \Delta_{\R^3} \Big( \frac{1}{|\x|} \Big) = 8 \pi M\: \delta^3(\x) \:. \]
Employing this formula in~\eqref{MRO}, we can carry out the $\x$-integral to obtain the result.

We remark that, as an alternative which avoids working with distributions, one can mollify~$V$
near the origin. Then the integral in~\eqref{MRO} can be evaluated with the help of the Gauss
divergence theorem to obtain
\begin{align*}
{\Mass}(R_0) &=
\frac{1}{18}\:\delta^2\, \s_2 \int_{R=R_0} \hat{\x}_\alpha \partial_\alpha V|_\x \: d\mu_R(\x) + \O \Big( \frac{M \delta^3}{R_0} \Big) \\
&= \frac{1}{18}\:\delta^2\, \s_2 \: 4 \pi R^2\: \frac{2M}{R^2} + \O \Big( \frac{M \delta^3}{R_0} \Big) \:,
\end{align*}
which again gives the result.
We finally point out that this mollification method is also helpful for clarifying the error term in~\eqref{MRO}.
Namely, mollifying the potential such that it vanishes inside a ball of radius~$R_0/2$, it becomes clear
that integrating the error term~$\O(1/|\x|^3)$ over the annulus~$R_0/2 < |\x| < R_0$ indeed gives
the error term~$\O(1/R_0)$.
\QED

\subsection{Analysis of Alignments}
In this section we shall compute the first two summands in~\eqref{Aalphadef}
in the Schwarzschild geometry and complete the proof of Theorem~\ref{pmtschwarz}.

\begin{Lemma} \label{lemmaA0} In Schwarzschild coordinates, the alignment vector field~$A^{(0)}$
defined in~\eqref{Aalphadef} has the form
\[ A^{(0)}_\alpha(\x) 
= \frac{1}{24}\: \delta^2 \, \s_2\: \partial_\alpha V(\x)\ + \O \Big( \frac{M \delta^3}{|\x|^3} \Big) \:. \]
\end{Lemma}
\Proof
Using~\eqref{vx} we obtain
\begin{align*}
B_\alpha(\x, \bxi) 
\,&\!:= \bxi_\alpha \,D_{1,\bv} \L\Big( \bzeta - \frac{\bxi}{2}, \bzeta + \frac{\bxi}{2} \Big) \\
&= -\frac{1}{4}\: \bxi_\alpha \int_{-\infty}^\infty \epsilon(\tau)\: d\tau \;h_{\beta \gamma}
\big|_{\bzeta + (\tau -\frac{1}{2}) \bxi}\: \frac{\partial}{\partial \bxi_\beta} \big( \bxi_\gamma\, \L[\bxi] \big) \:.
\end{align*}
We take the divergence,
\begin{align*}
&\frac{\partial}{\partial \bzeta_\alpha} B_\alpha(\x, \bxi)  = -\frac{1}{4}\: \bxi_\alpha \int_{-\infty}^\infty \epsilon(\tau)\: d\tau \;\partial_\alpha h_{\beta \gamma}
\big|_{\bzeta + (\tau -\frac{1}{2}) \bxi}\: \frac{\partial}{\partial \bxi_\beta} \big( \bxi_\gamma\, \L[\bxi] \big) \\
&= -\frac{1}{4} \int_{-\infty}^\infty \epsilon(\tau)\: d\tau \; \frac{d}{d\tau} h_{\beta \gamma}
\big|_{\bzeta + (\tau -\frac{1}{2}) \bxi}\: \frac{\partial}{\partial \bxi_\beta} \big( \bxi_\gamma\, \L[\bxi] \big)
= \frac{1}{2} h_{\beta \gamma}
\big|_{\bzeta -\frac{1}{2} \bxi}\: \frac{\partial}{\partial \bxi_\beta} \big( \bxi_\gamma\, \L[\bxi] \big) \:.
\end{align*}
Integrating over~$\bxi$ and using~\eqref{Aalphadef} gives
\begin{align*}
\frac{\partial}{\partial \bzeta_\alpha} A^{(0)}_\alpha(\x)
&= \frac{1}{2} \int_N h_{\beta \gamma}
\big|_{\bzeta -\frac{1}{2} \bxi}\: \frac{\partial}{\partial \bxi_\beta} \big( \bxi_\gamma\, \L[\bxi] \big)\: d^3\bxi \\
&= \frac{1}{4} \int_N \partial_\beta h_{\beta \gamma}
\big|_{\bzeta -\frac{1}{2} \bxi}\: \bxi_\gamma\, \L[\bxi]\: d^3\bxi
\overset{\eqref{divspace}}{=}
-\frac{1}{4} \int_N \partial_\alpha V|_{\bzeta -\frac{1}{2} \bxi}\: \bxi_\alpha\, \L[\bxi]\: d^3\bxi \:.
%\label{divform2}
\end{align*}
We now expand the potential in powers of~$\bxi$,
\begin{align}
\frac{\partial}{\partial \bzeta_\alpha} A^{(0)}_\alpha(\x)
&=-\frac{1}{4} \int_N \partial_\alpha V(\bzeta)\: \bxi_\alpha\, \L[\bxi]\: d^3\bxi \notag \\
&\quad\, +\frac{1}{8} \int_N \partial_{\alpha \beta} V(\bzeta)\: \bxi_\alpha\, \bxi_\beta \:\L[\bxi]\: d^3\bxi 
+ \O \Big( \frac{M \delta^3}{|\x|^3} \Big) \label{interror} \\
&\!\!\!\overset{\eqref{linLcomp}}{=} \frac{1}{24}\: \delta^2 \, \s_2 \: \Delta_{\R^3} V(\bzeta) \: d^3\bxi 
+ \O \Big( \frac{M \delta^3}{|\x|^3} \Big) \:.
\end{align}
Applying the Gauss divergence theorem and using spherical symmetry, we obtain the result.

We finally remark that, in order to integrate the error term over the ball to obtain~\eqref{interror}
in a clean way, exactly as explained at the end of the proof of Lemma~\ref{prpmassdirect},
one can again smoothen~$V$ such that it vanishes inside the ball of radius~$|x|/2$.
\QED

\begin{Lemma} \label{lemmaA1}
In Schwarzschild coordinates, the vector field~$A^{(1)}$ defined in~\eqref{Aalphadef} has the form
\[ A^{(1)}_\alpha(\x) = \frac{1}{72}\: \delta^2 \, \s_2\: \partial_\alpha V(\x) + \O \Big( \frac{M \delta^3}{|\x|^3} \Big) \:. \]
\end{Lemma}
\Proof
Evaluating~\eqref{Aalphadef} for~$k=1$ gives
\begin{align*}
A^{(1)}_\alpha(\bzeta)
&=\frac{1}{4\,3!}
\int_N \bxi_\alpha\: \bigg( \bxi\: \frac{\partial}{\partial \bzeta} \bigg)^2
D_{1,\bv} \L\Big( \bzeta - \frac{\bxi}{2}, \bzeta + \frac{\bxi}{2} \Big)\: d^3\bxi \\
&= -\frac{1}{4}\: \frac{1}{4\,3!}
\int_Nd^3\bxi\: \bxi_\alpha\: \bigg( \bxi\: \frac{\partial}{\partial \bzeta} \bigg)^2
\int_{-\infty}^\infty \epsilon(\tau)\: d\tau \;h_{\beta \gamma}
\big|_{\bzeta + (\tau -\frac{1}{2}) \bxi}\: \frac{\partial}{\partial \bxi_\beta} \big( \bxi_\gamma\, \L[\bxi] \big) \\
&= -\frac{1}{96}
\int_N d^3\bxi\:
\bxi_\alpha \int_{-\infty}^\infty \epsilon(\tau)\: d\tau \; \frac{d^2}{d\tau^2} h_{\beta \gamma}
\big|_{\bzeta + (\tau -\frac{1}{2}) \bxi}\: \frac{\partial}{\partial \bxi_\beta} \big( \bxi_\gamma\, \L[\bxi] \big) \\
&= \frac{1}{48}
\int_N \bxi_\alpha \:\bxi_\delta\, \partial_\delta h_{\beta \gamma}
\big|_{\bzeta -\frac{1}{2} \bxi}\: \frac{\partial}{\partial \bxi_\beta} \big( \bxi_\gamma\, \L[\bxi] \big) \:d^3\bxi \\
&= -\frac{1}{48}
\int_N \bxi_\delta\, \partial_\delta h_{\alpha \gamma}
\big|_{\bzeta -\frac{1}{2} \bxi}\: \bxi_\gamma\, \L[\bxi] \:d^3\bxi \\
&\quad\: -\frac{1}{48}
\int_N \bxi_\alpha \:\partial_\beta h_{\beta \gamma}
\big|_{\bzeta -\frac{1}{2} \bxi}\: \bxi_\gamma\, \L[\bxi] \:d^3\bxi \\
&\quad\: +\frac{1}{96}
\int_N \bxi_\alpha \:\bxi_\delta\, \partial_{\delta \beta} h_{\beta \gamma}
\big|_{\bzeta -\frac{1}{2} \bxi}\: \bxi_\gamma\, \L[\bxi] \:d^3\bxi \\
&= -\frac{1}{24} \int_N \bxi_\alpha \:\partial_\beta h_{\beta \gamma}
\big|_{\bzeta -\frac{1}{2} \bxi}\: \bxi_\gamma\, \L[\bxi] \:d^3\bxi + \O \Big( \frac{M \delta^3}{|\x|^3} \Big) \\
&\!\!\!\overset{\eqref{linLcomp}}{=} -\frac{1}{72}
\int_N \partial_\gamma h_{\alpha \gamma}
\big|_{\bzeta}\: |\bxi|^2\: \L[\bxi] \:d^3\bxi + \O \Big( \frac{M \delta^3}{|\x|^3} \Big) \\
&\!\!\!\overset{\eqref{divspace}}{=}
\frac{1}{72}\: \delta^2 \, \s_2\: \partial_\alpha V(\bzeta) + \O \Big( \frac{M \delta^3}{|\x|^3} \Big) \:.
\end{align*}
This concludes the proof.
\QED

\begin{Lemma} \label{lemmaA1inner} For an inner solution~$\u=(\div \bu, \bu)$,
\[ A_\alpha^{(1)}(\bzeta) =
\frac{1}{72}\: \delta^2\, \s_2 \,\Delta_{\R^3} \bu_\alpha(\bzeta) 
+ \frac{1}{36}\: \delta^2\, \s_2 \, \partial_\alpha \div \bu(\bzeta) + \O\Big(\frac{\delta^3}{\ell_\macro^3} \Big) \:. \]
\end{Lemma}
\Proof We again evaluate~\eqref{Aalphadef} for~$k=1$,
\begin{align*}
A_\alpha^{(1)}(\bzeta) &= \frac{1}{24}
\int_N \bxi_\alpha\: \bigg( \bxi\: \frac{\partial}{\partial \bzeta} \bigg)^2
\nabla_{1,\u} \L\Big( \bzeta - \frac{\bxi}{2}, \bzeta + \frac{\bxi}{2} \Big)\: d^3\bxi \\
&= \frac{1}{24}
\int_N \bxi_\alpha\: \bigg( \bxi\: \frac{\partial}{\partial \bzeta} \bigg)^2
\Big( \div \bu(\x) + \bu_\beta(\x) \: \frac{\partial}{\partial \x_\beta} \Big) \L[\bxi]\: d^3\bxi \\
&= \frac{1}{24} \int_N \bxi_\alpha\: \bxi_\gamma \bxi_\delta \: \partial_{\gamma \delta}
\div \bu(\x) \: \L[\bxi]\: d^3\bxi \\
&\quad\: -\frac{1}{24}
\int_N \bxi_\alpha\: \bxi_\gamma \bxi_\delta \: \partial_{\gamma \delta} \bu_\beta(\x) \:
\frac{\partial}{\partial \bxi_\beta} \L[\bxi]\: d^3\bxi \\
&= -\frac{1}{24}
\int_N \bxi_\alpha\: \bxi_\gamma \bxi_\delta \: \partial_{\gamma \delta} \bu_\beta(\bzeta) \:
\frac{\partial}{\partial \bxi_\beta} \L[\bxi]\: d^3\bxi +\O\Big(\frac{\delta^3}{\ell_\macro^3} \Big) \\
&= \frac{1}{24} \int_N \bxi_\gamma \bxi_\delta \:\partial_{\gamma \delta} \bu_\alpha(\bzeta) \:
\L[\bxi]\: d^3\bxi \\
&\quad\: +\frac{1}{12} \int_N \bxi_\alpha\: \bxi_\delta \:\partial_{\gamma \delta} \bu_\gamma(\bzeta) \:
\L[\bxi]\: d^3\bxi +\O\Big(\frac{\delta^3}{\ell_\macro^3} \Big) \:.
\end{align*}
Expanding in powers of~$\bxi$ gives the result.
\QED

\Proof[Proof of Theorem~\ref{pmtschwarz}] \label{proofpmtschwarz}
We first compute different contributions to the total mass using
the results of Lemmas~\ref{lemmaA0} and~\ref{lemmaA1} in the formula of Corollary~\ref{cormass}.
\begin{align*}
&\Mass^{(0)} (R_0) := \int_{r=R_0} A^{(0)}_\alpha(\x)\: \hat{\x}_\alpha\: d\mu_{R_0}(\x)
= \frac{1}{24}\: \delta^2 \, \s_2 \int_N \Delta_{\R^3} V(\x) \: d^3\x + \O \Big( \frac{M \delta^3}{R_0} \Big) \\
&= -\frac{M}{12}\: \delta^2 \, \s_2 \int_N \Delta_{\R^3} \Big( \frac{1}{|\x|} \Big) \: d^3\x
+ \O \Big( \frac{M \delta^3}{R_0} \Big) \\
&= -\frac{M}{12} \: \delta^2 \, \s_2\int_N (- 4 \pi)\: \delta^3(\x) \: d^3\x + \O \Big( \frac{M \delta^3}{|\x|^3} \Big)
= \frac{\pi}{3}\: \delta^2 \, \s_2\: M + \O \Big( \frac{M \delta^3}{R_0} \Big) \\
&\Mass^{(1)} (R_0) := \int_{r=R_0} A^{(1)}_\alpha(\x)\: \hat{\x}_\alpha\: d\mu_{R_0}(\x)
= \frac{1}{72} \: \delta^2 \, \s_2 \int_N \Delta_{\R^3} V(\x) \: d^3\x + \O \Big( \frac{M \delta^3}{R_0} \Big) \\
&= -\frac{M}{36}\: \delta^2 \, \s_2 \int_N \Delta_{\R^3} \Big( \frac{1}{|\x|} \Big) \: d^3\x
+ \O \Big( \frac{M \delta^3}{R_0} \Big) \\
&= -\frac{M}{36}\: \delta^2 \, \s_2 \int_N (- 4 \pi)\: \delta^3(\x) \: d^3\x + \O \Big( \frac{M \delta^3}{|\x|^3} \Big)
= \frac{\pi}{9}\: \delta^2 \, \s_2\: M + \O \Big( \frac{M \delta^3}{R_0} \Big) \:.
\end{align*}

In order to choose linear alignment, we must add an inner solution~$\u$.
More precisely, comparing Lemma~\ref{lemmaA0} with Lemma~\ref{lemmaA0inner}, one sees that
we must choose
\[ %\label{ucomp}
\u_\alpha(\x) = -\frac{1}{24\,\s}\:\delta^2\, \s_2\: \partial_\alpha V(\x)
+ \O \Big( \frac{M \delta^3}{|\x|^3} \Big) \:. \]
Then~$A^{(0)}$ and therefore also~$\Mass^{(0)}$ vanish.
According to Lemma~\ref{lemmaA1inner}, the inner solution changes~$A^{(1)}$
only to the order~$\O\big(\delta^3/|\x|^3)$. Therefore, the above formula for~$\Mass^{(1)} (R_0)$
still holds with alignment, giving~\eqref{Massaligned}.

It remains to show that the higher order contributions can be absorbed into the error term.
More precisely, let us show that
\beq \label{err4}
\Mass^{(k)}(R_0) = \O \Big( \frac{M \delta^4}{R_0^2} \Big) \qquad \text{for~$k=2,3,\ldots$} \:.
\eeq
To this end, we note that, according to~\eqref{Aalphadef}, these contributions involve
at least five factors~$\bxi$ and at least four $\bzeta$-derivatives.
Proceeding as in the proofs of Lemmas~\ref{lemmaA0} and~\ref{lemmaA1},
using the specific form of the jets, one can integrate by parts once in~$\bxi$
and once in~$\tau$. After doing so, at least four factors~$\bxi$ and at least three $\beta$-derivatives
remain. Since the $\bzeta$-derivatives act on the potential~$V$, we obtain a scaling factor~$R_0^{-4}$.
Taking into account that the surface integral gives a scaling factor~$R_0^2$, we obtain~\eqref{err4}.
\QED

\subsection{Spatially Isotropic Coordinates}
The form of the linearized Schwarzschild metric~\eqref{linschwarz} becomes somewhat simpler
by transforming to spatially isotropic coordinates 
(see also~\cite[Exercise~4 in~\S100]{landau2} or~\cite[Problem~1 in Section~6]{wald}).
On the other hand, the volume form is no longer preserved, giving rise to jets with scalar components.
This is why we decided to carry out all computations in Schwarzschild coordinates.
Nevertheless, we now explain how the linearized Schwarzschild metric in
spatially isotropic coordinates can be described with jets. This formulation might be
of advantage for future computations in asymptotically flat spacetimes.

In order to transform to spatially isotropic coordinates, one introduces
the new radial variable~$\tilde{r}$ by
\beq \label{isotrans}
r = \Big( 1 + \frac{M}{2 \tilde{r}} \Big)^{2}\:\tilde{r} \:, \qquad dr = 
\Big( 1 + \frac{M}{2 \tilde{r}} \Big) \Big( 1 - \frac{M}{2 \tilde{r}} \Big) \: d\tilde{r} \:.
\eeq
Then the line element becomes
\[ ds^2 = \left( \frac{ \displaystyle 1 - \frac{M}{2 \tilde{r}}}{ \displaystyle 1 + \frac{M}{2 \tilde{r}}} \right)^2\! dt^2
- \Big( 1 + \frac{M}{2 \tilde{r}} \Big)^4 \:\big( d\tilde{r}^2 + \tilde{r}^2 d\vartheta^2 + \tilde{r}^2 \sin^2 \vartheta\: d\varphi^2 \big) \:. \]
Again using the abbreviation~\eqref{Vdef}, the linearized metric is given by
\beq \label{liniso}
h'_{00}(x) = V(\x)\:,\quad h'_{\alpha \beta} = V(\x)\: \delta_{\alpha \beta}
\eeq

As already explained for ultrastatic spacetimes in Section~\ref{secultralin} (see~\eqref{infcoordultra}
and~\eqref{infmetricultra}), in the formalism of linearized gravity, one has
the freedom to perform infinitesimal coordinate transformations
\beq \label{infcoord}
x'^i = x^i + \zeta^i(\x)
\eeq
This transforms the linearized metric according to
\beq \label{infmetric}
h'_{jk} = h_{jk} - \partial_j \zeta_k - \partial_k \zeta_j \:.
\eeq
This freedom can be understood as the gauge freedom of the gravitational field.
It is most convenient to use this gauge freedom to arrange the gauge condition
\[ \partial^j h_{jk} = \frac{1}{2}\: \partial_k h \qquad \text{with} \qquad h := h^i_i \:. \]
This gauge condition means that we are working in {\em{harmonic coordinates}}
(for details see~\cite[\S107]{landau2}).
The spatially isotropic coordinates are indeed harmonic, as one sees directly from the computation
(see also~\cite[\S105]{landau2})
\begin{align*}
h'(x) &= h'_{00}(x) - \sum_{\alpha=1}^3 h'_{\alpha \alpha}(x) = -2V(\x) = 2 \:\frac{2M}{R} \\
 \partial^j h'_{jk}(x) &= -\sum_{\alpha=1}^3 \partial_\alpha h'_{\alpha k}(x)
= \delta_{\alpha k} \:\partial_\alpha \Big( \frac{2M}{R} \Big) = \partial_k \Big( \frac{2M}{R} \Big)
= \frac{1}{2}\: \partial_k h'(x) \:.
\end{align*}

The transition from Schwarzschild coordinates to spatially isotropic coordinates is
described by the infinitesimal coordinate transformation~\eqref{infcoord} with
\beq \label{zetaiso}
\bzeta_0(x) = 0 \qquad \text{and} \qquad \bzeta_\alpha(x) = M \: \frac{x_\alpha}{R}\:,
\eeq
as one sees by linearizing~\eqref{isotrans} or, alternatively, from~\eqref{infmetric} using that
\beq \label{zetah}
 \partial_\alpha \bzeta_\beta(x) + \partial_\beta \bzeta_\alpha(x)
= \frac{2M}{R} \Big( \delta_{\alpha \beta} - \frac{x_\alpha x_\beta}{R^2} \Big)
= -V(\x) \big( \delta_{\alpha \beta} - \hat{\x}_\alpha \hat{\x}_\beta \big) \:.
\eeq

The first variation of the Lagrangian can again be described with the help of~\eqref{delL}
and~\eqref{dL1}, \eqref{dL2}. Using~\eqref{liniso}, we obtain
\begin{align*}
\delta \L(\x, \y) &= -\frac{1}{2} \int_0^1 d\tau \;h'_{00}
\big|_{\tau y + (1-\tau)\, x}\: \L[\bxi] -\frac{1}{2} \int_0^1 d\tau \;h'_{\alpha \beta}
\big|_{\tau y + (1-\tau)\, x}\: \bxi_\alpha\: \frac{\partial}{\partial \y_\beta} \L[\bxi] \\
&= -\frac{1}{2} \int_0^1 d\tau \;V|_{\tau \y + (1-\tau)\, \x}\: \L[\bxi]
-\frac{1}{2} \int_0^1 d\tau \;V\big|_{\tau \y + (1-\tau)\, \x}\: \bxi_\alpha\: \frac{\partial}{\partial \y_\alpha}
\L[\bxi] \\
&= -\frac{1}{2} \int_0^1 d\tau \;V|_{\tau \y + (1-\tau)\, \x}\: \Big( 1 + \bxi_\alpha\: \frac{\partial}{\partial \y_\alpha} \Big) \L[\bxi] \:.
\end{align*}
Consequently, we introduce the jet derivatives
\begin{align}
D_{1,\bv'} \L(\x, \y) &:= -\frac{1}{4} \int_{-\infty}^\infty \epsilon(\tau)\: d\tau\;
V|_{\tau \y + (1-\tau)\, \x}\: \Big( 1 - \bxi_\alpha\: \frac{\partial}{\partial \x_\alpha} \Big) \L[\bxi] \label{vpx} \\
D_{2,\bv'} \L(\x, \y) &:= -\frac{1}{4} \int_{-\infty}^\infty \epsilon(1-\tau)\: d\tau\;
V|_{\tau \y + (1-\tau)\, \x}\: \Big( 1 + \bxi_\alpha\: \frac{\partial}{\partial \y_\alpha} \Big) \L[\bxi] \:. \label{vpy}
\end{align}
Moreover, we must take into account that the volume form changes. This gives rise to
a scalar component of the jet given by
\beq \label{bpdef}
b'(\x) = \delta \sqrt{|\det g'|} = \frac{1}{2}\: h'(\x) = -V(\x)\:.
\eeq

In the next lemma we verify in detail that the jets~$\v$ and~$\v'$
differ by an inner solution describing the infinitesimal coordinate transformation~\eqref{zetaiso}.
\begin{Lemma} $\;$ The diffeomorphism jet~$\v=(0,\bv)$ (with~$\bv$ according to~\eqref{vx} and~\eqref{vy})
and the jet~$\v'=(b, \bv')$ (with~$b$ according to~\eqref{bpdef} and~$\bv'$ as in~\eqref{vpx} and~\eqref{vpy})
satisfy the relation
\[ %\label{vvp}
\v'  = \v + \u \:, \]
where~$\u$ is the inner solution corresponding to the vector field~$\bzeta$ in~\eqref{zetaiso}, i.e.\
\beq \label{uform}
\u = (\div \bzeta, \bzeta) = \big( -V(\x), M \hat{\x} \big) \:.
\eeq
\end{Lemma}
\Proof We first compute the divergence of~$\bzeta$,
\[ \div \bzeta(x) = \partial_\alpha \Big( M \: \frac{x_\alpha}{R} \Big) = \frac{M}{R}\: (-1+3) = \frac{2M}{R} = -V(\x) \:, \]
proving the last equality in~\eqref{uform}.

It suffices to prove that
\beq \label{vmvp}
\big( \nabla_{1,\v'} - D_{1,\bv}  \big) \L(\x,\y) = \nabla_{1,\u} \L(\x,\y) \:,
\eeq
because the corresponding relation for the $\y$-derivatives acting is obtained by the
replacements~$\x \leftrightarrow \y$, $\bxi \rightarrow -\bxi$ and~$\tau \rightarrow 1-\tau$.
We first consider those terms on the left which involve derivatives of the Lagrangian.
According to~\eqref{vpx} and~\eqref{vx}, they can be written as
\begin{align*}
A &:= \frac{1}{4} \int_{-\infty}^\infty \epsilon(\tau)\: d\tau \;\Big( V|_{\tau \y + (1-\tau)\, \x}\: \bxi_\beta - h_{\alpha \beta}\big|_{\tau y + (1-\tau)\, x}\: 
\bxi_\alpha \Big) 
\frac{\partial}{\partial \x_\beta} \big( \, \L[\bxi] \big) \\
&\!\overset{\eqref{hsch}}{=} \frac{1}{4} \int_{-\infty}^\infty \epsilon(\tau)\: d\tau \;\Big( V(\z) \,
\big( \delta_{\alpha \beta} - \hat{\z}_\alpha\, \hat{\z}_\beta \big) \Big) \Big|_{z=\tau \y + (1-\tau)\, \x}\: 
\bxi_\alpha \: \frac{\partial}{\partial \x_\beta} \big( \, \L[\bxi] \big) \\
&\!\!\overset{\eqref{zetah}}{=} -\frac{1}{4} \int_{-\infty}^\infty \epsilon(\tau)\: d\tau \;\big( \partial_\alpha \bzeta_\beta + \partial_\beta \bzeta_\alpha \big)
\big|_{\tau \y + (1-\tau)\, \x}\: 
\bxi_\alpha \: \frac{\partial}{\partial \x_\beta} \big( \, \L[\bxi] \big) \:.
\end{align*}
Since the Lagrangian is spherically symmetric, its derivative~$\partial_{\x_\beta} \L(x,y)$
is proportional to~$\bxi_\beta$. Therefore,
\[ \big( \partial_\alpha \bzeta_\beta + \partial_\beta \bzeta_\alpha \big)
\big|_{\tau \y + (1-\tau)\, \x}\: 
\bxi_\alpha \: \frac{\partial}{\partial \x_\beta} \L[\bxi]  = 
2\, \bxi_\alpha (\partial_\alpha \bzeta_\beta) \: \frac{\partial}{\partial \x_\beta} \L[\bxi] \:. \]
We thus obtain
\begin{align*}
A &=-\frac{1}{2} 
\int_{-\infty}^\infty \epsilon(\tau)\: d\tau \;\frac{d}{d\tau} \bzeta_\beta \big|_{\tau \y + (1-\tau)\, \x}\: 
\frac{\partial}{\partial \x_\beta} \L[\bxi]
= \bzeta_\beta(\x)\: \frac{\partial}{\partial \x_\beta} \L[\bxi] = D_{1,\bu} \L(\x,\y)\:.
\end{align*}
Taking into account the contributions to the left of~\eqref{vmvp} where the Lagrangian is not
differentiated, we obtain from\eqref{bpdef}, \eqref{vpx} and~\eqref{vx}, 
\begin{align*}
&\big( \nabla_{1,\v'} - D_{1,\bv} \big) \L(\x,\y)
= A - V(\x)\: \L(\x,\y) \\
& + \frac{1}{4} \int_{-\infty}^\infty \epsilon(\tau)\: d\tau\; V|_{\tau \y + (1-\tau)\, \x}\:  \L(\x,\y)
- \frac{1}{4} \int_{-\infty}^\infty \epsilon(\tau)\: d\tau \;h_{\alpha \beta}
\big|_{\tau y + (1-\tau)\, x}\: \delta_{\alpha \beta}\, \L(\x,\y) \big) \\
&\!\!\!\overset{\eqref{hsch}}{=} A  - V(\x)\: \L(\x,\y) = \big( -V(\x) + D_{1,\bzeta} \big) \: \L(\x,\y)
\overset{\eqref{uform}}{=} \nabla_{1,\u} \L(\x,\y) \:.
\end{align*}
This concludes the proof.
\QED

\subsection{A Spatial Integral Related to the Total Mass}
In~\cite[Section~6.5]{pmt} the total mass was expressed in terms of a spatial integral,
\beq \label{MRdef}
{\Mass} = 2 \pi \lim_{R \rightarrow \infty} R^2 \, {\mathfrak{P}}(R)
\eeq
with
\beq \label{frakPdef}
{\mathfrak{P}}(R) := \int_N (r-R)\: \big(D_{1,\bv} - D_{2,\bv} \big) \L(\x, \y)\: d^3\y \:.
\eeq
In order to derive this representation, we employed a method used previously in~\cite[Lemma~5.5]{noether}
and~\cite[Lemma~5.2]{action} which consists in integrating the conserved surface layer integral
over the radius and taking its mean. More precisely, using the abbreviation
\[ A(r, r') = r^2\, r'^2\int_{S^2} d\omega \int_{S^2} d\omega' \; \big(D_{1,\bv} - D_{2,\bv} \big) \L(\x, \y) \:, \]
we obtain the identity
\begin{align}
{\Mass}(R) &= \frac{1}{L} \int_{R_{\min}}^R dr \int_R^{R+L} dr'\: (r'-R)\: A(r, r') \label{t1a} \\
&\quad\:+ \frac{1}{L} \int_{R_{\min}}^R dr \int_{R+L}^\infty dr'\: L\: A(r, r') \label{t2a} \\
&\quad\:+ \frac{1}{L} \int_R^{R+L} dr \int_r^{R+L} dr'\: (r'-r)\: A(r, r') \label{t3a} \\
&\quad\:+ \frac{1}{L} \int_{R}^{R+L} dr \int_{R+L}^\infty dr'\: (L-r+R)\: A(r, r') \:. \label{t4a}
\end{align}
In the limit~$L \rightarrow \infty$, the summands~\eqref{t1a} and~\eqref{t2a} obviously vanish.
Moreover, the inner integral~\eqref{t3a} can be shown to converge to~$2 \pi r^2\, {\mathfrak{P}}(r)$.
Also taking the limit~$R \rightarrow \infty$, we obtain the identity
\beq \label{t5}
{\Mass} = 2 \pi \lim_{r \rightarrow \infty} r^2\: {\mathfrak{P}}(r)
+ \lim_{R \rightarrow \infty} \lim_{L \rightarrow \infty}
\frac{1}{L} \int_{R}^{R+L} dr \int_{R+L}^\infty dr'\: (L-r+R)\: A(r, r') \:.
\eeq
If the remaining surface layer integral is bounded uniformly in~$L$, the last summand vanishes,
giving~\eqref{MRdef}. This is indeed the case in the previous applications in~\cite{noether, action}
as well as for the jet~$\v$ considered in~\cite{pmt}. 
However, for the jet~$\v$ considered here,
the surface layer integral in~\eqref{t5} turns out to be linearly divergent in~$L$, so that
the last summand is non-zero. In simple terms, this can be understood from the fact that
the jet~$\v$ described by~\eqref{vx} and~\eqref{vy} does not have the necessary decay conditions at infinity
(in~\cite{pmt} this issue is avoided with a scaling argument, which however does not capture
the higher orders in~$\delta$ correctly; see~\cite[Proposition~6.3]{pmt} and the corresponding footnote).
In order to explain in detail why the last summand in~\eqref{t5} is indeed non-zero, we
now compute~${\mathfrak{P}}(r)$.
\begin{Prp} \label{prpmassalt}
Evaluating the spatial integral~\eqref{MRdef} for~$\x$ on a sphere of radius~$R>0$,
the spatial integral~\eqref{frakPdef} can be computed to obtain
\[ \mathfrak{P}(R) =  \frac{M}{R^2}\: \delta^2\, \s_2 + \O \Big( \frac{M \delta^3}{R^3} \Big) \:. \]
\end{Prp} \noindent
Comparing with the result of Proposition~\ref{prpmassdirect}, one sees
explicitly that~\eqref{t5} is violated, because the prefactors are different.

The remainder of this appendix is devoted to the proof of Proposition~\ref{prpmassalt}.
First, from~\eqref{vx} and~\eqref{vy}, we find
\begin{align*}
\big( D_{1,\bv} - D_{2,\bv} \big) \L(\x, \y)
&= -\frac{1}{4} \int_{-\infty}^\infty \epsilon(\tau)\: d\tau \;h_{\alpha \beta}
\big|_{\tau y + (1-\tau)\, x}\: \frac{\partial}{\partial \y_\beta} \big( \bxi_\alpha\, \L[\bxi] \big) \\
&\quad\; +\frac{1}{4} \int_{-\infty}^\infty \epsilon(1-\tau)\: d\tau \;h_{\alpha \beta}
\big|_{\tau y + (1-\tau)\, x}\: \frac{\partial}{\partial \y_\beta} \big( \bxi_\alpha\, \L[\bxi] \big) \\
&= -\frac{1}{2}\, \bigg( \int_1^\infty - \int_{-\infty}^0 \bigg) \:d\tau \;h_{\alpha \beta}
\big|_{\tau y + (1-\tau)\, x}\: \frac{\partial}{\partial \y_\beta} \big( \bxi_\alpha\, \L[\bxi] \big) \:.
\end{align*}
As a consequence,
\begin{align}
&{\mathfrak{P}}(R) \notag \\
&= -\frac{1}{2}\, \bigg( \int_1^\infty - \int_{-\infty}^0 \bigg) \:d\tau 
\int_N d^3\y\: (r-R)\;h_{\alpha \beta}
\big|_{\tau y + (1-\tau)\, x}\: \frac{\partial}{\partial \y_\beta} \big( \bxi_\alpha\, \L[\bxi] \big) \notag \\
&= \frac{1}{2}\, \bigg( \int_1^\infty - \int_{-\infty}^0 \bigg) \:d\tau 
\int_N d^3\y\: \hat{\y}_\beta \;h_{\alpha \beta}
\big|_{\tau y + (1-\tau)\, x}\: \bxi_\alpha\, \L[\bxi] \label{t1b} \\
&\quad\: +\frac{1}{2}\, \bigg( \int_1^\infty - \int_{-\infty}^0 \bigg) \:d\tau 
\int_N d^3\y\: (r-R)\; \tau\: \partial_\beta h_{\alpha \beta}
\big|_{\tau y + (1-\tau)\, x}\: \bxi_\alpha\, \L[\bxi] \:. \label{t2b}
\end{align}
In~\eqref{t2b} we can apply~\eqref{divspace} and integrate by parts,
\begin{align*}
\eqref{t2b} &= -\frac{1}{2}\, \bigg( \int_1^\infty - \int_{-\infty}^0 \bigg) \:d\tau 
\int_N d^3\y\: (r-R)\; \tau\: \frac{d}{d\tau} V |_{\tau y + (1-\tau)\, x}\:\L[\bxi] \\
&= \frac{1}{2} \int_N d^3\y\: (r-R)\: V(\y)\: \L[\bxi] \\
&\quad\: +\frac{1}{2}\, \bigg( \int_1^\infty - \int_{-\infty}^0 \bigg) \:d\tau 
\int_N d^3\y\: (r-R)\;V|_{\tau \y + (1-\tau)\, \x}\: \L[\bxi] \:.
\end{align*}
In order to simplify~\eqref{t1b}, we use that, according to~\eqref{linschwarz} and~\eqref{Vdef},
\begin{align*}
 \partial_\alpha \Big( \frac{\z_\beta}{|\z|} \Big) &= \frac{\delta_{\alpha \beta}}{|\z|}
- \frac{\z_\alpha \z_\beta}{|\z|^3} \\
h_{\alpha \beta}(\z) &= -\frac{2M}{|\z|}\: \hat{\z}_\alpha\, \hat{\z}_\beta
= -2M\: \frac{\z_\alpha\, \z_\beta}{|\z|^3} \\
&= 2 M\: \partial_\alpha \Big( \frac{\z_\beta}{|\z|} \Big) - 2 M\: \frac{\delta_{\alpha \beta}}{|\z|} 
= -\partial_\alpha \big( \z_\beta\: V(\z) \big) + \delta_{\alpha \beta}\: V(\z) \:. %\label{habrel}
\end{align*}
It follows that
\begin{align*}
\eqref{t1b} &= -\frac{1}{2}\, \bigg( \int_1^\infty - \int_{-\infty}^0 \bigg) \:d\tau 
\int_N d^3\y\: \hat{\y}_\beta \;\frac{d}{d\tau} \big( \z_\beta\: V(\z) \big)
\big|_{\tau y + (1-\tau)\, x}\: \L[\bxi] \\
&\quad\: +\frac{1}{2}\, \bigg( \int_1^\infty - \int_{-\infty}^0 \bigg) \:d\tau 
\int_N d^3\y\: \hat{\y}_\alpha\, \bxi_\alpha \;V |_{\tau y + (1-\tau)\, x}\: \, \L[\bxi] \\
&= \frac{1}{2} 
\int_N d^3\y\: \hat{\y}_\alpha \;\Big( \y_\alpha\: V(\y) + \x_\alpha\: V(\x) \Big)\: \L[\bxi] \\
&\quad\:+\frac{1}{2}\, \bigg( \int_1^\infty - \int_{-\infty}^0 \bigg) \:d\tau 
\int_N d^3\y\: \hat{\y}_\alpha\, \bxi_\alpha \;V |_{\tau y + (1-\tau)\, x}\: \, \L[\bxi] \:.
\end{align*}
Adding all the contributions, we obtain
\begin{align}
{\mathfrak{P}}(R)
&= \frac{1}{2}\, \bigg( \int_1^\infty - \int_{-\infty}^0 \bigg) \:d\tau 
\int_N d^3\y\: (r-R)\;V|_{\tau \y + (1-\tau)\, \x}\: \L[\bxi] \label{ta} \\
&\quad\: +\frac{1}{2}\, \bigg( \int_1^\infty - \int_{-\infty}^0 \bigg) \:d\tau 
\int_N d^3\y\: \hat{\y}_\alpha\, \bxi_\alpha \;V |_{\tau y + (1-\tau)\, x}\: \, \L[\bxi] \label{tb} \\
&\quad\: +\frac{1}{2} \int_N d^3\y\: (r-R)\: V(\y)\: \L[\bxi] \label{tc} \\
&\quad\: +\frac{1}{2} 
\int_N d^3\y\: \hat{\y}_\alpha \;\Big( \y_\alpha\: V(\y) + \x_\alpha\: V(\x) \Big)\: \L[\bxi] \:. \label{td}
\end{align}

In the next step we Taylor expand in powers of~$\bxi$.
\begin{Lemma}
\begin{align}
{\mathfrak{P}}(R)
&= \bigg( \int_1^\infty - \int_{-\infty}^0 \bigg) \:d\tau 
\int_N d^3\y\: \frac{\x \bxi}{|\x|} \;V|_{\tau \y + (1-\tau)\, \x}\: \L[\bxi] \label{ua} \\
&\quad\: +\frac{3}{4}\: \bigg( \int_1^\infty - \int_{-\infty}^0 \bigg) \:d\tau 
\int_N d^3\y\: \bigg( \frac{\bxi^2}{|\x|} - \frac{( \x \bxi)^2}{|\x|^3}
\bigg)\;V|_{\tau \y + (1-\tau)\, \x}\: \L[\bxi] \label{ub} \\
&\quad\: + \s\: |\x|\, V(\x) -\frac{1}{6\, |\x|}\: \delta^2\, \s_2\: V(\x) + \O \Big( \frac{M \delta^3}{|\x|^3} \Big) \:. \label{uc}
\end{align}
\end{Lemma}
\Proof
A direct computation yields
\begin{align*}
r-R &= |\y| - |\x| = |\x + \bxi| - |\x|
= \sqrt{ \x^2 + 2 \x \bxi + \xi^2} - |\x| \\
&= \frac{1}{2\, |\x|} \big( 2 \x \bxi + \bxi^2 \big) -\frac{1}{8\, |\x|^3}\: \big( 2 \x \bxi \big)^2 + \O\big( \bxi^3 \big) \\
&= \frac{\x \bxi}{|\x|} +\frac{\bxi^2}{2\, |\x|} -\frac{( \x \bxi)^2}{2\, |\x|^3} + \O\big( \bxi^3 \big) \\
\hat{\y}_\alpha\, \bxi_\alpha &= \frac{\y \bxi}{|\y|} = \frac{(\x+\bxi) \bxi}{|\x + \bxi|}
= \frac{\x \bxi}{|\x|} +  \frac{\bxi^2}{|\x|} - \frac{( \x \bxi)^2}{|\x|^3} + \O\big( \bxi^3 \big) \\
\hat{\y}_\alpha\, \y_\alpha &= r = |\x| + \big( r - R \big) \\
&= |\x| + \frac{\x \bxi}{|\x|} +\frac{\bxi^2}{2\, |\x|} -\frac{( \x \bxi)^2}{2\, |\x|^3} + \O\big( \bxi^3 \big) \\
\hat{\y}_\alpha\, \x_\alpha &= \hat{\y}_\alpha \, \big( \y_\alpha - \bxi_\alpha \big) \\
&= |\x| + \frac{\x \bxi}{|\x|} +\frac{\bxi^2}{2\, |\x|} -\frac{( \x \bxi)^2}{2\, |\x|^3}
- \Big( \frac{\x \bxi}{|\x|} +  \frac{\bxi^2}{|\x|} - \frac{( \x \bxi)^2}{|\x|^3} \Big) + \O\big( \bxi^3 \big) \\
&= |\x| - \frac{\bxi^2}{2\, |\x|} +\frac{( \x \bxi)^2}{2\, |\x|^3} + \O\big( \bxi^3 \big) \:.
\end{align*}
Hence
\begin{align*}
&\eqref{ta} + \eqref{tb} \\
&= \bigg( \int_1^\infty - \int_{-\infty}^0 \bigg) \:d\tau 
\int_N d^3\y\: \bigg( \frac{\x \bxi}{|\x|} +\frac{3}{4}\: \frac{\bxi^2}{|\x|} - \frac{3}{4}\:\frac{( \x \bxi)^2}{|\x|^3}
+ \O\big( \bxi^3 \big) \bigg)\;V|_{\tau \y + (1-\tau)\, \x}\: \L[\bxi] \\
&\eqref{tc} + \eqref{td} \\
&= \frac{1}{2} \int_N d^3\y\: \big(r-R + \hat{\y}_\alpha \, \y_\alpha \big)\: V(\y)\: \L[\bxi]
+\frac{1}{2} 
\int_N d^3\y\: \hat{\y}_\alpha\, \x_\alpha\: V(\x)\: \L[\bxi] \\
&=\frac{1}{2} \int_N d^3\y\: \Big(  |\x|  + 2\:\frac{\x \bxi}{|\x|} +\frac{\bxi^2}{|\x|} -\frac{( \x \bxi)^2}{|\x|^3} \Big)\: V(\y)\: \L[\bxi] \\
&\quad\: +\frac{1}{2} 
\int_N d^3\y\:\Big( |\x| - \frac{\bxi^2}{2\,|\x|}+ \frac{(\x \bxi)^2}{2 \,|\x|^3} \Big)\: V(\x)\: \L[\bxi] + 
+ \O \Big( \frac{M \delta^3}{|\x|^3} \Big) \\
&= \s\: |\x|\, V(\x) + \frac{1}{2} \: |\x| \int_N d^3\y \:\frac{1}{2}\: \bxi_\alpha \bxi_\beta \:\partial_{\alpha \beta} V(\x)\:
\L[\bxi] +\int_N d^3\y\: \frac{\x \bxi}{|\x|} \:\bxi_\alpha \:\partial_\alpha V(\x)\: \L[\bxi] \\
&\quad\: +\frac{1}{2} \int_N d^3\y\: \Big( \frac{\bxi^2}{2\,|\x|} -\frac{( \x \bxi)^2}{2\,|\x|^3} \Big)\: V(\x)\: \L(\x,\y) 
+ \O \Big( \frac{M \delta^3}{|\x|^3} \Big) \\
&= \s\: |\x|\, V(\x) + \frac{1}{12}\: \delta^2\, \s_2 \: |\x| \:\Delta_{\R^3} V(\x) +
\frac{1}{3\,|\x|}\: \delta^2\, \s_2\: \x_\alpha \partial_\alpha V(\x) \\
&\quad\: +\frac{1}{6\, |\x|}\: \delta^2\, \s_2\: V(\x) + \O \Big( \frac{M \delta^3}{|\x|^3} \Big) \:.
\end{align*}
We now use that, away from the origin, the potential is harmonic and homogeneous of degree minus one, i.e.\
\[ \Delta_{\R^3} V(\x) = 0 \qquad \text{and} \qquad  \x_\alpha \partial_\alpha V(\x) = -V(\x) \:. \]
We thus obtain
\[ \eqref{tc} + \eqref{td} = \s\: |\x|\, V(\x) -\frac{1}{6\, |\x|}\: \delta^2\, \s_2\: V(\x) + \O \Big( \frac{M \delta^3}{|\x|^3} \Big) \:. \]
Collecting all the terms gives the result.
\QED

Our next task is to compute the $\tau$-integrals in~\eqref{ua} and~\eqref{ub}.
This will be accomplished in the following two lemmas.
\begin{Lemma} For any~$\x$ on a sphere of radius~$R>0$,
\begin{align*}
&\bigg( \int_1^\infty - \int_{-\infty}^0 \bigg) \:d\tau 
\int_N d^3\y\: \frac{\x \bxi}{|\x|}\;V|_{\tau \y + (1-\tau)\, \x}\: \L(\x,\y) \\
&= -\s \,|\x| \, V(\x) + \frac{1}{6}\: \delta^2\, \s_2 \: \frac{1}{|\x|} \:V(\x) + \O \Big( \frac{M \delta^3}{|\x|^3} \Big) \:.
\end{align*}
\end{Lemma}
\Proof For any~$\x \in B_R(0)$, we compute the following divergence
using integration by parts,
\begin{align*}
&\frac{\partial}{\partial \x_\alpha}
\bigg( \int_1^\infty - \int_{-\infty}^0 \bigg) \:d\tau 
\int_N d^3\y\: \bxi_\alpha\;V|_{\tau \y + (1-\tau)\, \x}\: \L(\x,\y) \\
&=-3 \:\bigg( \int_1^\infty - \int_{-\infty}^0 \bigg) \:d\tau 
\int_N d^3\y\: V|_{\tau \y + (1-\tau)\, \x}\: \L(\x,\y) \\
&\quad\: +\bigg( \int_1^\infty - \int_{-\infty}^0 \bigg) \:d\tau 
\int_N d^3\y\: (1-\tau)\: \frac{d}{d\tau} V|_{\tau \y + (1-\tau)\, \x}\: \L(\x,\y) \\
&\quad\: - \bigg( \int_1^\infty - \int_{-\infty}^0 \bigg) \:d\tau 
\int_N d^3\y\: \bxi_\alpha\;V|_{\tau \y + (1-\tau)\, \x}\: \frac{\partial}{\partial \y_\alpha} \L(\x,\y) \\
&=%-3 \:\bigg( \int_1^\infty - \int_{-\infty}^0 \bigg) \:d\tau \int_N d^3\y\: V|_{\tau \y + (1-\tau)\, \x}\: \L(\x,\y) \\ &\quad\: +
\bigg( \int_1^\infty - \int_{-\infty}^0 \bigg) \:d\tau 
\int_N d^3\y\: (1-\tau)\: \frac{d}{d\tau} V|_{\tau \y + (1-\tau)\, \x}\: \L(\x,\y) \\
%&\quad\: + 3 \:\bigg( \int_1^\infty - \int_{-\infty}^0 \bigg) \:d\tau 
%\int_N d^3\y\: V|_{\tau \y + (1-\tau)\, \x}\: \L(\x,\y) \\
&\quad\: + \bigg( \int_1^\infty - \int_{-\infty}^0 \bigg) \:d\tau 
\int_N d^3\y\: \tau\: \frac{d}{d\tau} V|_{\tau \y + (1-\tau)\, \x}\: \L(\x,\y) \\
&= \bigg( \int_1^\infty - \int_{-\infty}^0 \bigg) \:d\tau 
\int_N d^3\y\: \frac{d}{d\tau} V|_{\tau \y + (1-\tau)\, \x}\: \L(\x,\y) \\
&= - \int_N d^3\y\: \big( V(\y) + V(\x) \big)\: \L(\x,\y) \\
&= - 2 \s\, V(\x) - \int_N d^3\y\: \frac{1}{2}\: \bxi_\alpha \bxi_\beta \partial_{\alpha \beta} V(\x)\: \L(\x,\y) \\
&= - 2 \s\, V(\x) - \frac{1}{6}\: \delta^2\, \s_2 \: \Delta_{\R^3} V(\x) + \O \Big( \frac{M \delta^3}{|\x|^3} \Big) \:,
\end{align*}
where in the last steps we again performed an in powers of~$\xi$
and used~\eqref{linLcomp}.

Applying the Gauss divergence theorem and spherical symmetry, we conclude that
\begin{align}
& \bigg( \int_1^\infty - \int_{-\infty}^0 \bigg) \:d\tau 
\int_N d^3\y\: \bxi_\alpha\;V|_{\tau \y + (1-\tau)\, \x}\: \L(\x,\y) \notag \\
&= -\s \,\x_\alpha \, V(\x) - \frac{1}{6}\: \delta^2\, \s_2 \: \partial_\alpha V(\x) + \O \Big( \frac{M \delta^3}{|\x|^3} \Big) \:.
\label{tomollify}
\end{align}
Note that we used that
\[ \partial_\alpha \big( \x_\alpha \, V(\x) \big) = 3\, V(\x) + \x^\alpha \partial_\alpha V(\x)
= 2\, V(\x) \]
(where in the last step we used again that the potential is homogeneous of degree minus one).
Multiplying by~$\hat{\x}_\alpha$ gives
\begin{align*}
&\bigg( \int_1^\infty - \int_{-\infty}^0 \bigg) \:d\tau 
\int_N d^3\y\: \frac{\x \bxi}{|\x|}\;V|_{\tau \y + (1-\tau)\, \x}\: \L(\x,\y) \\
&= -\s \,|\x| \, V(\x) - \frac{1}{6}\: \delta^2\, \s_2 \: \frac{\x_\alpha}{|\x|}\: \partial_\alpha V(\x) + \O \Big( \frac{M \delta^3}{|\x|^3} \Big) \:.
\end{align*}
We finally use again that~$V$ is homogeneous of degree minus one.

We remark that the reader who feels uneasy with the singularity of~$V$ at the origin can
again smoothen~$V$ near the origin. For a spherically symmetric smooth function~$W(|\x|)$, the divergence equation
\[ \partial_\alpha u_\alpha(\x) = W(|\x|) \]
has the explicit solution
\[ u_\alpha(\x) = \frac{\x_\alpha}{2\, |\x|^3} \int_0^r r^2\: W(r)\: dr \:. \]
Evaluating the integral for the mollified Newtonian potential, one sees that the mollification can be
removed, giving~$u_\alpha(\x) = -\x_\alpha V(\x)$ for~$|\x|=R$. This gives a clean justification of the first
summand in~\eqref{tomollify}.
\QED

\begin{Lemma}
\begin{align*}
&\frac{3}{4}\: \bigg( \int_1^\infty - \int_{-\infty}^0 \bigg) \:d\tau 
\int_N d^3\y\: \bigg( \frac{\bxi^2}{|\x|} - \frac{( \x \bxi)^2}{|\x|^3}
\bigg)\;V|_{\tau \y + (1-\tau)\, \x}\: \L(\x,\y) \\
&=-\frac{1}{2}\: \delta^2\, \s_2\: \frac{1}{|\x|}\: V(\x)+ \O \Big( \frac{M \delta^3}{|\x|^3} \Big) \:.
\end{align*}
\end{Lemma}
\Proof Applying the transformations~\eqref{fliptrafo1} and~\eqref{fliptrafo2}, one sees that
\[ \int_{-\infty}^\infty \epsilon(\tau) \:d\tau 
\int_N d^3\y\: \bigg( \frac{\bxi^2}{|\x|} - \frac{( \x \bxi)^2}{|\x|^3}
\bigg)\;V|_{\tau \y + (1-\tau)\, \x}\: \L(\x,\y) = 0 \:. \]
Hence
\begin{align*}
&\bigg( \int_1^\infty - \int_{-\infty}^0 \bigg) \:d\tau 
\int_N d^3\y\: \bigg( \frac{\bxi^2}{|\x|} - \frac{( \x \bxi)^2}{|\x|^3}
\bigg)\;V|_{\tau \y + (1-\tau)\, \x}\: \L(\x,\y) \\
&=-\int_0^1d\tau \int_N d^3\y\: \bigg( \frac{\bxi^2}{|\x|} - \frac{( \x \bxi)^2}{|\x|^3}
\bigg)\;V|_{\tau \y + (1-\tau)\, \x}\: \L(\x,\y) \\
&=-\int_N d^3\y\: \bigg( \frac{\bxi^2}{|\x|} - \frac{( \x \bxi)^2}{|\x|^3}
\bigg)\;V(\x)\: \L(\x,\y) + \O \Big( \frac{M \delta^3}{|\x|^3} \Big) \:.
\end{align*}
Using again~\eqref{linLcomp} gives the result.
\QED

Collecting all the terms, we obtain
\begin{align*}
{\mathfrak{P}}(R)
&= -\s \,|\x| \, V(\x) + \frac{1}{6}\: \delta^2\, \s_2 \: \frac{1}{|\x|} \:V(\x)
-\frac{1}{2}\: \delta^2\, \s_2\: \frac{1}{|\x|}\: V(\x) \\
&\quad\: + \s\: |\x|\, V(\x) -\frac{1}{6\, |\x|}\: \delta^2\, \s_2\: V(\x) + \O \Big( \frac{M \delta^3}{|\x|^3} \Big) \\
&= -\frac{1}{2}\: \delta^2\, \s_2\: \frac{1}{|\x|}\: V(\x) + \O \Big( \frac{M \delta^3}{|\x|^3} \Big)
= \frac{M}{|\x|^2}\: \delta^2\, \s_2 + \O \Big( \frac{M \delta^3}{|\x|^3} \Big)\:.
\end{align*}
Using this result in~\eqref{MRdef} concludes the proof of Proposition~\ref{prpmassalt}.

\Thanks{{\em{Acknowledgments:}}
We are grateful to the referee for helpful comments and suggestions.
%%We are grateful to ... for helpful discussions.
%We are grateful to the ``Universit\"atsstiftung Hans Vielberth'' for generous support.
%N.K.'s research was also supported by the NSERC grant RGPIN~105490-2018.

\vspace*{1em}
\noindent

%\Felix{Nehme das noch heraus.}%
%{\bf{Declarations:}} Data sharing not applicable to this article as no datasets were generated or analyzed during the current study. The authors have read and accept the Publisher’s Data Use Privacy Policy and the Aries Privacy Policy. The authors comply to the highest scientific and ethical standards.
%
%The research leading to these results received funding from the NSERC grant RGPIN~105490-2018. No other funds, grants or other support was received. The authors have no relevant financial or non-financial interests to disclose. The authors have no competing interests or conflicts of interest
%to declare that are relevant to the content of this article.

%\bibliographystyle{amsplain}
%\bibliography{../../aarbeit/felix}
\providecommand{\bysame}{\leavevmode\hbox to3em{\hrulefill}\thinspace}
\providecommand{\MR}{\relax\ifhmode\unskip\space\fi MR }
% \MRhref is called by the amsart/book/proc definition of \MR.
\providecommand{\MRhref}[2]{%
  \href{http://www.ams.org/mathscinet-getitem?mr=#1}{#2}
}
\providecommand{\href}[2]{#2}

\end{document}